\documentclass[useAMS]{rsos}

\usepackage{upgreek}

\begin{document}

\title{Celebrating 30 Years of Science from the James Clerk Maxwell Telescope}

\author{
Ian Robson$^{1,2}$, Wayne S. Holland$^{1,2}$ and Per Friberg$^{3}$}

\address{$^{1}$UK Astronomy Technology Centre, Royal Observatory, Blackford Hill, Edinburgh, EH9 3HJ, UK\\
$^{2}$Institute for Astronomy, University of Edinburgh, Royal Observatory, Blackford Hill, Edinburgh, EH9 3HJ, UK\\
$^{3}$East Asian Observatory, 660 N. A`oh\={o}k\={u} Place, University Park, Hilo, HI 96720, USA}

\subject{Observational astronomy}


\corres{Ian Robson\\
\email{ian.robson@stfc.ac.uk}}


\begin{abstract} 

The James Clerk Maxwell Telescope (JCMT) has been the world's most successful single dish telescope at submillimetre wavelengths since it began operations 
in 1987. From the pioneering days of single-element photometers and mixers, through the first modest imaging arrays, leading to the state-of-the-art 
wide-field camera SCUBA-2 and the spectrometer array HARP, the JCMT has been associated with a number of major scientific discoveries. Famous for the 
discovery of ``SCUBA'' galaxies, which are responsible for a large fraction of the far-infrared background, to the first images of huge discs of cool 
debris around nearby stars, possibly giving us clues to the evolution of planetary systems, the JCMT has pushed the sensitivity limits more than any other 
facility in this most difficult of wavebands in which to observe. Now approaching the 30th anniversary of the first observations the telescope continues to 
carry out unique and innovative science. In this review article we look back on just some of the scientific highlights from the past 30 years.

\end{abstract}



\begin{fmtext}
\end{fmtext}


\maketitle

\section{Introduction and overview}

The James Clerk Maxwell Telescope (JCMT) has been the world's premier single-dish submillimetre telescope since its opening in 1987. At 13,425\,ft above 
sea-level on Mauna Kea in Hawaii it has the benefit of being above much of the water vapour that restricts ground-based, submillimetre astronomy to a few 
narrow ``windows'' through which observations are possible. It was a purpose-built facility with a 15\,m diameter, high-surface-accuracy primary mirror 
that feeds the incoming radiation to a receiver cabin at the Cassegrain focus behind the main dish and also to two Nasmyth platforms at each side of the 
elevation bearings.

\vskip 1mm

The scientific output of any astronomical facility is a combination of many factors: the intrinsic efficiency of the design, location and weather, 
instrumentation capability and reliability, operability and the effectiveness of the data-reduction software. Throughout the life of the JCMT, the 
observatory management strove to provide the world's best instrumentation, improve the reliability of the facility and to maximise the scientific output 
for the funding agencies and astronomical community through science-ranked, weather-dependant, queue-based flexible scheduling of observations. Indeed, the 
three funding agencies of the UK (55\%), Canada (25\%) and the Netherlands (20\%) deserve credit for their continued support of the developments of the 
facility and instrumentation even in times of severe financial pressure on home budgets.

\vskip 1mm

This article discusses the scientific output of the JCMT, doing so through the eyes of the instrumentation suite as this provides one of the most coherent 
ways of seeing how technology has driven the scientific capability and new discoveries emanating from the facility. We have subdivided this treatment into 
four sections: the three continuum instrument epochs of UKT14, SCUBA, SCUBA-2, and the heterodyne spectrometers, culminating in HARP. Within these 
sections the science output is naturally grouped into astronomical themes. Whilst it has not been possible to include all the science from the 30 years of 
operations, the science highlights readily stand-out.

\vskip 1mm

\begin{figure}[!h]
\centering
\includegraphics[width=120mm]{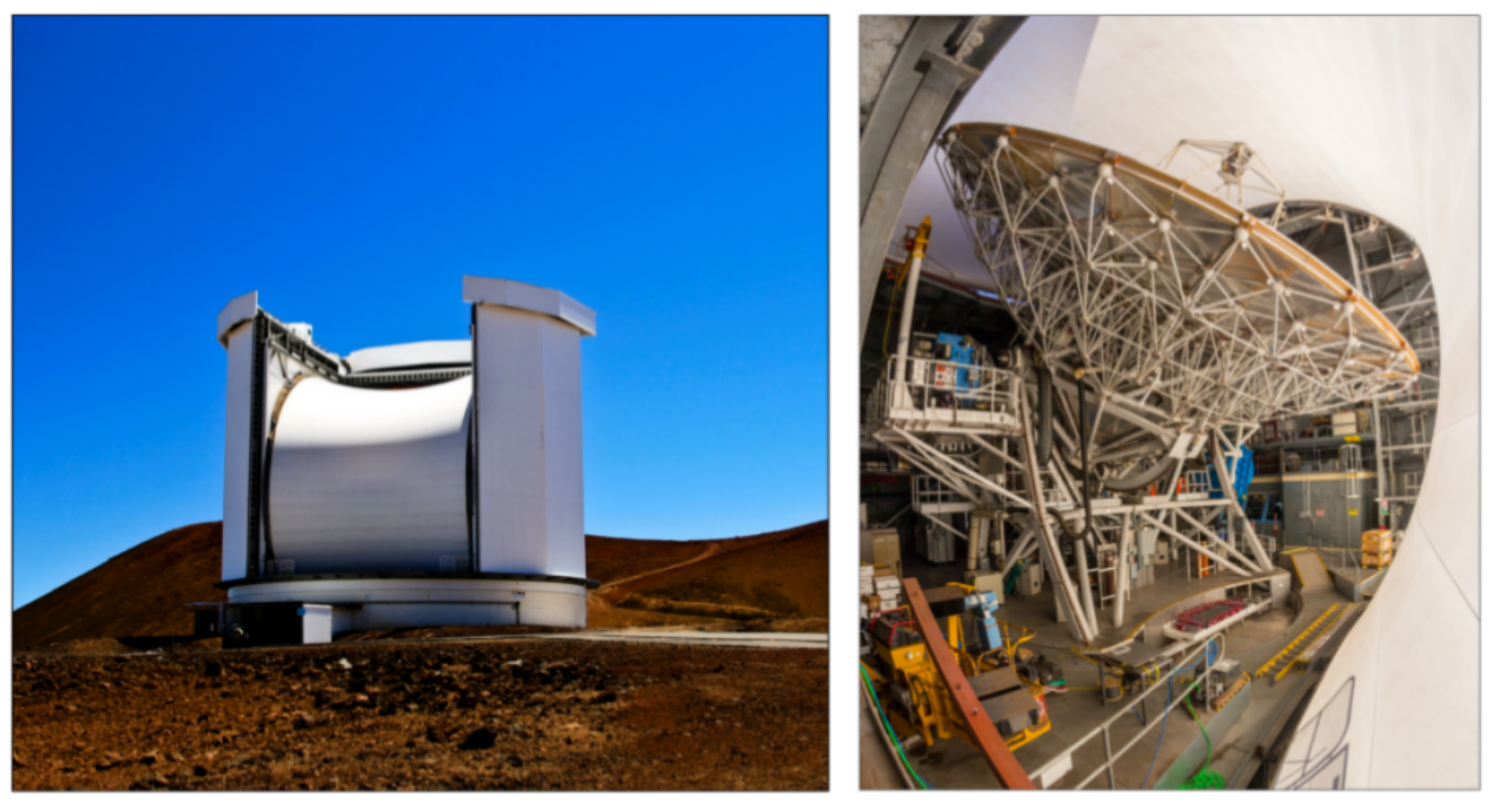}

\caption{(left) The JCMT in its usual mode of operation with a protective membrane in place (courtesy of Royal Observatory Edinburgh) (right) Inside the 
enclosure showing the main antenna and secondary mirror unit. The HARP heterodyne array is seen on the platform to the left, and SCUBA-2 on the right side 
(courtesy of East Asian Observatory). }

\label{fig:jcmt_images}
\end{figure}

\vskip 1mm

In terms of a timeline, the story can be broken down into some very clear regimes. In the continuum, we begin with the single-pixel, common-user 
bolometer, UKT14 \cite{Duncan1990} which reigned supreme in the world from 1988 until the arrival of SCUBA in 1997. SCUBA, the Submillimetre Common-User 
Bolometer Array \cite{Holland1999}, was the world's first submillimetre ``camera'' with 128 pixels in two arrays, one operating at a primary wavelength 
of 850\,$\upmu$m with 37 pixels and one at 450\,$\upmu$m with 91 pixels. It is not an overstatement to say that SCUBA brought about the ``submillimetre 
revolution'' in astronomy and the science highlights from this instrument are described in Section 3. The outstanding success of SCUBA immediately showed 
the need for a next-generation camera, one with many more pixels to provide a larger field-of-view and with improved sensitivity. This resulted in 
SCUBA-2 \cite{Holland2013}, a revolutionary instrument that was fraught with technological challenges. Nevertheless, it became operational on the 
telescope in 2011, and with over 5000 pixels at each of the main SCUBA wavelengths it heralded the onset of large-scale mapping of the submillimetre sky. 
SCUBA-2 continues in operation to this day and Section 4 describes the major inroads of science this instrument has brought, moving from the snapshots of 
discrete objects or mapping tiny areas of sky, to large-scale imaging, resulting in statistically significant samples of objects and addressing evolution 
across many scale-sizes and cosmological timescales.

\vskip 1mm

In terms of heterodyne spectroscopy the JCMT had a slower start and went through a number of iterations of single- and dual-pixel instruments operating 
in most of the submillimetre and near-millimetre atmospheric windows. The most successful of these was the 350\,GHz (850\,$\upmu$m) receiver RxB3 
\cite{Parker1991,Avery1992}, a dual-channel instrument receiving orthogonal linear polarisations from the same position on the sky, and which operated on 
the telescope between 1997 and 2006. The Digital Autocorrelating Spectrometer (DAS) \cite{Bos1986} was used in conjunction with many of the early 
heterodyne mixers. The arrival of the 350\,GHz 16-element Heterodyne ARray Program (HARP) receiver \cite{Buckle2009} heralded the ability to carry out 
high resolution spectroscopy over large areas of sky. This instrument was introduced in 2007 and came with a new digital spectral correlator, the Auto 
Correlation Spectrometer Imaging System (ACSIS) \cite{Buckle2009}. The heterodyne suite of instruments has proven to be very successful over the years 
and some of the major science results are described in Section 5.

\vskip 1mm

The eventual move to flexible scheduling of observations, based on the science priority of the proposal and the weather at the telescope, led to an 
enormous increase in productivity and in subsequent scientific impact of the facility \cite{Robson2002}. Although flexible scheduling was somewhat 
resisted by the users at the start, the eventual implementation meant that no longer were top-ranked proposals at the risk of being blighted by poor 
weather that happened to coincide with the fixed schedule of their observing run, but they would be undertaken throughout the semester when the weather 
was best suited to the scientific requirements. The JCMT was one of the first observatories to bring about this innovation and was a major operational 
advantage for SCUBA and the instruments that followed. Finally, the importance of an easily-accessible and user-friendly archive was duly recognised, 
particularly when the volume of data being generated increased significantly (e.g. with the introduction of SCUBA). Ths led to the creation of the JCMT 
Science Archive (JSA) \cite{Gaudet2008}, hosted at the Canadian Data Archive Centre \cite{Economou2011}. The JSA is designed to increase the productivity 
of the telescope by making not only the raw data, but also science-quality reduced images available to the JCMT and wider astronomy communities. This 
will allow the astronomers of the future to interrogate the data to explore, for example, time-dependent phenomena over the lifetime of the telescope.

\section{Scientific results from UKT14}

Although only a single-pixel device, UKT14 came with many improvements over its predecessors. It was constructed by the Royal Observatory Edinburgh and 
from the start it was designed as a ``common-user'' instrument and crucially came with a user-friendly data reduction software suite. It was originally 
designed for and operated on the United Kingdom Infrared Telescope (UKIRT) but when moved to the JCMT brought more than a 100 times increase in sensitivity 
over UKIRT along with an increase of a factor of 4 in angular resolution, with beam sizes (full-width at half-maximum) of 14 arcseconds at 
850\,$\upmu$m and 6 -- 7 arcseconds at 350/450\,$\upmu$m. The instrument had very carefully designed optics to minimise stray radiation and a range of 
filters to select the atmospheric ``windows'' allowing photometric observations to be made from 2\,mm to 350\,$\upmu$m. This was very important for studies 
of spectral energy distributions, albeit the most often used filters tended to be in the most stable windows in the submillimetre at 800 and 450\,$\upmu$m. 
It was a very sensitive photometer, the composite germanium bolometer being cooled to 0.35\,K by liquid He$^3$, and capable of detecting point-sources down 
to a level of $\sim$8\,mJy at 800\,$\upmu$m in one hour of observing time. UKT14 turned out to be critically important: it was state-of-the-art in the 
early days of the JCMT; it opened up a range of new science ventures for submillimetre study; but perhaps most importantly, it introduced a whole new 
generation of astronomers to the field, many of whom were not necessarily submillimetre astronomy experts. Indeed, these were the same astronomers who 
would go on to make revolutionary discoveries with the introduction of SCUBA.

\vskip 1mm

The science output from UKT14 was indeed huge, both in extent and depth. According to the JCMT Annual Reports, over 180 papers in refereed journals 
contained data from UKT14 before it was superseded by SCUBA in 1996. This is a staggering 44\% of all the JCMT refereed papers over the same period, 
showing the dominance of UKT14 and continuum science in the first decade of the JCMT. The science topics ranged from observations of comets in the Solar 
System to high-redshift galaxies, mostly detecting the thermal emission from cold dust grains. There was a long-standing and very successful 
program of monitoring the flaring emission from blazars, however, which originated from non-thermal emission from relativistic electrons. Because ground-based 
submillimetre astronomy was still in its infancy, UKT14 was also used extensively for assessing observing techniques, studying the atmospheric extinction 
and identifying calibration sources, all of which would provide the sound basis for visiting astronomers to build upon \cite{Stevens1994a}. In compiling 
examples of UKT14 science, we have tried to show the breadth of the different astrophysical topics opened up by this remarkably versatile instrument, and 
so present a wide but relatively shallow selection of topics, albeit selected mainly through citation indexes.

\subsection{Solar System studies}

One of the early observations with UKT14 was the first detection of a comet at submillimetre wavelengths in 1989 by Jewitt \& Luu \cite{Jewitt&Luu1990}. 
They found that the emission of comet P/Brorsen-Metcalf could be modelled by a population of transient, large grains with a total mass of 
$\sim$10$^9$\,kg, which could have been produced by some form of breakdown of part of the refractory mantle of the comet. Later observations of the comet 
Hyakutake in 1996 by Jewitt \& Matthews \cite{Jewitt&Matthews1997} found that from 1.1\,mm to 350\,$\upmu$m the emission can be described as thermal 
emission from large ($\sim$1\,mm) dust grains in the coma and a resulting total mass of around 2 $\times$ 10$^9$\,kg. The spectral index indicates that 
the opacity factor is similar to that found in the circumstellar discs of young stars. Remarkably, a small map was made at 800\,$\upmu$m, which showed 
that the emission was consistent with the steady emission of solid particles from the cometary nucleus on timescales less than 1 day. A major study by 
Redman and co-workers of the asteroid 4-Vesta in 1989 \cite{Redman1992} showed that the submillimetre emission might originate from a form of dusty, 
porous regolith. Furthermore, unlike the single-peaked rotational light-curve in the optical, the millimetre light-curve was seen to be double-peaked, 
indicating that it was most likely dominated by the triaxial shape of the asteroid. A major investigation was undertaken by Griffin \& Orton 
\cite{Griffin&Orton1993} who measured the emission from Uranus and Neptune from 2\,mm to 350\,$\upmu$m. These precise data allowed the brightness 
temperatures of the planetary atmospheres to be calculated with greater accuracy (with uncertainties of $<$2\,K) based on data from Mars, the primary 
calibration source at submillimetre wavelengths. These new values enabled Uranus to become a valuable calibrator for submillimetre observations, both for 
ground- and space-based facilities. Addressing both calibration purposes and intrinsic properties of asteroids, M\"{u}ller and Lagerros \cite{Muller1998} 
used JSA data on 1-Ceres, 2-Pallas, 4-Vesta, 532-Herculina, 10-Hygiea, 106-Dione and 313-Chaldaea to determine thermal models for the asteroids as well 
as defining new far-infrared (far-IR) photometric standards to be used by the ISOPHOT instrument on the \emph{ISO} satellite. This demonstrated the value 
and accessibility of the JCMT science archive. The work also built on a programme of observations that produced a major publication by Redman, Feldman \& 
Matthews \cite{Redman1998} in which the spectral energy distribution was obtained for seven asteroids, five of which were non-metallic and two were 
metallic. The data showed that there was a notable range of physical properties of the surfaces, even for the non-metallic bodies.

\vskip 1mm

To conclude this section Stern, Weintraub \& Festou \cite{Stern1993} succeeded where many had previously failed and detected Pluto at 1.3\,mm and 
800\,$\upmu$m, deducing a surface temperature of 30 -- 44\,K with a most probable range of 35 -- 37\,K. This range is significantly lower than had been 
predicted from radiative equilibrium models and from other observations and showed that the methane features in Pluto's spectrum were from solid, rather 
than gas-phase, absorptions, demonstrating that Pluto's atmosphere is dominated by nitrogen or carbon monoxide rather than methane.

\subsection{Star formation}

\begin{figure}[!h]
\centering
\includegraphics[width=120mm]{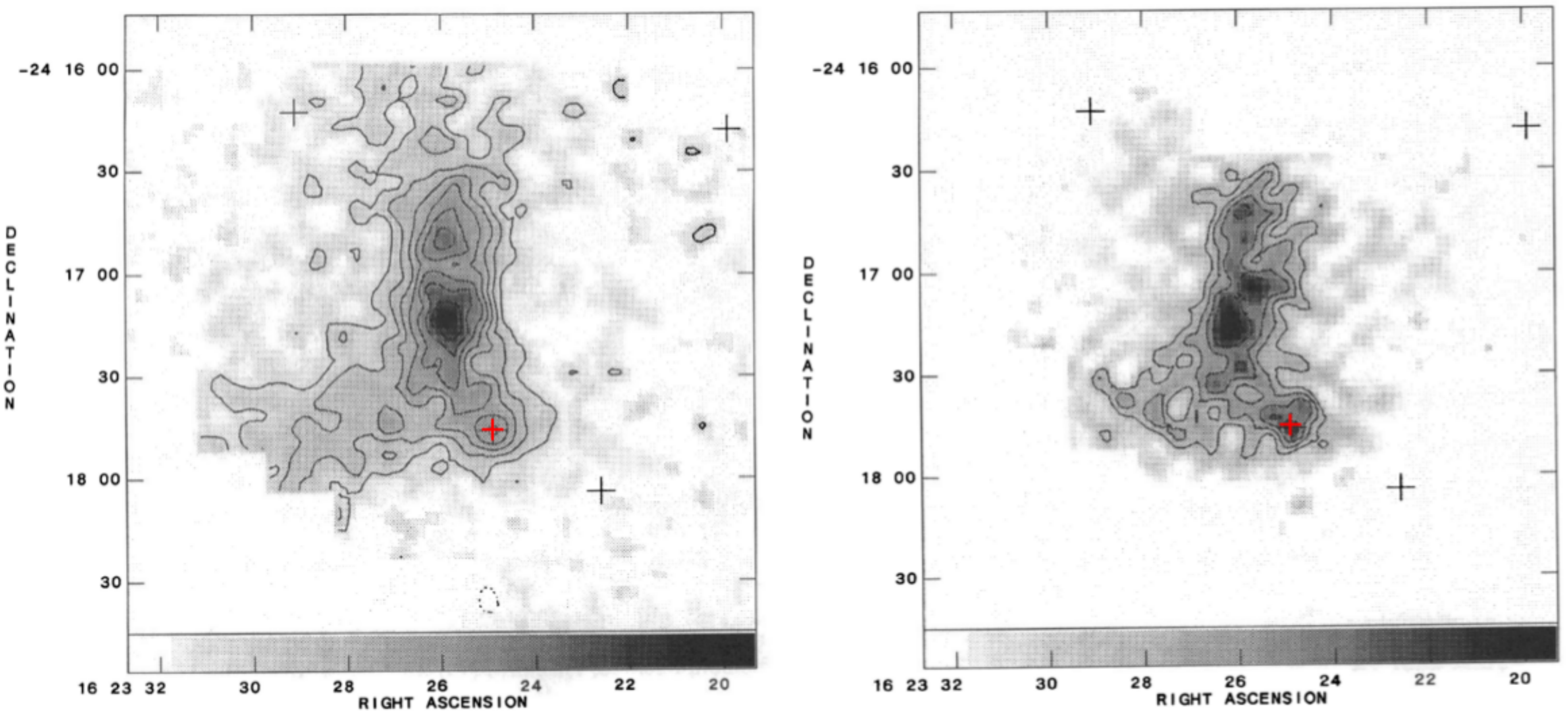}

\caption{The $\rho$ Ophiuchus A molecular cloud at 800\,$\mu$m (left) and 450\,$\mu$m (right). The red cross towards the lower right of the 
cloud extent marks the position of the Class 0 protostar, VLA 1623. Figure from Andr\'{e}, Ward-Thompson \& Barsony \cite{Andre1993}.}

\label{fig:RhoOph_UKT14}
\end{figure}

The very early stages of star formation, detecting the emission from cold dust, turned out to be one of the key areas for UKT14 study and was one of the 
most popular series of targets resulting in many publications over the period. One of the most spectacular set of observations and indeed, the most cited 
UKT14 result, came from maps by Andr\'{e}, Ward-Thompson \& Barsony \cite{Andre1993} of the $\rho$ Ophiuchus molecular cloud in which the protostellar 
source VLA 1623 was proposed as a new category, ``Class 0'', as the earliest phase in the star formation sequence. The observations of the core of the 
cloud at 800\,$\upmu$m and 450\,$\upmu$m took 18 hours of integration over three nights of excellent and stable weather. The results are shown in 
Fig.~\ref{fig:RhoOph_UKT14}. The emission shows four compact clumps with masses $\sim$1 solar masses (M$_\odot$) embedded in a ridge of about 
15\,M$_\odot$. VLA 1623 is the coldest clump with a temperature estimated to be 15 -- 20\,K, appearing to have no central heat source and not detected by 
the \emph{IRAS} satellite. The mass was estimated to be 0.6\,M$_\odot$ with a luminosity of around the same as the Sun. These observations showed the 
potential of the submillimetre for uniquely being able to study the earliest phases of star formation and this paper was a landmark in the field.

\vskip 1mm

\vskip 1mm

The observations of $\rho$ Oph were rapidly followed up with the next most cited UKT14 paper, in which Ward-Thompson and co-workers \cite{Ward-Thompson1994} 
made observations of 21 cold molecular cores in dark clouds with no infrared source; the so-called ``starless cores''. These clumps have insufficient 
bolometric luminosity to be typical of a ``Class 1'' protostar and a crucial discovery was that these cores differed from those that had an \emph{IRAS} 
far-IR detection in that they are all more diffuse and less centrally peaked. The clumps had densities 10$^5$ to 10$^6$\,cm$^{-3}$ but the density 
profile was inconsistent with the $r^{-2}$ or $r^{-3/2}$ profiles predicted by standard theory and instead were more consistent with magnetic support. The 
authors concluded that these submillimetre bright, dark cores are indeed pre-protostellar and are in the very earliest stages prior to protostellar 
collapse.

\vskip 1mm

Saraceno et al. \cite{Saraceno1996} conducted a series of observations of a sample of 45 Class I and Class 0 young stellar objects (YSOs) at 1.3\,mm 
using UKT14 and the Swedish-ESO Submillimetre Telescope (SEST) telescope and made a number of important conclusions regarding their evolutionary 
sequencing. These included that the evolution of a protostar was mainly controlled by the mass of both the central object and circumstellar material and 
that the Class 0 sources were indeed the earliest stages of star formation yet observed. Also, the Class I sources showing outflow had dynamical 
timescales exceeding ten thousand years and that they were probably in the deuterium burning phase where they spend most of their lifetime accreting 
material. On the other hand, the Class I sources with no outflow behave like Class II sources with no outflow, and the authors suggest that these are 
most probably Class II sources suffering high extinction from foreground emission. Further examples of early phases of star formation can be found in the 
search for protostellar cores in Bok globules \cite{Launhardt1997} and in \emph{IRAS} sources \cite{Zinnecker1992,Sandell1991}.

\vskip 1mm

Although the Orion complex of clouds and star formation were a notable source of study, perhaps surprisingly they hardly feature in the highly-rated 
citations of UKT14. Chini and co-workers \cite{Chini1997} produced probably the most definitive study of submillimetre emission from OMC-2 and OMC-3 
detecting six probable Class 0 protostars as well as describing the general dust morphology and temperature of these complexes. To conclude this section on 
protostars, the value of using UKT14 with its associated polarimeter \cite{Flett&Murray1991} to detect the polarisation properties of protostars was 
demonstrated by Holland et al. \cite{Holland1996} who observed the polarised 800\,$\mu$m emission from aligned dust grains in the prototypical Class 0 
source VLA 1623 and in Sharples 106-IR, a high-mass, YSO along with its associated protostar S\,106-FIR. For VLA 1623, the magnetic 
field was found to be almost exactly perpendicular to the highly collimated CO outflow, suggesting that the outflow is not collimated by the magnetic 
field. However, for the S\,106 region, the situation was more complex, and it was clear that more extensive imaging would be the way forward, and that 
would have to wait for SCUBA.

\subsection{Stars and discs}

The study of nearby stars and their associated circumstellar discs attracted a lesser degree of attention, probably because most of the observations turned 
out to be very difficult and at the limit of the capability of UKT14. On the other hand, the observations provided unique insights into a number of 
phenomena. Mannings \& Emerson \cite{Mannings&Emerson1994} observed six T Tauri stars to investigate the dusty, circumstellar discs surrounding these stars 
in the early stages of stellar evolution using UKT14's full filter set of 2\,mm to 350\,$\upmu$m. For the optically thin sources, the spectral indexes 
indicated that the dust grains were larger than found in the interstellar medium, suggesting that grain growth in the protoplanetary discs had already occurred 
and was ongoing. The authors estimated that the rate of growth was of order 10$^6$ M$_\odot$ per year.

\vskip 1mm

The Vega phenomenon (excess thermal emission above that expected from the the stellar photosphere at far-IR wavelengths) was first discovered by the 
\emph{IRAS} satellite and subsequently investigated by several UKT14 observational campaigns. The breakthrough was made by Zuckerman \& Becklin 
\cite{Zuckerman&Becklin1993} who made the first detections of excess 800\,$\upmu$m emission from the stars Vega, $\beta$ Pictoris and Fomalhaut. Sylvester 
and co-workers \cite{Sylvester1996} followed up with observations from 2\,mm to 450\,$\upmu$m in a major paper describing the far-infrared emission from a 
large sample of stars including Vega. Nine stars were detected at millimetre/submillimetre wavelengths and the data suggested that the surrounding dust was 
generally more likely to be in a ring rather than a spherical cloud and that they were composed of larger grains than found in the interstellar medium. 
Although there remained some uncertainty about the precise evolutionary state of all of the stars (some may have been younger than main sequence), 
nevertheless, these important papers pointed the way forward to some of the earliest and most important observations that SCUBA would make (see Section 3).

\vskip 1mm

The extended dusty envelope of five highly evolved stars (including one planetary nebula) were observed by Knapp, Sandell \& Robson \cite{Knapp1993} who 
found that the spectral index was just less than unity for all the sources, irrespective of whether the envelope of the star was carbon or oxygen rich. The 
gas-to-dust ratio was calculated to be around 100. Observations of one of the sources (CRL618) suggested that it might be slowly variable, which was 
important as one of the primary aims of this programme was to investigate whether these sources might be suitable as calibrators in the 
millimetre/submillimetre region. Finally, addressing the latest stages of stellar evolution Williams and co-workers \cite{Williams1997} used UKT14 as part 
of a multifrequency study of the Wolf-Rayet system WR 147 and showed that the presence of non-thermal emission between the two stars was most probably dues 
to colliding winds in the system.

\subsection{The Galactic Centre}

The Galactic Centre region was a difficult target for UKT14 due to the complex emission over an extended region and the southerly declination. Dent and 
collaborators \cite{Dent1993} made maps of the 10 -- 20\,pc region at 1100\,$\upmu$m and 800\,$\upmu$m, as well as the inner region at 450\,$\upmu$m. The 
2\,pc inner ring was clearly detected as was the dust emission from three giant molecular clouds, which seemed to be connected by a ridge of thermal 
emission. The 2\,pc ring revealed a two-component structure in the submillimetre: northern and southern emission, which were bounded by the radio continuum 
spiral arms. The initial results were followed up by Zylka et al. \cite{Zylka1995} who made maps with UKT14 at 800\,$\upmu$m, 600\,$\upmu$m and 
450\,$\upmu$m of the 2 arcminute region surrounding the strong radio source Sgr A*. This was one of the rare examples of the difficult-to-calibrate 
600\,$\upmu$m data being used from UKT14, especially in a map. A number of important conclusions were derived from these observations: that warm dust 
emission was definitely responsible for most of the far-IR emission from the region and that the heating of the dust was not from the central 
supermassive black hole but from a cluster of hot and luminous stars in the central parsec region.

\subsection{Dust emission in external galaxies}

Although the \emph{IRAS} satellite had detected strong dust emission from many relatively nearby galaxies, the low surface brightness and the small size of 
the JCMT beam in comparison made UKT14 detections relatively difficult. Chini et al. \cite{Chini1995} carried out observations at 800\,$\upmu$m and 
400\,$\upmu$m of seven of the 32 spiral galaxies that were previously mapped at 1.3\,mm by the IRAM 30\,m telescope. It was found that the sample split into 
two, with one half being dominated by cold interstellar dust with a temperature of $\sim$20\,K, while for the rest, much colder dust ($\sim$10\,K) 
dominated. The ratio of the infrared luminosity to the gas mass turned out to be equivalent to the star formation rate in the Milky Way. Fich \& Hodge 
\cite{Fich&Hodge1993} observed a sample of 22 early-type galaxies detected by \emph{IRAS} and managed to detect six of them with tight upper limits being 
obtained on a further eight. These allowed upper limits to be determined for the dust temperatures, the upper limit being principally because of the very 
extended size of the \emph{IRAS} beam compared to that of UKT14 on the JCMT. Depending on the value of the emissivity, these temperatures lay between 20\,K 
and 40\,K.

\vskip 1mm

\begin{figure}[!h]
\centering

\includegraphics[width=120mm]{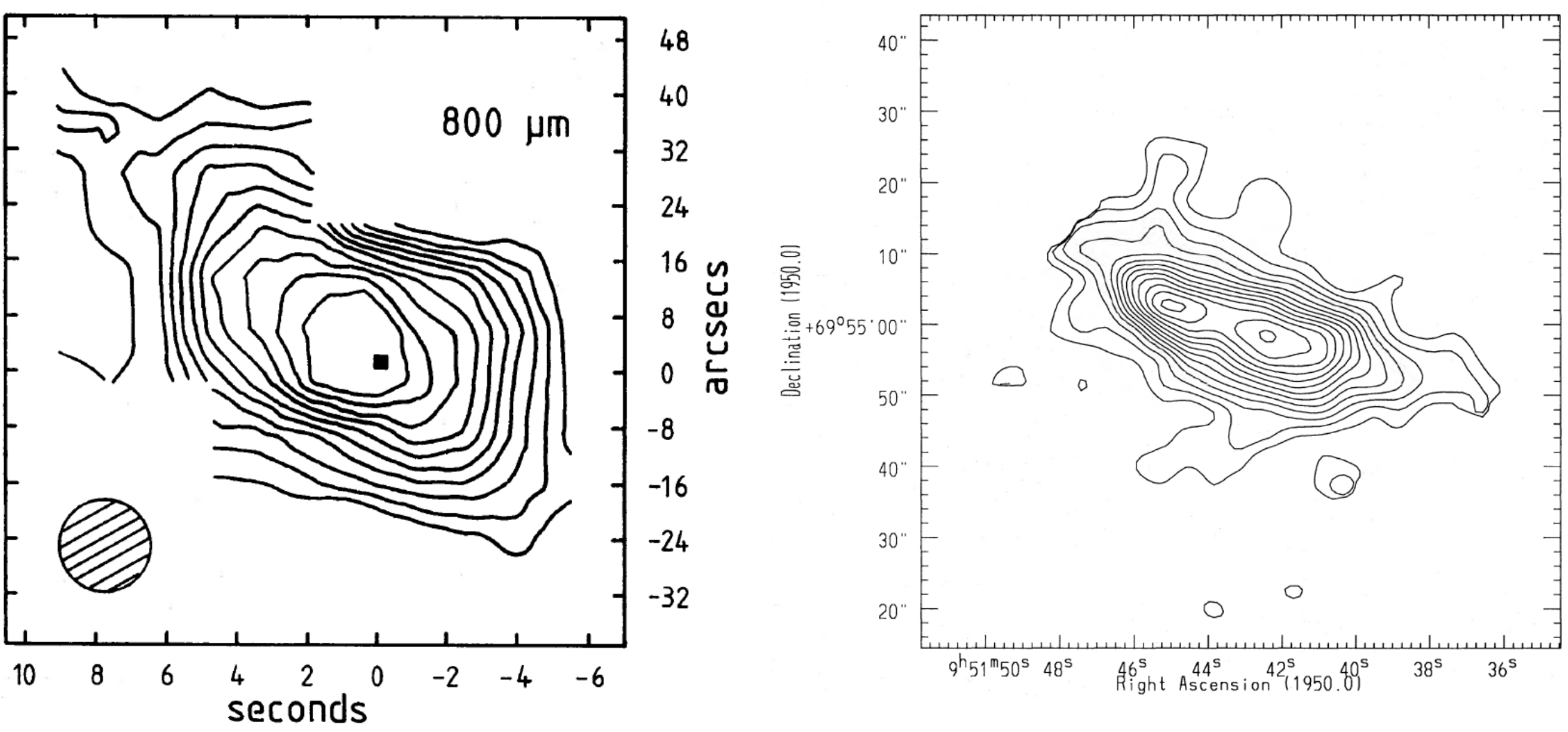}

\caption{UKT14 imaging of the starburst galaxy M82 at 800\,$\upmu$m (left) and at 450\,$\upmu$m (right). The higher angular resolution afforded by the 
450\,$\upmu$m observations revealed the double-peaked central structure for the first time at submillimetre wavelengths. Figures from Hughes et al. 
\cite{Hughes1990},\cite{Hughes1994}.}

\label{fig:M82_UKT14}
\end{figure}

Mapping of nearby galaxies was a difficult proposition, requiring the most stable and driest conditions and so the results were relatively sparse given the 
fickle nature of matching observing conditions to requirements with the fixed-date scheduling at the time. However, two programmes stand out. Hughes, Gear 
\& Robson \cite{Hughes1990} followed up their earlier maps of the nearby starburst galaxy M82 at 1100\,$\upmu$m and 800\,$\upmu$m by succeeding in making a 
diffraction-limited map of the 1.5\,kpc diameter nuclear regions at 450\,$\upmu$m \cite{Hughes1994}. The 9 arcsecond resolution of the map showed that the 
thermal emission from the central dust cloud was double in structure and the dust temperature was around 48\,K. Both the earlier 800\,$\upmu$m map and 
450\,$\upmu$m image are shown in Fig.~\ref{fig:M82_UKT14}. Hawarden and co-workers \cite{Hawarden1993} made an 800\,$\upmu$m map of the central region of the 
nearby radio galaxy NGC 5128 (Centaurus A); a difficult observation due to the very southerly declination of the source. The map was combined with 
photometry at all the UKT14 filters apart from 600\,$\upmu$m and the results showed that the non-thermal central emission was surrounded by a circumnuclear 
torus of dust. Farther out from the centre extended dust emission was observed and even farther out, the dark optical dust lanes in the galaxy were also 
detected.

\vskip 1mm

\vskip 1mm

Many programmes sought to detect submillimetre thermal emission from dust in radio quiet quasars and active galactic nuclei (AGN) and in 1992, Barvainis, 
Antonucci \& Coleman \cite{Barvainis1992} made the breakthrough by detecting the Seyfert I galaxy PG1434+590 and the gravitationally lensed quasar 1413+117 
(the ``Cloverleaf'') at 450\,$\upmu$m and 350\,$\upmu$m, the latter was also detected at 800\,$\upmu$m. The measured spectral indexes favoured thermal 
emission from dust but non-thermal, synchrotron emission could not be completely ruled out. At a redshift of 2.546 the strong suggestion was that this was 
the emission from typical far-IR galaxies red-shifted into the submillimetre. This was a milestone observation. At the same time, Clements et al. 
\cite{Clements1992} detected the high-redshift ($z$ = 2.286) \emph{IRAS} galaxy 10214+4724 at 800\,$\upmu$m and 450\,$\upmu$m. The authors concluded that the 
dust-enshrouded Seyfert model and the primeval galaxy model were both excluded by their observations but the submillimetre emission was consistent with a 
massive starburst of around 100\,M$_\odot$ per year. However, the observations were unable to determine whether this particular starburst was responsible 
for the formation of a significant fraction of the stars in the galaxy and indeed, much of the dust probably existed from previous bursts of star 
formation. This work was immediately followed up by Hughes, Robson, Dunlop \& Gear \cite{Hughes1993} who detected three out of a sample of ten 
\emph{IRAS}-selected radio-quiet quasars at 800\,$\upmu$m and 450\,$\upmu$m. In this case the very steep spectral index confirmed that the emission was 
indeed thermal radiation from warm dust and conclusively ruled-out any significant non-thermal synchrotron contribution. The dust grain temperature was 
estimated to be between 45 -- 60\,K.

\vskip 1mm

\emph{IRAS} detections provided the sources for many samples of submillimetre observations and a number of multi-wavelength investigations of ultraluminous 
\emph{IRAS} galaxies (ULIRGs) were undertaken using UKT14. An example of one of these was by Rigopoulou, Lawrence \& Rowan-Robinson \cite{Rigopoulou1996} 
who detected 9 out of the ten brightest ULIRGs at 350\,$\upmu$m, 450\,$\upmu$m, 800\,$\upmu$m and 1.1\,mm. One of the key aims of this programme was to 
determine whether the far-IR emission was powered by thermal emission from dust in a starburst (as in the above), or by accretion onto a central 
supermassive black hole. The observations showed that for these galaxies, the submillimetre emission was consistent with thermal emission. Although two of 
the objects might have housed an AGN, nevertheless, their very weak X-ray emission implied a very large optical depth and so again argued strongly against 
accretion being responsible for the far infrared luminosity.

\vskip 1mm

Moving on to radio galaxies, in 1994 Dunlop and collaborators \cite{Dunlop1994} investigated the concept that the stars in elliptical galaxies were 
believed to have been formed in a rapid bust of star formation in the early Universe. They used UKT14 observations to search for evidence in the form of 
thermal emission from dust from the anticipated starburst phase in these distant galaxies; this same dust making them very faint for optical studies. From 
a sample of six high redshift radio galaxies they were successful in detecting the redshift 3.8 galaxy, 4C41.17, at 800\,$\upmu$m with a strong upper limit 
at 450\,$\upmu$m. These two points were sufficient to establish thermal emission from dust at a temperature around 40\,K and that the dust mass was >10 
times that found in corresponding low-redshift radio galaxies. The dust luminosity was $\sim$5 $\times$ 10$^{13}$ solar luminosities (L$_\odot$), 
corresponding to a starburst of a few thousand M$_\odot$ per year. The authors concluded that the observations were consistent with around 10\% of the 
eventual mass of the galaxy still to be converted into stars and that it was possible that we were witnessing the end-product of star formation that had 
begun at much higher redshifts. However, it was not possible to say more about the formation epoch of elliptical galaxies per se. This was probably the 
earliest UKT14 observation relating to cosmology and the epoch of galaxy formation and early evolution.

\vskip 1mm

These exciting high-redshift studies were continued and the context of unambiguously identifying dust emission from relatively low signal-to-noise 
observations were discussed in detail by Hughes, Dunlop and Rawlings \cite{Hughes1997}, reviewing the observations to date, including some recent and very 
faint detections at 800\,$\mu$m from their on-going programme. This was an important overview, especially in the context of the upcoming SCUBA instrument.

\subsection{Flat spectrum radio sources and blazars}

Extensive programmes of observations of flat-spectrum radio sources, in particular blazars, were a feature of UKT14 on the JCMT, following on from earlier 
work on UKIRT. These observations sought to answer specific questions for these relativistic-jet dominated sources: to determine the ``snapshot'' spectral 
energy distribution from radio to gamma-ray regions; to monitor and determine the variability behaviour; to test theoretical models such as the 
``shock-in-jet'' model of Marscher \& Gear \cite{Marscher&Gear1985}. Being bright, these sources were readily observable in many weather conditions and 
also required only short integration times. They were also a feature of the Discretionary Time available to the Director of the telescope. They featured 
prominently in many coordinated multi-wavelength campaigns from the radio to gamma-rays.

\vskip 1mm

Stevens et al. \cite{Stevens1994} published the last in the series of multi-frequency observations of blazars in 1994 which covered 17 blazars at 
wavelengths from 800\,$\upmu$m to 13\,mm. Good agreement was found between the variability and the shock-in-jet models and it was found that the flares in 
the BL Lac objects tended to reach a maximum at a longer wavelength than those of the optically violently variable (OVV) quasars. This might indicate a 
stronger shock in the former objects. The data also showed that the flaring behaviour was complex, with multiple maxima and flickering being present and 
that ``clean'' flares were relatively rare, all of which indicated greater temporal sampling being needed.

\vskip 1mm

Gear and collaborators \cite{Gear1994} further investigated the differences between the BL Lacs and OVVs through a large sample of 22 of the former and 24 
of the latter. Quasi-simultaneous data were obtained across a wide wavelength range, which showed that the overall spectral shape was relatively consistent 
across all the sources. While this indicated that the same basic mechanism was at work in both classes, however, as noted above, a clear difference was 
again found in the millimetre-region spectra indicating a subtle difference in the jet properties between the two classes of flat-spectrum radio source. The 
authors suggested that the parent sources for the two classes might be the Fanaroff-Riley Class I and Class II sources, in which case submillimetre 
polarimetry would be an acid test for the future.

\vskip 1mm

This was eventually undertaken using the UKT14 polarimeter to observe 26 flat-spectrum radio sources at 1.1\,mm and 800\,$\upmu$m \cite{Nartallo1998}. 
Although a significant level of linear polarisation was detected in most sources (of order 10 -- 15\%) the magnetic field seemed less well-ordered on 
sub-parsec scales than on parsec scales and in the most highly ordered cases it was perpendicular to the jet axis. No significant difference was found 
between the BL Lac and the flat-spectrum quasars and whilst the emission from many of the most highly-polarised sources could be well-fitted by 
shock-in-jet models, for most sources this was not the case. Conical shock models seemed to be the best descriptor for the diverse emission from the jets 
in the sample.

\vskip 1mm

Extensive multi-frequency observing campaigns from radio to gamma-rays were carried out on particular sources, for example, results from the quasar 3C279 
were reported by Hartman and co-workers between 1996 and 2001 \cite{Hartman1996, Hartman2001}. These extensive observations showed that the variability was 
very complex, with different correlations being seen for different flares. The spectra could be best modelled with a relativistic electron dominated jet 
with gamma-ray production arising through a combination of synchrotron self-Compton and external Compton processes. Interestingly, when 3C279 was in its 
high state, the gamma-ray luminosity dominated everything else by at least a factor of ten.

\vskip 1mm

Another very popular quasar for multi-frequency observations was 3C273, typified by the work by Robson et al. \cite{Robson1993} who reported on a four-year 
observing campaign from infrared through centimetre wavelengths during which a number of flares were seen. A period of relative inactivity allowed the 
quiescent spectrum to be obtained, from which flaring behaviour could be subtracted to determine the flare emission itself. Caution was noted in that the 
behaviour of a flare was critically dependent on the temporal overlap of the observations at differing wavelengths but where these were simultaneous the 
infrared emission preceded that at longer wavelengths and there was distinct evidence for the evolution of the turnover of the flare to propagate to longer 
wavelengths. The emission between the infrared and 2\,mm wavelengths was commensurate with a single synchrotron component associated with the innermost 
part of the relativistic jet or the injection zone itself. A major paper by Turler et al. in 1999 \cite{Turler1999} reported on thirty years of 
multi-wavelength monitoring of 3C273 showing the complex behaviour of this source and providing a database for emission modelling purposes.

\vskip 1mm

With the demise of UKT14 and the introduction of SCUBA, these programmes tended to fall in popularity due to the high impact of imaging science compared to 
single-pixel photometry. However, the work exploring dust emission from radio quiet galaxies and AGNs at cosmological redshifts was ideally suited to 
SCUBA, as will be seen in the next section.


\section{The first camera arrays: the SCUBA era}

SCUBA, the Submillimetre Common-User Bolometer Array, built by the Royal Observatory Edinburgh, was in the late 1990's the most versatile and powerful of a 
new generation of submillimetre cameras \cite{Holland1999}. It uniquely combined a sensitive dual-waveband imaging array with a three-band photometer, and 
had a sensitivity background-limited by the emission from the Mauna Kea atmosphere and telescope at all observing wavelengths from 350\,$\upmu$m to 2\,mm. 
The increased sensitivity and array size mean that SCUBA mapped $\sim$10,000 times faster than UKT14 to the same signal-to-noise. Most importantly, SCUBA 
was a facility instrument, open to the world community of users, and was provided with an unprecedented high level of user support.

\vskip 1mm

The dual-camera system consisted of a short-wavelength (SW) array of 91 pixels and a long-wavelength (LW) array of 37 pixels. Each array had approximately 
the same field-of-view on the sky (2.3\,arcmin in diameter) and could be used simultaneously by means of a dichroic beamsplitter. The SW array was optimised 
for operation at 450\,$\upmu$m (but could also be used at 350\,$\upmu$m), whilst the LW array was optimised for 850\,$\upmu$m (with observations at 
750\,$\upmu$m and 600\,$\upmu$m also possible). The array pixels were arranged in a close-packed hexagon, with the photometric pixels positioned around the 
outside of the LW array. The detectors were cooled to $\sim$100\,mK using a dilution refrigerator, ensuring close to a factor of 10 increase in sensitivity 
per pixel over UKT14. SCUBA was delivered to the JCMT in April 1996, and first light on the telescope was obtained in July. After extensive commissioning, 
the first astronomical data for the community were taken in May 1997 using two modes of operation: photometry and ``jiggle-mapping'', the latter using 
novel movement of the secondary mirror to create a fully-sampled image. The final major mode of data acquisition, ``scan-mapping'', was released in 
February 1998. Although issues with the filter drum meant that SCUBA became a fixed 450/850\,$\upmu$m imager by the early 2000's, the 
popularity of the instrument remained very high during its entire lifetime.

\subsection{New perspectives on galaxy formation and evolution}

In late 1997 SCUBA made several monumental discoveries, particularly in the area of galaxy formation and evolution. Capitalising on a spectacular period of 
good weather on Mauna Kea (the El Ni\~{n}o event of late 1997/early 1998) observations revealed a population of galaxies responsible for at least part of 
the far-IR background, detected a number of high redshift galaxies and provided new insights into galaxy evolution. In the next section we summarise 
just a few of these high profile discoveries.

\subsubsection{``SCUBA galaxies''} 

The most vigorously star-forming galaxies in the nearby Universe are also those in which dust obscuration is the most significant. It was long suspected, 
therefore, that the early evolution of galaxies would take place inside shrouds of dust. The first deep SCUBA maps outside of the Galactic Plane 
immediately confirmed this suspicion, revealing a large population of hitherto unknown, star-forming galaxies. This discovery was reported by Smail, Ivison 
\& Blain \cite{Smail1997} in a series of targeted observations towards lensed galaxy clusters, exploiting the amplification of all background sources by 
the clusters. The authors concluded that the observed source counts needed a significant increase in the number density of star-forming galaxies in the 
high redshift Universe and suggest that previous optical surveys may have underestimated the star formation density by a large factor. This work 
was the first peer-reviewed paper to emerge from SCUBA observations.

\vskip 1mm

\begin{figure}[!h]
\centering

\includegraphics[width=120mm]{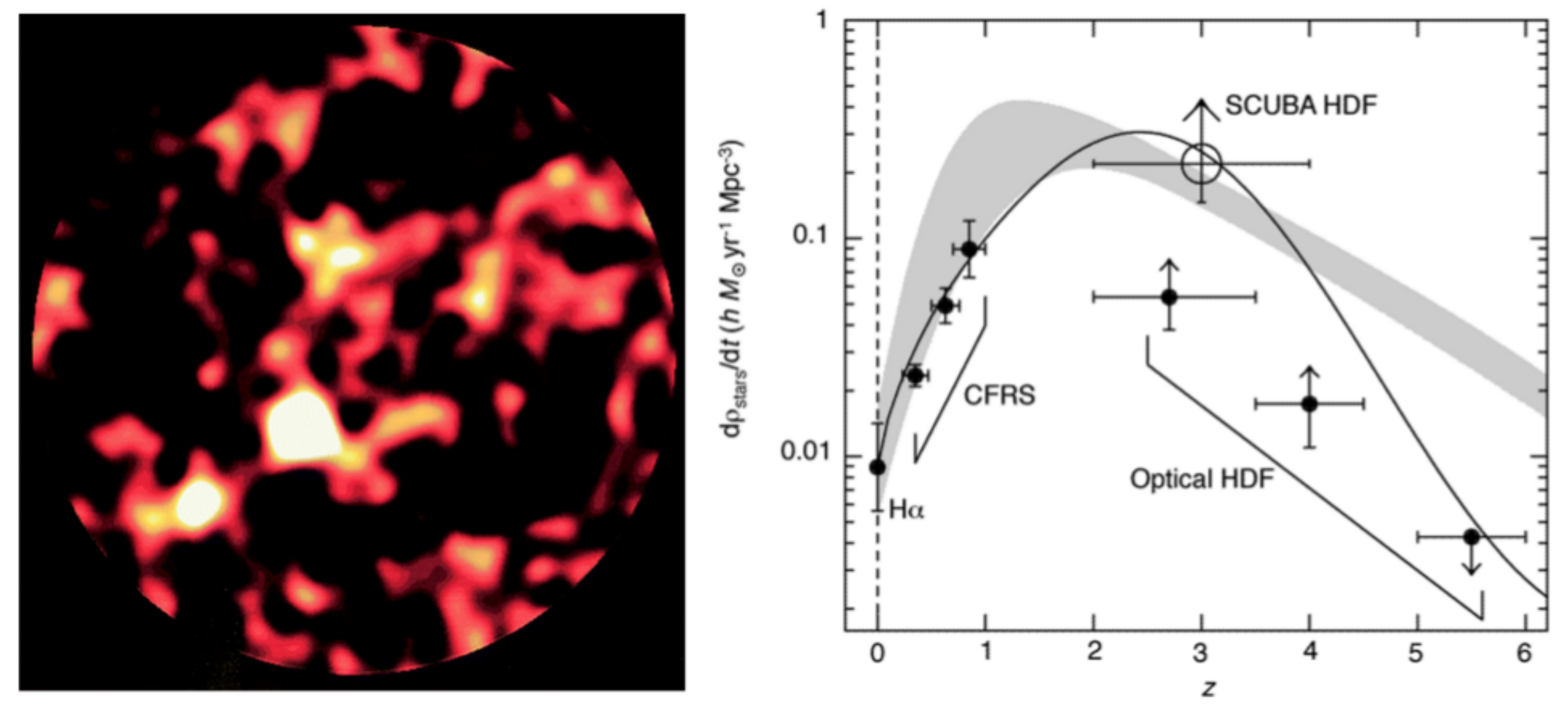}

\caption{(left) The SCUBA 850\,$\mu$m image of the Hubble Deep Field; (right) The global star-formation history of the Universe. The mean 
star formation rate in the Universe is plotted against redshift, implying a rate a factor of 5 higher based on the SCUBA observations compared to values 
obtained from previous optical measurements \cite{Hughes1998}. Figures from Hughes et al. \cite{Hughes1998}.}

\label{fig:HDF_SCUBA}
\end{figure}

Subsequent unbiased (``blank-field'') surveys by groups led by Barger \cite{Barger1998}, Hughes \cite{Hughes1998} and Eales \cite{Eales1999} confirmed that 
the surface density of submillimetre sources was several orders of magnitude above that expected for a non-evolving galaxy population. The conclusion was 
that strongly star-forming galaxies must have been substantially more common in the early Universe than they are today. Having an instrument with hitherto 
unprecedented imaging capability and sensitivity meant that SCUBA could maximise the use of good-weather periods for statistically significant wide and 
deep surveys. For example, 14 nights of the some of the best weather seen on Mauna Kea were used to produce the deepest ever submillimetre image to date. 
The 850\,$\upmu$m image of the Hubble Deep Field (HDF) by Hughes and co-workers \cite{Hughes1998} reached a 1$\sigma$ noise limit of 0.7\,mJy/beam at 
850\,$\upmu$m over an area of around 5 \,arcmin$^2$. It was concluded that the radiation from the five most significant detections in this iconic image, as 
shown in Fig.~\ref{fig:HDF_SCUBA} (left), accounted for 30 -- 50\% of the previously unresolved background emission in the HDF area. The star formation 
rate implied from these redshift 2 -- 4 galaxies was a factor of five higher than that inferred from optical observations (right panel of 
Fig.~\ref{fig:HDF_SCUBA}). The paper describing this seminal discovery has at the time of writing (June 2017) reached over 1000 citations.

\vskip 1mm

\vskip 1mm

Blain, Smail, Ivison \& Kneib \cite{Blain1999} went on to conclude that these first deep submillimetre surveys confirmed a large population of dusty 
galaxies was missing from optical inventories of star formation activity. Further support for this was obtained with the submillimetre detection of an 
extremely red galaxy, HR10, at $z$ = 1.4 by Cimatti and collaborators \cite{Cimatti1998} and the radio galaxy, 8C1435+635 at $z$ = 4.25 by Ivison et al. 
\cite{Ivison1998}. The former is a relatively common class of galaxy previously thought to consist of very old, quiescent ellipticals, but which SCUBA 
revealed to comprise young, star-forming systems similar to local ultraluminous \emph{IRAS} galaxies \cite{Cimatti1998, Dey1998}. The distant, 
submillimetre-selected galaxies discovered by Smail, Ivison \& Blain \cite{Smail1997} were also shown to resemble ULIRGs, at least in the rest-frame 
ultraviolet/optical, with a similar proportion of mergers \cite{Sanders1996, Smail1998}. The diversity of SCUBA-selected galaxies was first shown from 
observations of the distant, galaxies detected in the field of the massive cluster lens Abell 1835 by Ivison et al. \cite{Ivison2000}. One galaxy showed 
almost pure starburst characteristics, whilst the others had varying degrees of AGN activity. The study showed that although almost identical spectral 
energy distributions are seen for many galaxies, they often exhibit strikingly different optical/UV spectral characteristics. It was concluded that 
optical/UV spectral classifications can hence be misleading when applied to distant, highly-obscured galaxies, and that other means of determining the 
various contributions to the overall energy budget of submillimetre galaxies (and hence to the far-IR extragalactic background) are needed.

\subsubsection{Extragalactic surveys go deeper and wider}

By the early 2000's deep extragalactic surveys had become more ambitious and included the 3\,hr and 10\,hr fields of the Canada-UK Deep Submillimetre Survey 
\cite{Eales1999, Lilly1999}, the 8\,mJy survey of the ELAIS N2 and Lockman-Hole E fields \cite{Scott2002} and wider map of the HDF North region 
\cite{Borys2003}. In this latter work Borys and co-workers mapped 165 arcmin$^2$ of the region surrounding the HDF detecting 19 sources at $>$4$\sigma$ 
significance, and concluded that the number of galaxies detected accounted for approximately 40\% of the 850\,$\upmu$m submillimetre background. Moreover, 
the nature of the galaxies uncovered in these surveys was becoming clearer, with critical measurements such as the determination of a median redshift of 
2.4 from radio measurements reported by Chapman et al. \cite{Chapman2003}.

\vskip 1mm

\begin{figure}[!h]
\centering

\includegraphics[width=130mm]{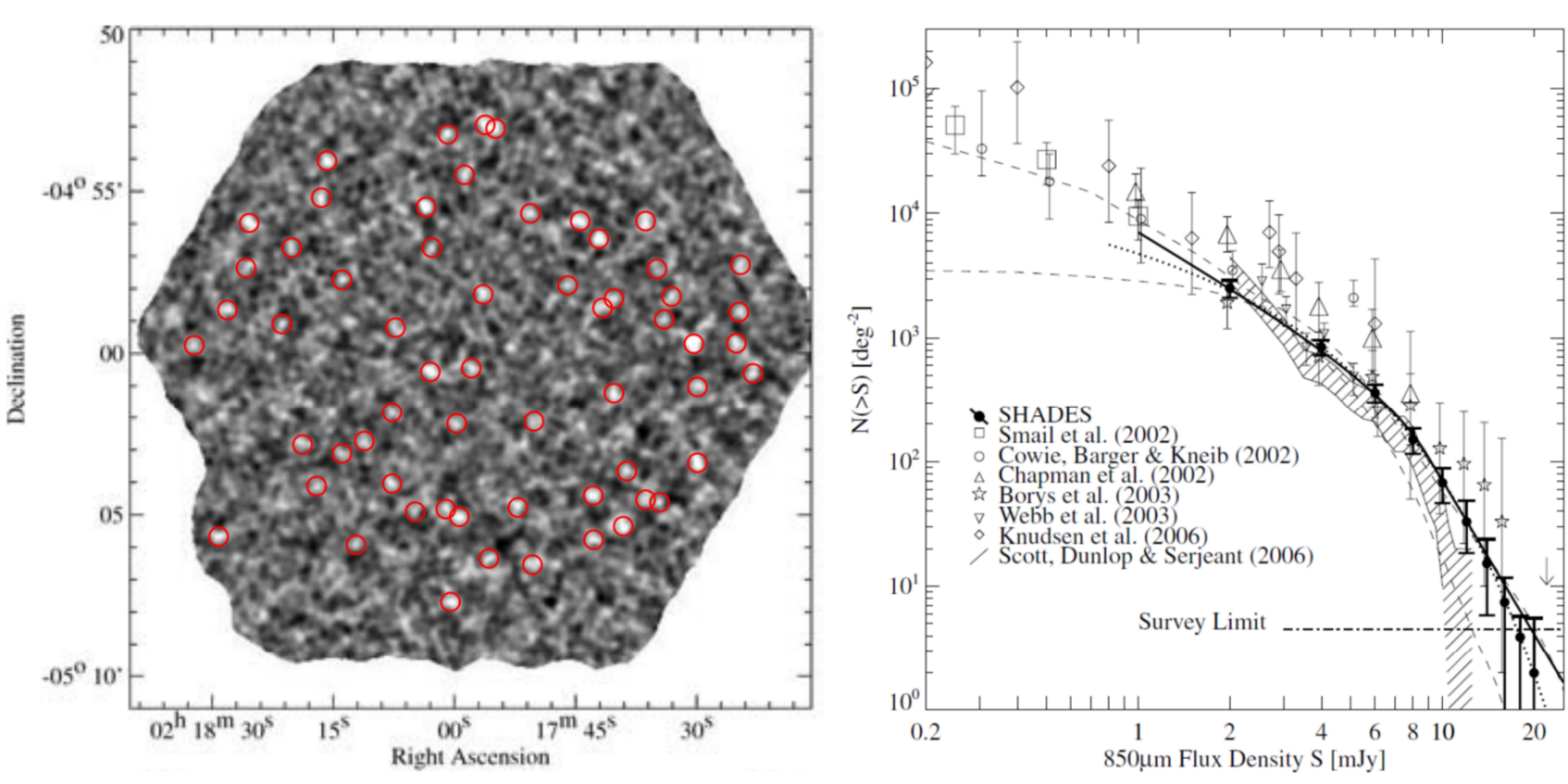}

\caption{(left) The 850\,$\mu$m SCUBA image of the 406 arcmin$^2$ SXDF/UDS SHADES field, identifying some 60 sources; (right) Cumulative 
combined SHADES source counts compared to previous estimates. Figures from Coppin et al. \cite{Coppin2006}.}

\label{fig:SHADES_SCUBA}
\end{figure}

Towards the end of 2002, the first data were taken in what was to be the most ambitious extragalactic survey undertaken to date at the JCMT. This major, 
collaborative survey was called SHADES (the SCUBA HAlf Degree Extra-galactic Survey) and aimed to cover 0.5 degree$^2$ to a 4$\sigma$ detection limit of 
8\,mJy/beam at 850\,$\upmu$m. SHADES was motivated by many science drivers, particularly the desire to clarify the number density, redshift distribution, and 
clustering properties of the bright submillimetre-selected galaxy population. To make further progress in this field required a large and complete sample 
of 850\,$\upmu$m sources (analogous to the 3C radio source sample, which ultimately revolutionised extragalactic radio astronomy). The main issue 
(particularly for the non-cosmologists that used the JCMT!) was that the survey would require approximately one third of the usable UK time on the 
telescope over the subsequent 3 years. The resulting 850\,$\upmu$m maps of the Lockman Hole and SXDF/UDS fields (the latter is shown in the left panel of 
Fig.~\ref{fig:SHADES_SCUBA}) formed the largest submillimetre imaging survey of meaningful depth ever undertaken to date, and provided a uniquely powerful 
resource for the study of the bright submillimetre galaxy population. The results from this survey, reported by Coppin and collaborators \cite{Coppin2006}, 
included a new sample of 120 sources and a definitive measurement of the source number counts in the 1 -- 10\,mJy range (as shown in the right panel of 
Fig.~\ref{fig:SHADES_SCUBA}), resolving some of 20 -- 30\% of the far-IR background.

\vskip 1mm

\vskip 1mm

\subsubsection{New insights on radio galaxies and AGN}

The study of galaxies with AGN was also revolutionised by SCUBA. Early use of the jiggle-mapping mode led to the discovery of SMM02399-0136 by Ivison et 
al. \cite{Ivison1998b}, a hyper-luminous galaxy at $z$ $\sim$ 2.8 hosting an AGN. Such galaxies could not be easily detected in conventional AGN/QSO 
surveys, so the presence of SMM02399-0136 in the very first submillimetre image of the distant Universe suggested that estimates of the prevalence of AGN 
may require substantial revision. The unprecedented sensitivity of SCUBA's photometry mode allowed the study of radio selected and optically selected AGN 
to move from the pioneering world of bare detections to the reliable extraction of physical parameters. For the high-redshift radio galaxy 8C 1435+635, 
Ivison et al. \cite{Ivison1998} presented 450\,$\upmu$m and 850\,$\upmu$m detections of sufficient quality to infer that the formative starbursts of such 
massive ellipticals may still be in progress at $z$ $\sim$ 4. Observations of a sample of radio galaxies by Archibald and co-workers \cite{Archibald2001}, 
spanning a range of redshifts between 1 and 5, showed that the submillimetre luminosity of radio galaxies is primarily a function of redshift, and 
furthermore may be representative of massive ellipitcals in general. The authors concluded that the observed increase in submillimetre detection rate and 
characteristic luminosity with redshift is due to the increasing youthfulness of the stellar population of radio galaxies in their sample.

\vskip 1mm

The steep-spectrum, narrow-line radio galaxy 53W002 was especially interesting as it had been shown to contain an over-density of compact, Ly-$\alpha$ 
emission-line galaxies at $z$ $\sim$ 2.4. SCUBA observations of the 53W002 field by Smail et al. \cite{Smail2003} uncovered four luminous submillimetre 
galaxies. By matching the submillimetre source position using an astrometrically-precise 1.4\,GHz map one of these sources was shown to be coincident with 
a Lyman-$\alpha$-selected galaxy at $z$ = 2.39, 330\,kpc away from the radio galaxy in projection. This confirmed the presence of ultraluminous, dusty 
galaxies in the over dense structure around 53W002 at a look-back time of $\sim$11\,Gyrs. SCUBA galaxies, as the progenitors of massive elliptical 
galaxies, should therefore be strongly clustered in the highest density regions of the distant universe.

\subsubsection{Gamma-ray bursts: galaxy evolution at high redshifts}

An alternative method for studying the characteristics and evolution of galaxies at high redshift is to use gamma-ray bursts (GRBs). SCUBA pioneered early 
observations both of the host galaxy and of the afterglow from GRBs, speculating that the hosts were early starburst galaxies in contrast to the 
previously SCUBA-selected galaxies, which tended to host populations of more evolved stars. The implication was that the submillimetre surveys, which have 
certain selection biases, miss a fraction of the cosmic star formation, which can be possibly recovered by observations of GRB hosts. Although the 
observations were somewhat of a struggle (due to low flux levels), and only a few host galaxies were detected in the submillimetre \cite{Berger2003}, the 
results paved the way for future studies, particularly from space missions such as \emph{SWIFT}, \emph{Spitzer} and \emph{Herschel}. For some bursts the 
early afterglow (hours to weeks following the burst itself) peaks in emission in the submillimetre. By tracking the evolving afterglow emission across the 
entire spectrum, Smith and co-workers \cite{Smith2001} showed that it was possible to study aspects such as the types of shocks involved, whether the 
outflow has a jet or spherical geometry, and to also investigate the geometry of the surrounding medium (uniform versus prior stellar wind).

\subsubsection{Probing large-scale structure: The Sunyaev-Zel'dovich effect}

One of the most versatile probes of large-scale structure of the Universe is the Sunyaev-Zel'dovich (SZ) effect -- the distortion of the cosmic microwave 
background (CMB) radiation through inverse Compton scattering by high energy electrons in galaxy clusters. This distortion produces a characteristic 
``increment'' in the CMB temperature above frequencies of around 200\,GHz, an effect that had only been measured in eight clusters before SCUBA made 
observations of a further two compact galaxy clusters, thereby providing a significant addition to this field of study. Constraining the full spectral 
shape of a cluster's SZ distortion allows separation of the thermal SZ effect, which is caused by the random motions of the cluster's electrons, from the 
kinetic effect, caused by the cluster's motion relative to the CMB rest frame. For the observations a large (180\,arcsecond) chop throw of the secondary 
mirror had to be employed to ensure that no significant SZ flux (a small amplitude signal) appeared in the reference beams. In addition, 450\,$\upmu$m data 
were used to remove the effects of atmospheric emission from the 850\,$\upmu$m data since standard, in-band atmospheric corrections would cancel the SZ 
intensity. The JCMT's high angular resolution also allowed rejection of possible point-source contaminants which plague SZ measurements with smaller aperture 
instruments. The results of this work by Zemcov and collaborators \cite{Zemcov2003} provided robust, high S/N measurements of the SZ increment towards the 
clusters Cl 0016+16 and MS 1054.4--0321.

\subsection{The nearby universe: cold dust and giant magnetic bubbles}

\begin{figure}[!h]
\centering

\includegraphics[width=130mm]{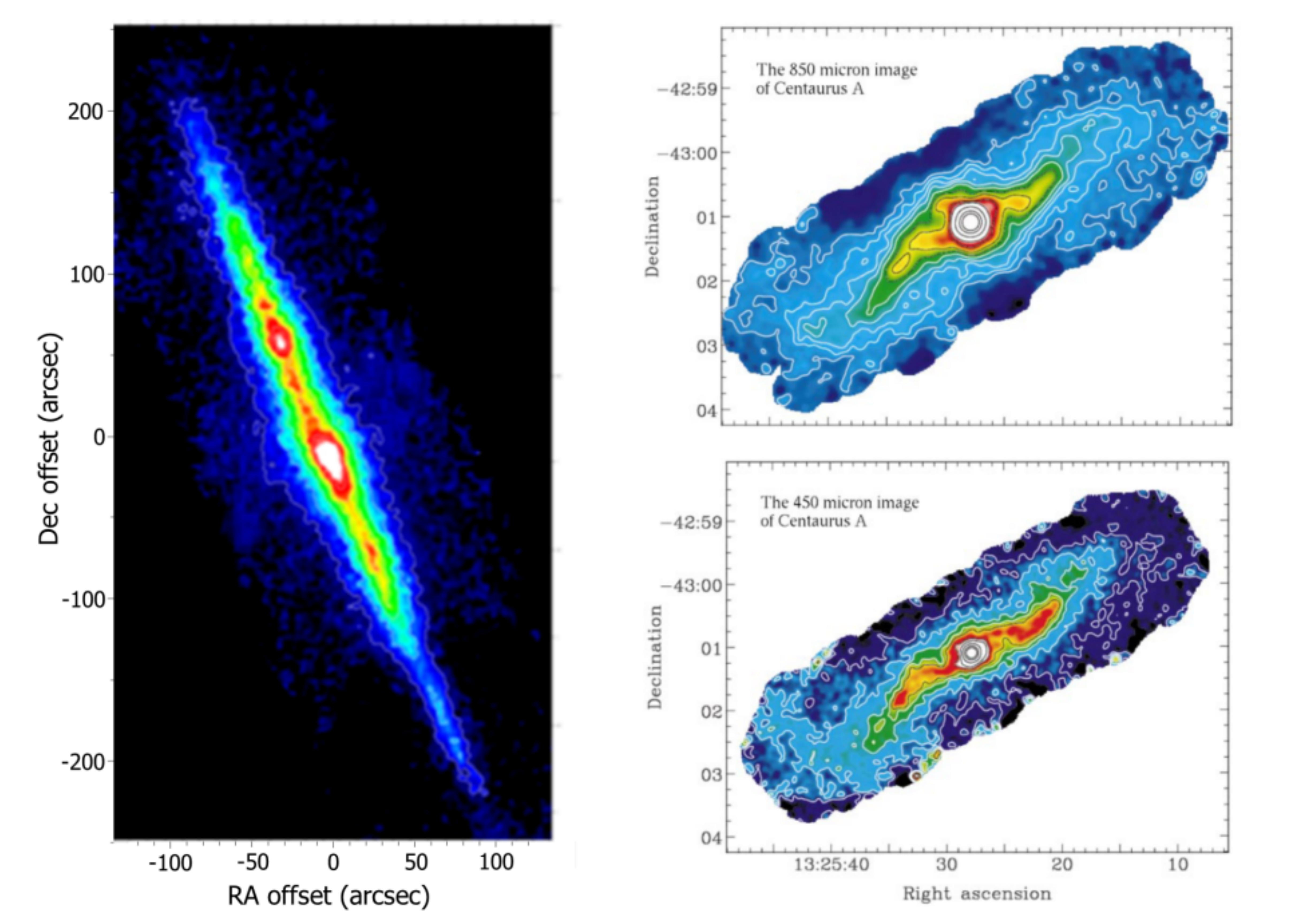}

\caption{(left) SCUBA 450\,$\upmu$m image of NGC 891 highlighting regions of cold dust extending to large galactic radii; (right) 850\,$\upmu$m (top) and 
450\,$\upmu$m (bottom) SCUBA mosaic image of the nearest giant elliptical galaxy Centaurus A, showing a warped ``S-shaped'' inner disc. 
The NGC 891 figure is adapted from Alton et al. \cite{Alton2001}, and Centaurus A from Leeuw et al. \cite{Leeuw2002}. }

\label{fig:NGC891+CenA_SCUBA}
\end{figure}

Another key area of research focused on utilising SCUBA's sensitivity and mapping capabilities to make the deepest images to date of the location of cold 
dust reservoirs in nearby spiral galaxies. The bulk of star formation activity in nearby spirals is often missed by IR studies, since most of the dust mass 
resides in cold, extended, low-surface brightness discs, often far from the galactic nucleus. Studies of nearby galaxies such as NGC891 (see the left 
panel of Fig.~\ref{fig:NGC891+CenA_SCUBA}) and NGC7331 by Alton and co-workers \cite{Alton 1998, Alton2001} revealed that up to 90\% of the total dust 
mass can be located within galactic discs at large radii. The images also detected spectacular dust ``chimneys'' escaping from the main absorption layer up 
to z-heights of nearly 2\,kpc. Further observations of cold dust emission in the ``Whirlpool Galaxy'' (M51) by Meijerink et al. \cite{Meijerink2005} 
showed that the 850\,$\upmu$m originated in an underlying exponential disc with a scale length of 5.5\,kpc. This reinforced the view that the submillimetre 
emission from spiral galaxy discs traces the total hydrogen column density (i.e. the sum of H$_2$ and HI).

\vskip 1mm

\vskip 1mm

Dunne and collaborators \cite{Dunne2000} observed 104 galaxies from the \emph{IRAS} Bright Galaxies sample to provide the first statistical survey of the 
submillimetre properties of the Local Universe. They made the first direct measurements of the submillimetre luminosity function, concluding that the slope 
of the function must flatten at luminosities values lower than in the survey. They postulated the existence of a population of ``cold'' galaxies ($<$25\,K) 
emitting strongly in the submillimetre that would have been missed from far-IR selected samples. By comparing the global galaxy properties with their 
submillimetre/far-IR properties, average gas-to-dust ratios of close to 600 were found, compared to the Galactic value of only 160. The conclusion was that 
most galaxies in the sample must contain a ``cold dust'' component with a temperature of $<$20\,K.

\vskip 1mm

The sensitivity of SCUBA was also used in earnest to make deep submillimetre images of the central 8 $\times$ 2\,kpc region of Centaurus A, the nearest 
giant elliptical galaxy. The remarkable images at 450\,$\upmu$m and 850\,$\upmu$m by Leeuw et al. \cite{Leeuw2002} revealed an unresolved central AGN core, 
an inner jet interacting with gas in a dust lane, an S-shaped inner disc of active star formation, and a colder outer disk. These images are shown in the 
right-hand panel of Fig.~\ref{fig:NGC891+CenA_SCUBA}. The results supported the theory that the inner disc material is consistent with a warped-disc 
model of tilted rings. Finally, with magnetic fields believed to play an important role in the star formation process in the central region of starburst 
galaxies, Greaves et al. \cite{Greaves2000} used the SCUBA polarimeter \cite{Greaves2003} to map the magnetic field morphology surrounding the inner 
regions of M82. The polarised dust emission showed that the major magnetic features found in M82 are ordered fields over scales of hundreds of parsecs 
within the torus, and an outer ``bubble-like'' field associated with the dusty halo, with a diameter of at least 1\,kpc.

\subsection{Large-scale mapping of the Galactic Centre}

\begin{figure}[!h]
\centering

\includegraphics[width=135mm]{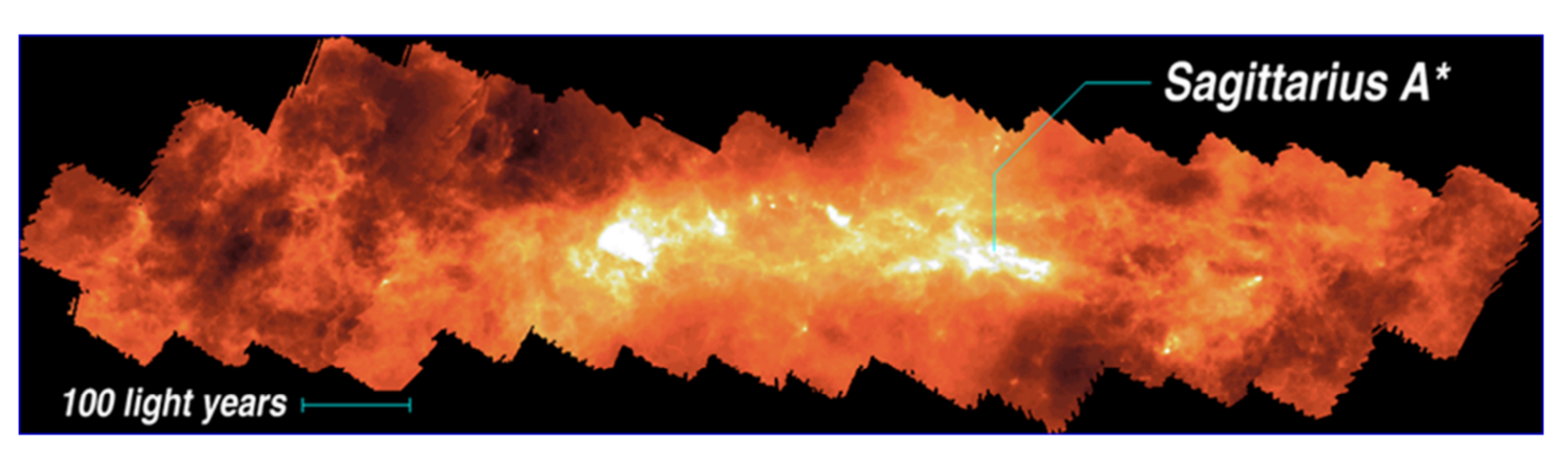}

\caption{The SCUBA 850 \,$\upmu$m image of Central Molecular Zone of the Galactic Centre, covering a region of 3 $\times$ 0.6 degrees including the cloud 
populations near SgrA*. Figure adapted from Pierce-Price et al. \cite{Pierce-Price2000}.}

\label{fig:GC_SCUBA}
\end{figure}

One of the most ambitious, large-scale projects undertaken with SCUBA was to map the Central Molecular Zone (CMZ) of the Galactic Centre over an extent of 
3 degrees in galactic longitude. This stunning data set by Pierce-Price and co-workers \cite{Pierce-Price2000}, shown in Fig.~\ref{fig:GC_SCUBA}, contains 
detailed information on both the warm cloud population near SgrA* (the non-thermal radio source at the centre of the Galaxy), and the circumnuclear disc, with a 
sensitivity RMS limit of approximately 30\,mJy/beam at 850\,$\upmu$m, equivalent to just a few M$_\odot$. There is clearly an extraordinary amount of 
structure, and such data are vital for understanding cloud evolution in a dynamic (sheared and rapidly-rotating) environment. The images also provide input 
to models of the starburst phenomenon in other galaxies as well as the periodic star-formation inferred in our own Galactic Centre. The SCUBA data 
represented the first optically thin map to trace essentially all the mass in the CMZ at high spatial resolution.

\vskip 1mm

\vskip 1mm

Aitken and collaborators \cite{Aitken2000} reported the first detections of linear polarisation from SgrA* at 750\,$\upmu$m, 850\,$\upmu$m, 1.35\,mm and 
2\,mm, confirming the contribution of synchrotron radiation. Large changes in the position angle between the submillimetre and millimetre measurements 
were observed and the best model to explain these changes was one in which the synchrotron radiation from the excess flux is self-absorbed in the 
millimetre but becomes optically thin in the submillimetre. The authors conclude that this suggests the flux originates from an extremely compact source of 
$\sim$2 Schwarzschild radii.

\subsection{Debris discs: the fallout of planetesimal collisions around stars}

\begin{figure}[!h]
\centering

\includegraphics[width=135mm]{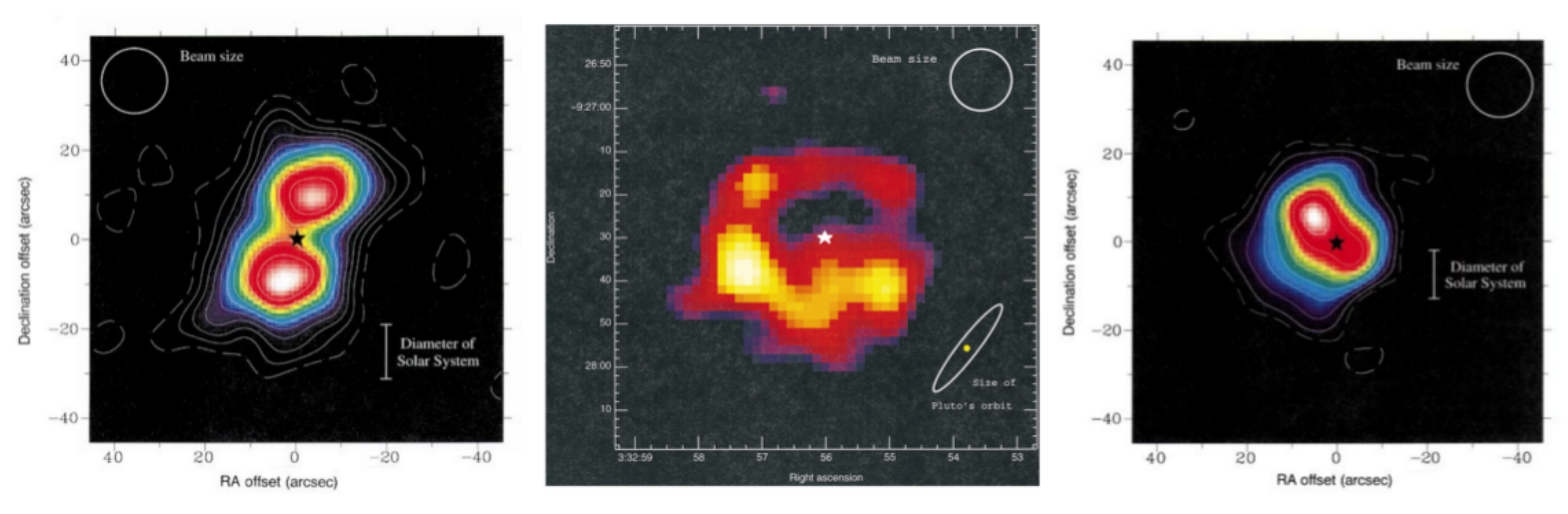}

\caption{A selection of 850\,$\upmu$m images of debris discs observed using SCUBA (left) Fomalhaut (centre) $\epsilon$ Eridani and (right) Vega. The 
projected diameter of the Solar System (size of Pluto's orbit) at the distance of each star is shown on each image. Figures from Holland et al. 
\cite{Holland1998} (Fomalhaut and Vega) and Greaves et al. \cite{Greaves1998} ($\epsilon$ Eridani).}

\label{fig:debrisdiscs_SCUBA}
\end{figure}

Observations with UKT14 had already shown an intriguing glimpse (via point-by-point photometry) of what was possible in terms of imaging the faint discs 
that surround many main sequence stars \cite{Zuckerman&Becklin1993}. The fact that such material exists suggests the presence of larger unobservable 
bodies in these systems, such as planets. SCUBA was well-suited to measure the low-level thermal emission from the dust grains in such discs. The results 
from the work of Holland and co-workers \cite{Holland1998} were spectacular, and included the first images of the debris discs around the well-known 
stars Fomalhaut and Vega. For example, around Fomalhaut, the peak flux in the map (see the left panel of Fig.~\ref{fig:debrisdiscs_SCUBA}) was seen to 
occur in two distinct regions, offset from the stellar position. The image is consistent with an edge-on torus (``doughnut-like'') structure of a size 
similar to our own Edgeworth-Kuiper Belt (EKB), and with a central cavity containing significantly less dust emission. The cavity is about the diameter 
of Neptune's orbit, and a possible explanation is that the region has been cleared of gas and dust by the formation of planetesimals \cite{Holland1998}. 
More recent observations by the \emph{Hubble Space Telescope (HST)} \cite{Kalas2005}, \emph{Herschel} \cite{Acke2012} and the Atacama Millimeter Array 
(ALMA) \cite{White2017}, at higher angular resolution, have shown that the disc is actually a thin ring, possibly shepherded by one or more planets.

\vskip 1mm

How typical is our Solar System architecture around other stars is one of the most fundamental questions in astronomy. Fomalhaut and Vega (see right panel of 
Fig.~\ref{fig:debrisdiscs_SCUBA}) are extremely luminous, relatively short-lived A-stars, and hence any planetary system that may exist around them would 
likely be very different from our Solar System. Further work with SCUBA therefore targeted G and K stars in an attempt to address the uniqueness of our 
Solar System architecture. The image of the debris disc around the young, nearby (only 3\,pc) $\epsilon$ Eridani by Greaves et al. \cite{Greaves1998}, 
shown in the centre panel of Fig.~\ref{fig:debrisdiscs_SCUBA}, revealed a dust ring peaking at 60\,AU from the star with a void of emission in the inner 
3\,AU radius \cite{Greaves1998}. Substructure, observed as asymmetries within the ring, was interpreted as possibly being due to perturbations by planets. 
Moreover, observations of the Sun-like G8 star $\tau$ Ceti by Greaves and colleagues \cite{Greaves2004} revealed a vast EKB-like disc. Modelling 
showed that the mass in colliding bodies up to 10\,km in size is around 1.2 Earth masses, compared with 0.1 Earth masses in the EKB, and hence the 
evolution around the two stars has been very different. One possibility is that $\tau$ Ceti has lost fewer comets from the outskirts of the system, 
compared with the Sun.

\vskip 1mm

\vskip 1mm

\subsection{Large-scale mapping of star-forming regions}

\begin{figure}[!h]
\centering

\includegraphics[width=110mm]{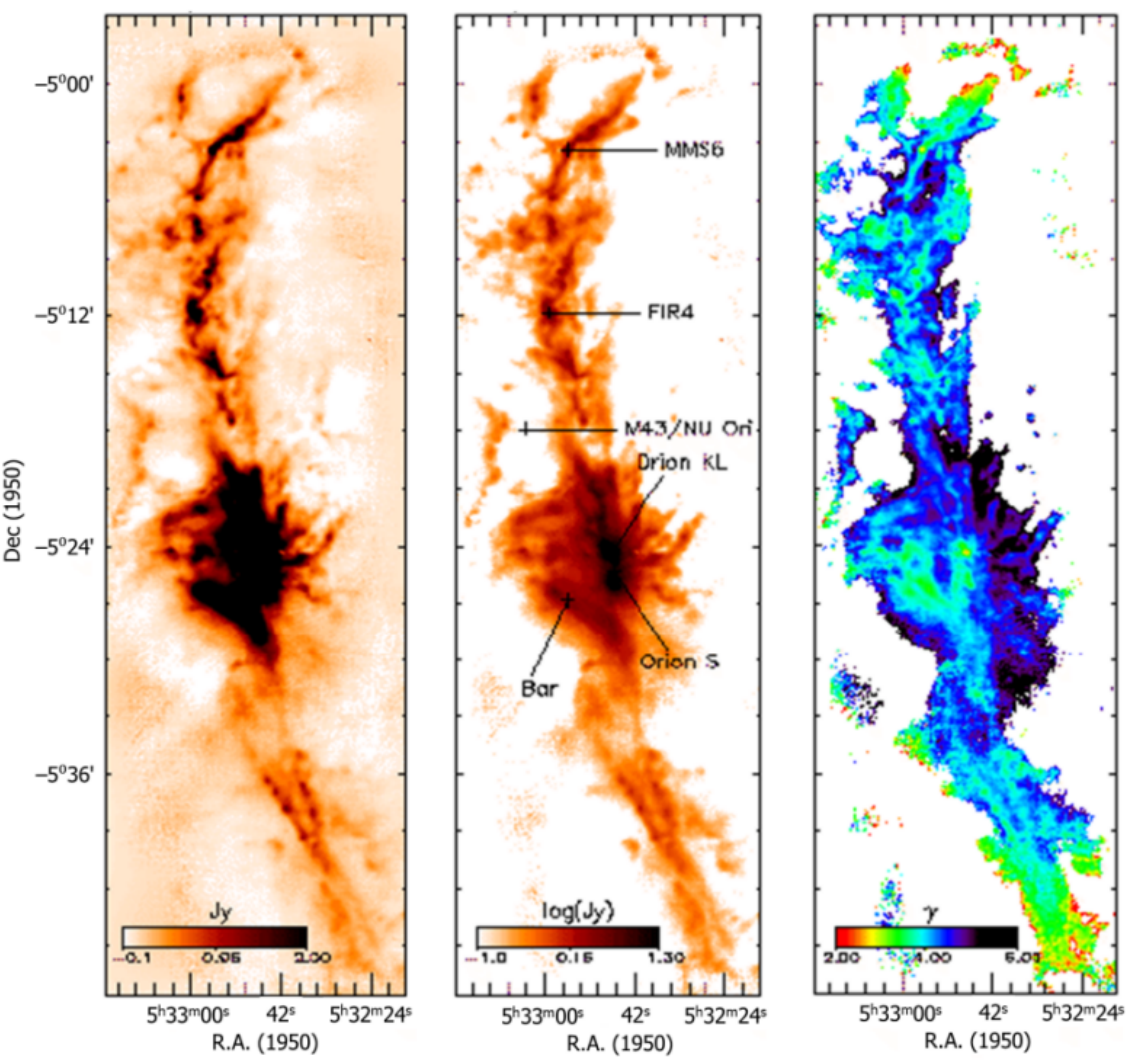}

\caption{SCUBA image of the Orion A molecular cloud: 850\,$\upmu$m image (left) log-intensity image at 850\,$\upmu$m (centre) and spectral index map 
(right), determined from the ratio of the 450\,$\upmu$m and 850\,$\upmu$m images. The main regions are highlighted in the centre image. Figure 
from Johnstone and Bally \cite{Johnstone&Bally1999}.}

\label{fig:OrionA_SCUBA}
\end{figure}

One of the key goals for SCUBA was to provide the capability to carry out large-scale (several degrees), high dynamic range imaging of star-forming regions 
in the Milky Way. One of the first regions to be imaged was the central region of the Orion A molecular cloud. Fig.~\ref{fig:OrionA_SCUBA} from Johnstone 
and Bally \cite{Johnstone&Bally1999} shows that the SCUBA images trace the morphology and spectral index of the optically thin emission from interstellar 
dust. The famous Orion ``bright bar'' is clearly seen in the image together with a chain of compact sources embedded in a narrow, high column-density 
filament that extends over the entire length of the map. The region is also believed to be a site of progressive star formation (from the south to the 
north), and so offers an opportunity to compare dust core properties (such as the spectral index - see right panel in Fig.~\ref{fig:OrionA_SCUBA}) over a 
range of evolutionary stages.

\vskip 1mm

\vskip 1mm

Further large-scale mapping of regions such as the $\rho$ Ophiuchus molecular cloud by Johnstone et al. \cite{Johnstone2000} used clump-finding algorithms 
to identify, and compare the properties of individual objects. Thus it became possible to determine the mass distribution of clumps based on submillimetre 
fluxes for the first time. In the case of $\rho$ Ophiuchus, the clumps spanned a mass range of 0.02 to 6.3 M$_\odot$ and the distribution was characterised 
by a broken power-law, $N(M) \propto M^{-\alpha}$, where $\alpha$ is typically 0.5 -- 1.5. As with other studies it was concluded that the observed clumps 
may represent a evolutionary stage, being fragments produced during the collapse of a larger and gravitationally unstable core within the cloud. The 
observations of the $\rho$ Oph A region of the cloud complex also highlighted the vast improvement in performance of SCUBA over UKT14. To achieve the same 
S/N as SCUBA over the 4 $\times$ 3 arcminute region shown in Fig.~\ref{fig:RhoOph_UKT14} would have taken UKT14 over 10,000 hours!

\subsection{The Holy Grail: protostars}

One of the long-standing challenges facing infrared and submillimetre astronomy is the understanding of the earliest stages of star formation. SCUBA readily 
demonstrated the power of deep imaging to discover new candidate protostars, as well as obtaining reliable statistics on the early stages of stellar 
evolution, including the protostellar Class 0 phase. Unbiased surveys of extended dark clouds, for example by Visser and co-workers \cite{Visser2001}, were 
also carried out to identify complete samples of protostellar condensations, allowing the measurement of star formation efficiencies, mass accretion rates 
and evolutionary lifetimes.

\vskip 1mm

Furthermore, as a result of SCUBA observations interest began to focus on the starless (or ``pre-stellar'') cores, which are significant in that they 
constrain the initial conditions of protostellar collapse. Over 40 such cores in the Orion molecular cloud were studied by Nutter and Ward-Thompson 
\cite{Nutter2007}, who concluded that the high-mass, core mass function (CMF) follows a roughly Salpeter-like slope, just like the initial mass function (IMF) 
seen in earlier studies. The deep SCUBA maps showed that the CMF turns over at $\sim$ 1.3\,M$_\odot$, about a factor of 4 higher than the completeness 
limit. This turnover, never previously observed and only revealed by the much deeper SCUBA maps, mimics the turnover seen in the stellar IMF at 
$\sim$0.1\,M$_\odot$. The low-mass side of the CMF is a power-law with an exponent of, 0.35 -- 0.2, which is consistent with the low-mass slope of the 
young cluster IMF of 0.3 -- 0.1. This shows that the CMF continues to mimic the shape of the IMF all the way down to the lower completeness limit of these 
data at $\sim$0.3\,M$_\odot$.

\vskip 1mm

\begin{figure}[!h]
\centering

\includegraphics[width=110mm]{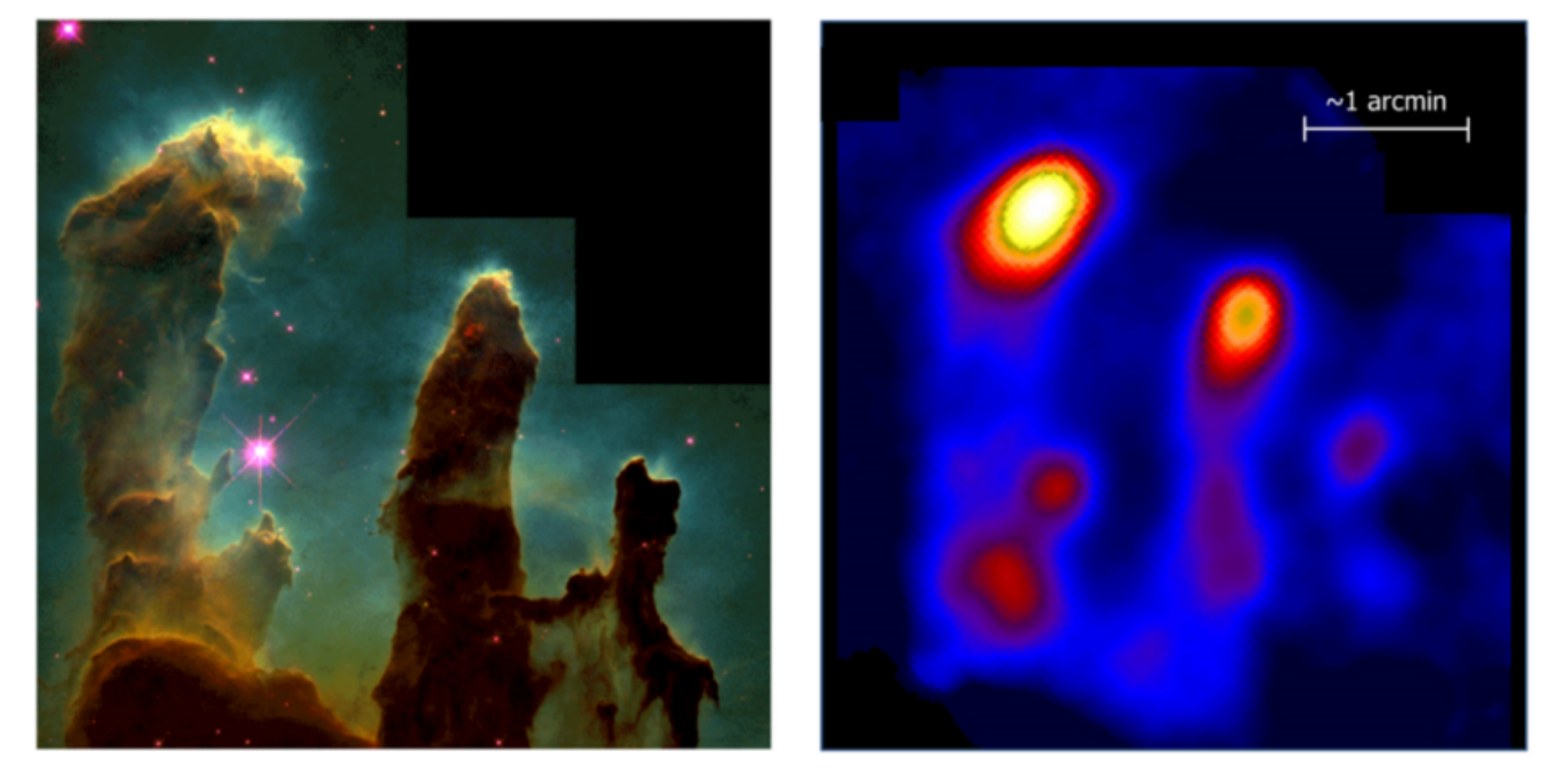}

\caption{(left) The Eagle Nebula (M16) as seen from the \emph{HST} (image courtesy of J. Hester and P. Scowen); (right) The SCUBA 450\,$\upmu$m image of 
the same region highlighting the bright thermal emission from the tips of the ``fingers''. SCUBA figure adapted from White et al. \cite{White1999}).}

\label{fig:M16_SCUBA}

\end{figure}

One of the the most spectacular images of the earliest stages of star formation came from SCUBA imaging by White and co-workers \cite{White1999} of the 
famous Eagle nebula (M16). As shown in the 450\,$\upmu$m SCUBA-2 image presented Fig.~\ref{fig:M16_SCUBA} some differences are immediately evident from 
the \emph{HST} optical image, particularly in terms of the dominant thermal emission from the tips of the ``fingers'' seen in the SCUBA map. The 
continuum spectra of these cores show that they are much cooler ($\sim$20\,K) than the surrounding molecular gas in each of the fingers. The results of a 
thermal and chemical model of the environment concluded that the fingers appear to have been formed after the primordial dense clumps in the original 
cloud were irradiated by light from its own OB stars. During the subsequent photoevaporative dispersal of the cloud the clumps shielded material lying 
behind it, facilitating the formation of the fingers. The absence of embedded IR sources or molecular outflows suggest that the cores at the tips of the 
fingers have the characteristics of the earliest stages of protostellar formation.

\vskip 1mm

\vskip 1mm

One of the most ambitious projects to be undertaken with SCUBA and it's polarimeter was an attempt to make the first observations of the magnetic field 
geometry in pre-stellar cores. By doing so such results would test the theoretical ideas about the way in which the field geometry affects the star 
formation process. The first published maps by Ward-Thompson et al. \cite{Ward-Thompson2000} for three cores revealed smooth, uniform polarisation vectors 
in the plane of the sky, showing no evidence of the type of geometry that might be expected in magnetically dominated stage of evolution. It was concluded 
that no current model of magnetically-regulated star formation could explain the existing observations.

\subsection{Solar System Science: Comet Hale-Bopp and the subsurface of Pluto}

It has been long suspected that large particles may be present in comets sufficient to dominate the total mass of the coma. Believed to be products of 
agglomerative growth in the proto-solar nebula these particles are the local analogues of the dust observed in the disks around many young stars. Some of 
the first evidence for this was published for the Comet Hale-Bopp by Jewitt and Matthews \cite{Jewitt&Matthews1999} based on SCUBA 850\,$\upmu$m 
observations. The dust coma surface brightness is well described by a steady-state outflow model, in which the dust density varies with the inverse square 
distance from the nucleus. Submillimetre observations have proved vital in studying the properties of these large particles; the data provide an estimate 
of the total mass, the dust mass production rate as a function of heliocentric distance, and the size of the particles in comparison with those in 
circumstellar discs.

\vskip 1mm

The subsurface of Pluto is known contain a reservoir of frozen volatiles but very little is known about it. Greaves, Whitelaw and Bendo \cite{Greaves2015} 
used archival light curves of the brightness of Pluto to probe just below the skin depth of the thermal changes over Pluto's day. With the light curve in 
the submillimetre differing significantly from those measured in the mid- and far-IR, in a region that is optically dark on the planet's surface, the 
suggestion is that the layers a few centimetres below the surface have not undergone any major temperature change. One possibility is that these regions 
could have a different emissivity, perhaps with a subsurface layer richer in nitrigen or methance ices than the surface. Results from the NASA \emph{New 
Horizons} probe concluded that the surface composition is suprisingly complex, with the nitrogen, methane and water-rich areas creating a puzzle for 
understanding Pluto's climate and geologic history.


\section{SCUBA-2: Wide-field imaging in the submillimetre becomes a reality}

Although SCUBA had made so many pioneering discoveries it was obvious by the turn of the century that an even more sensitive camera was required, 
specifically to allow wide-field surveys to be undertaken, in line with the planned work by satellites such as \emph{Herschel} in the far-IR. The project 
became known as SCUBA-2, and involved an international partnership between institutes in the UK, USA and Canada. As was the case with SCUBA, SCUBA-2 had 
two imaging arrays working simultaneously in the atmospheric windows at 450 and 850\,$\upmu$m, the vast increase in pixel count to over 10,000 meant that 
SCUBA-2 would map the sky 100 -- 150 times faster than SCUBA to the same signal-to-noise. SCUBA-2 was a major step forward in technology. It was the first 
astronomical camera to use superconducting transition-edge sensors in a time-domain multiplexed readout scheme. The instrument itself was also a major 
challenge having a liquid cryogen-free dilution refrigerator to cool the detector to $<$100\,mK and over 600\,kg of optics cooled to less than 4\,K. 
SCUBA-2 was delivered to the JCMT in April 2008 with two engineering sub-arrays (one quarter of the field-of-view at each wavelength), and eventually began 
science operations in late 2011 with fully populated, science-grade focal planes. In February 2012, SCUBA-2 began a series of unique legacy surveys for the 
JCMT community. These surveys took almost 3 years and the results provided complementary data to the shorter wavelength, shallower, larger area surveys 
from \emph{Herschel}. The SCUBA-2 surveys have also provided a wealth of information for further study with new facilities such as ALMA, and future 
possible telescopes such as \emph{SPICA} and ground-based large, single-aperture dishes.

\subsection{The first generation surveys with SCUBA-2}

The key scientific driver for SCUBA-2 was the ability to carry out large-scale surveys of the submillimetre sky to unprecedented depth. Six first 
generation ``legacy-style'' survey programmes were approved covering a very broad base, ranging from the studies of debris discs around nearby stars to 
galaxy populations and evolution in the early Universe. The SCUBA-2 element of these surveys was initially approved to run from February 2012 to September 
2014, with several benefiting from an extension to February 2015. In the next sections we briefly describe these surveys and summarise some of the key 
findings so far.

\subsubsection{Galactic plane survey}

\begin{figure}[!h] 
\centering 
\includegraphics[width=130mm]{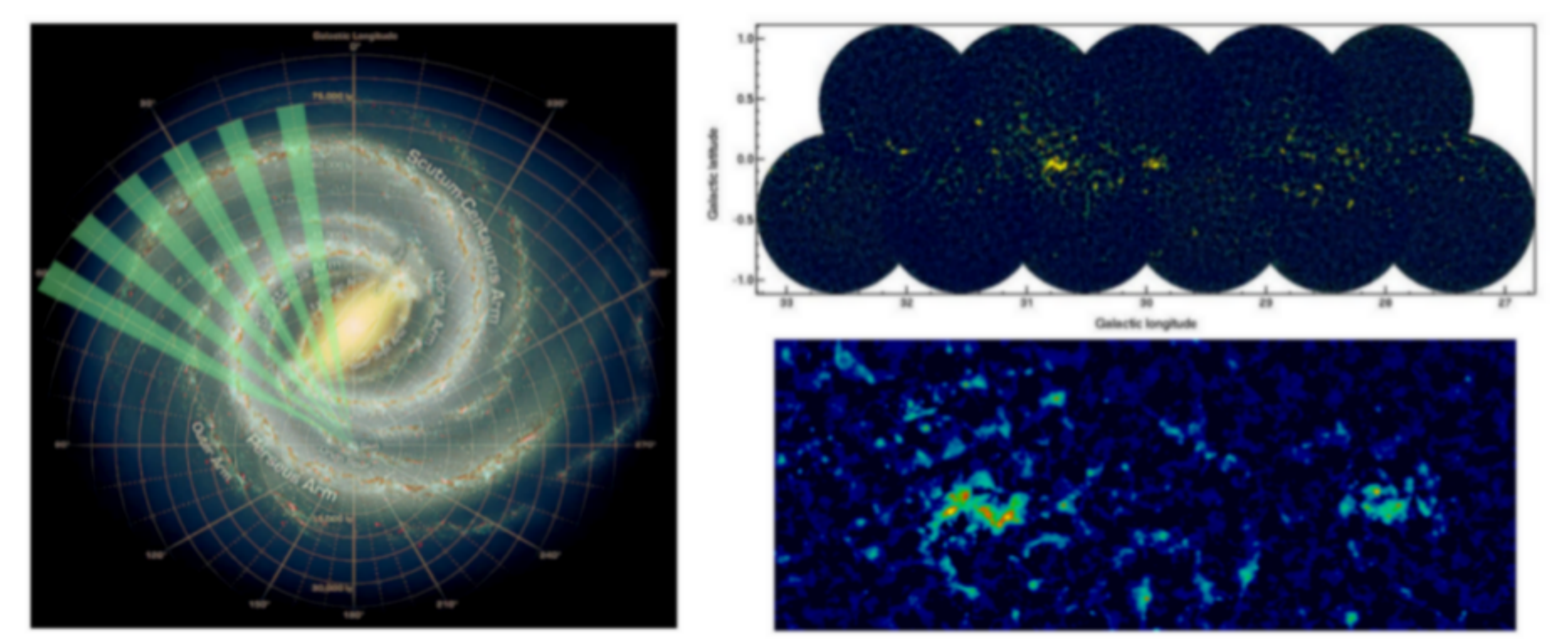} 

\caption{(left) The area covered by the JPS (green segments) overlaid on a sketch of the Milky Way by Robert Hurt \cite{Moore2015} (right top) The $l$ = 30 
degree full-field image of the Galactic Plane from JPS (right bottom) Highlighted is the W43 star-forming region from the $l$ = 30 image. SCUBA-2 figures 
from Eden et al. \cite{Eden2017}.}

\label{fig:JPS_SCUBA2} 

\end{figure}

The JCMT Plane Survey (JPS) sought to achieve a full census of star-formation activity in the plane of the Galaxy observable from JCMT to a detected mass 
limit of around 40 M$_\odot$ at the far edge of the Galaxy. The aims included examining triggered and large-scale star formation and to study the evolution 
of massive YSOs, infrared dark clouds and filaments, along with dust evolution and molecular cloud structure. Surveys of the Galactic Plane in the 
millimetre/submillimetre are currently the only approach to determine the relative importance of the physical processes that are likely to affect the star 
formation efficiency on Galactic scales ($>$1\,kpc, e.g., spiral density waves) and within individual molecular clouds (e.g., temperature and pressure). To 
achieve this, the JPS observed six fields along the Galactic Plane at longitudes of 10, 20, 30, 40, 50 and 60 degrees (see left panel of 
Fig.~\ref{fig:JPS_SCUBA2} with each field just over 5 $\times$ 1.7 degrees in area as described by Moore et al. \cite{Moore2015}. The argument was that 
large fractions of the Plane needed to be surveyed in order to account for the statistical distribution of cloud masses and YSO luminosities, as well as 
local variance.

\vskip 1mm

The final survey covers an area of approximately 50 degree$^2$ and achieved an average noise level of 7.2\,mJy/beam at 850\,$\upmu$m, when smoothed over a 
beam diameter. An example of one of the fields, highlighting the W43 star-forming region, is shown in the right panels of Fig.~\ref{fig:JPS_SCUBA2}. The 
survey is approximately 10$\times$ more sensitive than the complementary ATLASGAL survey carried out on the APEX telescope, which studied the inner 
Galactic Plane at 870\,$\upmu$m, covering galactic longitudes between 60 and 270 degrees. A catalogue of $\sim$7800 compact sources was generated from the 
JPS, and it was shown that these sources are responsible for 42\% of the total emission from the maps with the remaining flux lying in filamentary 
structures. One of the key outcomes of the survey, which also included large-scale observations with HARP (see section 5), is that the 
dominant scale of variations in star formation efficiency in the Galactic disc is that of individual molecular clouds as described in one of the first 
papers by Eden and collaborators \cite{Eden2015}, with spiral arms having only a relatively minor influence. Using the \emph{Herschel} 70\,$\upmu$m data, 
the survey team showed that 38\% of the sources detected show evidence of ongoing star formation \cite{Eden2017}. The JPS images and associated source 
catalogue represent a valuable resource for studying the role of environment and spiral arm structure on star formation in the Galaxy.

\subsubsection{Gould Belt survey}

The Gould Belt is a large ($\sim$1 kpc diameter) ring of molecular clouds and OB star associations that is inclined at $\sim$20$^{\circ}$ to the Galactic 
Plane. It is important for star formation studies as it contains most of the nearby low- and intermediate-mass star formation regions such as the Orion and 
Taurus-Auriga molecular clouds. The JCMT Gould Belt Survey (GBS) aimed to address several of the major unsolved questions in star formation: the evolution 
of pre- and protostellar cores, the origin of the IMF, and the link between star formation and molecular cloud properties \cite{Ward-Thompson2007}. The 
targets were molecular clouds within 500\,pc of the Sun where the angular resolution is high enough to separate individual pre/protostellar cores 
(0.1\,pc). The survey was awarded 612 hours of observing time, which included both SCUBA-2 and HARP observations of 14 nearby clouds, covering a total 
area of almost 700 degree$^2$. The improved resolution of the JCMT also allows for more detailed study of large-scale structures such as filaments, 
protostellar envelopes, extended cloud structure and morphology down to the Jeans length.

\vskip 1mm

\begin{figure}[!h] 

\centering 

\includegraphics[width=120mm]{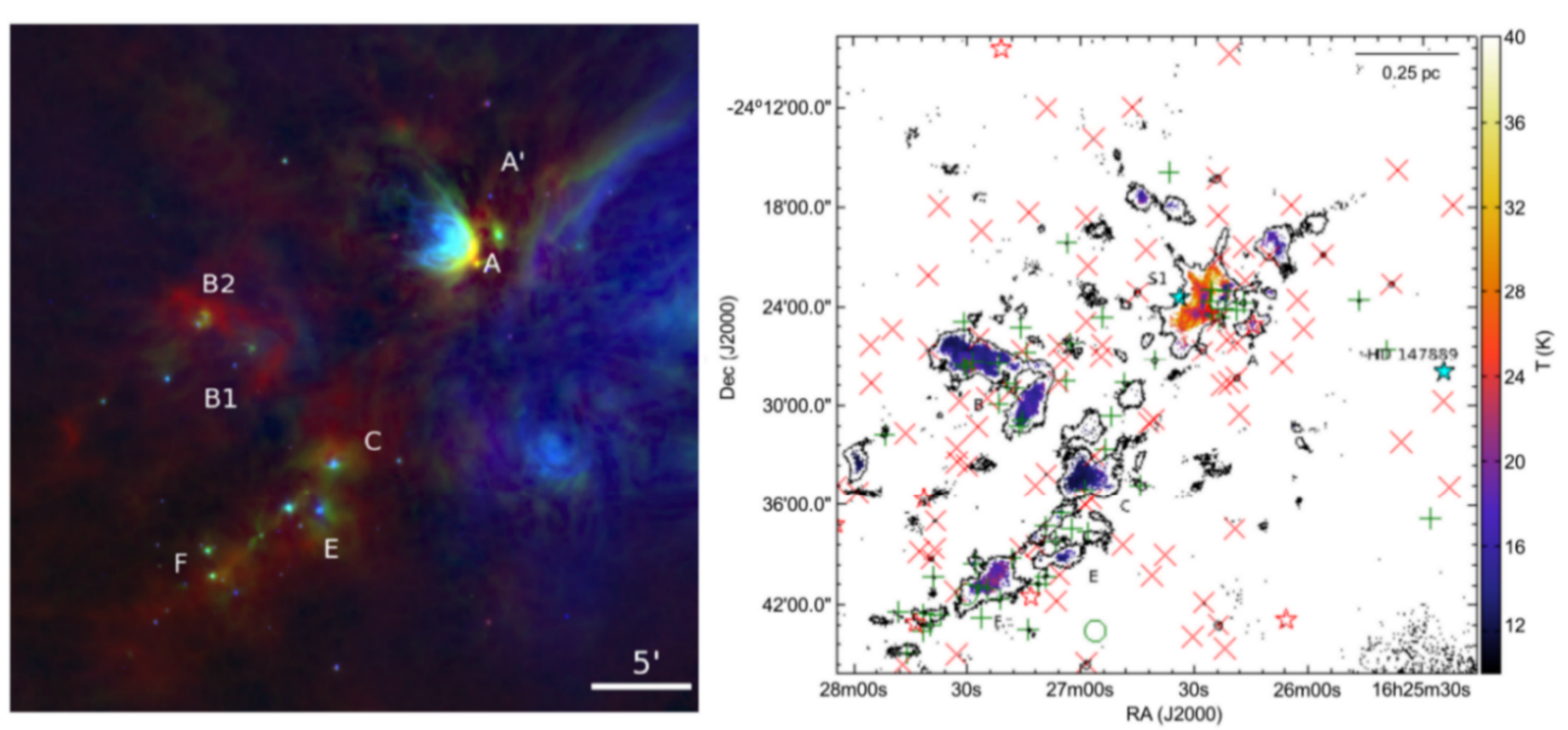} 

\caption{(left) Three-colour image of the L1668 region with the main molecular clouds labelled. The red, green and blue channels are from SCUBA-2, 
\emph{Herschel} 100\,$\upmu$m, and \emph{Spitzer} 8\,$\upmu$m, respectively; (right) A dust temperature map of the same region determined 
from SCUBA-2 450\,$\upmu$m and 850\,$\upmu$m flux ratios. Figures from Pattle et al. \cite{Pattle2015} (L1668 three-colour image) and 
Rumble \cite{Rumble2016} (dust temperature map).}

\label{fig:L1688_SCUBA2} 

\end{figure}

\vskip 1mm

The resulting maps of the Gould Belt molecular clouds are amongst the deepest ever undertaken with typical noise levels of 3--4\,mJy/beam at 850\,$\upmu$m 
over tens of square degrees. The uniqueness of the SCUBA-2 observations is that they predominantly trace cold, dense cores that are more likely to be 
pre-stellar than the more evolved clumps seen by \emph{Herschel}. For example, using observations from both SCUBA-2 and HARP, Pattle et al. 
\cite{Pattle2015} identified significant fractions of pressure-supported starless cores in Ophiuchus that are unlikely to ever become gravitationally 
bound. Furthermore, by combining SCUBA-2 data with shorter wavelength data from \emph{Spitzer} and \emph{Herschel} it is also possible to measure the 
variation in temperature along the line-of-sight, as shown in Fig.~\ref{fig:L1688_SCUBA2}, for the Ophiuchus L1688 cloud complex. Mairs and collaborators 
\cite{Mairs2016} presented a catalogue of sources from observations of the Southern Orion A cloud, showing that the larger-scale regions of emission within 
the cloud are often subdivided into smaller dense fragments that are usually invisible in shorter wavelength surveys. One of the key aims of the survey was 
to investigate the prestellar mass function in the various molecular clouds. Salji and collaborators \cite{Salji2015} determined that it peaked at 1.39 
M$_\odot$ in Orion A, revealing a star-forming efficiency of 14\% when compared the Orion nebula cluster IMF. Furthermore, the prestellar mass function was 
found to decay with a high-mass power-law exponent of 2.53, similar to the Salpeter IMF value of 2.35 for stars in the Solar neighbourhood. The extensive 
GBS data continues to be analysed and has already produced a number of new source catalogues, characterising thousands of cores and clumps in terms of 
their properties and evolutionary status.

\subsubsection{Nearby galaxy survey}

SCUBA pioneered some of the earliest observations of the extent of cold dust in nearby spiral galaxies (see Section 3b). As described by Wilson and 
collaborators \cite{Wilson2012} the JCMT Nearby Galaxies legacy survey (NGLS) aimed to use both SCUBA-2 and HARP to investigate both the physical 
properties of gas and dust in galaxies, along with the effect that galaxy morphology and unusual environments (such as metallicity) has on the properties of the 
dense ISM. The SCUBA-2 part of the survey was allocated 100 hours of observing time, with a goal of reaching a sensitivity level of 1.8\,mJy/beam at 
850\,$\upmu$m. A total of 48 spiral galaxies were observed, the majority of which came from the \emph{Spitzer} SINGS survey. One of the major findings was 
the presence of significant levels (up to 25\%) of CO in the centres of galaxies. Much of the analysis of the data is still underway; the emission from the 
majority of the galaxies is very weak, and recovering the flux on scales of several arcminutes has been a challenge for the data reduction. Combining the 
data with shorter wavelength data from \emph{Herschel} will also allow the dust temperature variation in these galaxies to be mapped across large galactic 
scales for the first time.

\subsubsection{Cosmology legacy survey}

The JCMT Cosmology Legacy Survey (S2CLS) sought to capitalise on the pioneering high redshift galaxy work undertaken by SCUBA and other early submillimetre 
cameras. Submillimetre galaxies (SMGs) are amongst the most luminous dusty galaxies in the Universe but their true nature remained unclear. It could be 
that they are the progenitors of the massive elliptical galaxies seen in the local Universe, or a short-lived phase of a more typical star-forming galaxy. 
As described by Geach et al. \cite{Geach2013} the key driver of the 850\,$\upmu$m survey was to deliver a sufficient number of galaxies to address this 
question by reliably measuring the clustering of the submillimetre population (providing valuable constraints on galaxy formation models) and to detect and 
study the (rare) progenitors of rich clusters. The 850\,$\upmu$m survey would also establish, unambiguously, the faint end of the counts of SMGs in this 
band. The goal of the deep 450\,$\upmu$m component of the survey was to resolve a significant fraction of the extragalactic background light at 
450\,$\upmu$m into individual galaxies by getting as close as possible to the confusion limit at this shorter wavelength (similar to that achieved with 
SCUBA for the HDF at 850\,$\upmu$m). The survey plan was to map an area of 10 deg$^2$ at 850\,$\upmu$m to a depth of 1$\sigma$ = 
1.5\,mJy/beam and 0.25 deg$^2$ at 450\,$\upmu$m to a depth of 1$\sigma$ = 1.2\,mJy/beam. To achieve this, several extragalactic survey fields (including 
the original SHADES fields from SCUBA) were to be mapped for which a wide range of ancillary data is available from other wavelengths. The wide-field 
850\,$\upmu$m included a number of well-studied fields, such as the UKISS-UDS, COSMOS, \emph{Akari}-Northern ecliptic pole, and Lockman Hole north regions, 
whilst the ultra-deep 450\,$\upmu$m maps were centred in the COSMOS and UDS fields. The S2CLS was the by far the largest of the first generation legacy 
surveys, and was awarded close to 1800 hours or 51\% of the total survey time over the 3 year period.

\vskip 1mm

\begin{figure}[!h] 

\centering 

\includegraphics[width=110mm]{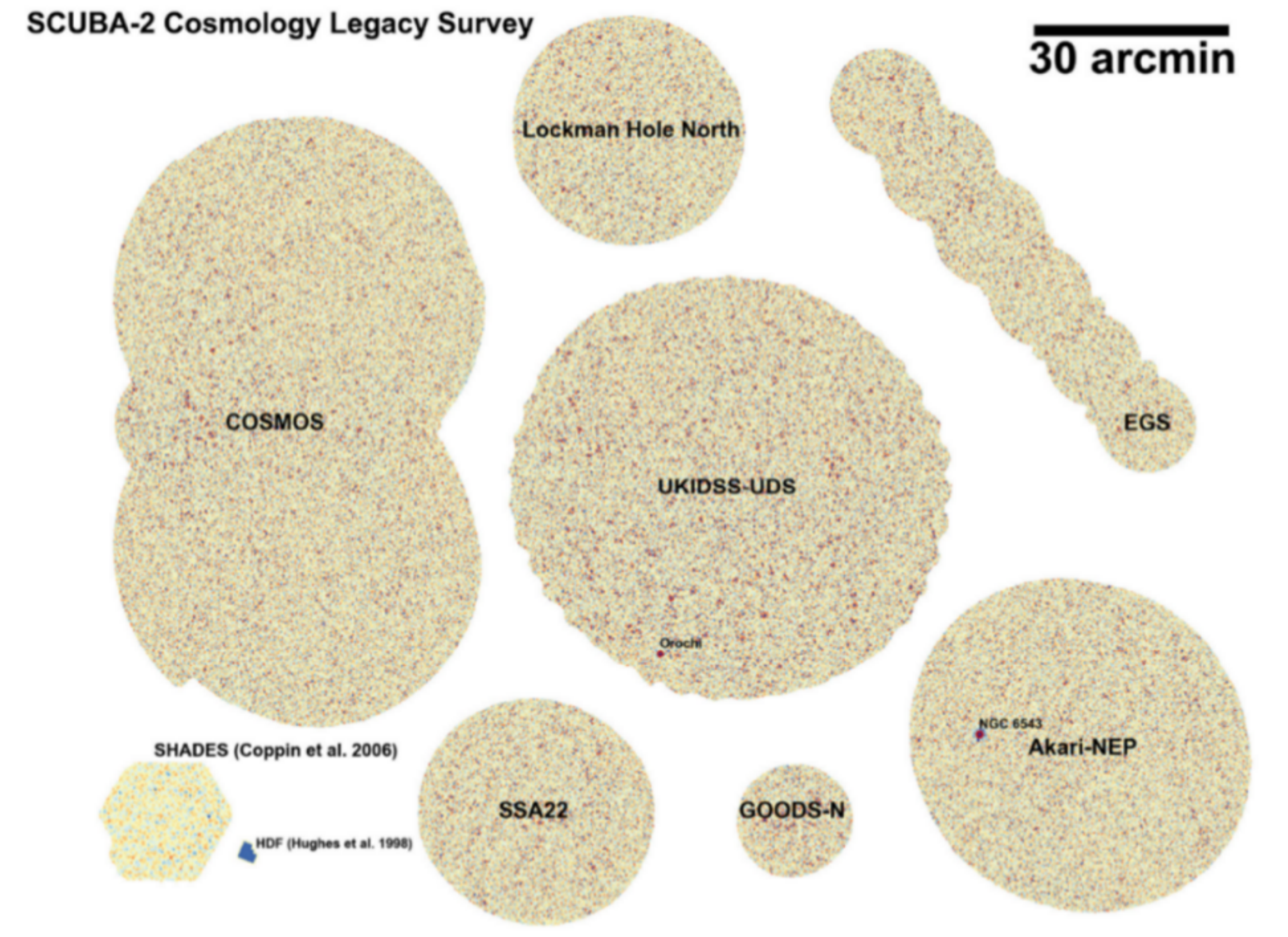} 

\caption{A montage of the 850\,$\upmu$m signal-to-noise maps from the seven extragalactic fields of the SCUBA-2 Cosmology Legacy survey. For comparison the 
previous work on the Hubble Deep Field and SHADES, undertaken with SCUBA, are also shown (see sections 3a(i) and 3a(ii), respectively). Figure from Geach 
et al. \cite{Geach2017}. }

\label{fig:S2CLS_SCUBA2} 

\end{figure}

\vskip 1mm

\begin{figure}[!h] 

\centering 

\includegraphics[width=125mm]{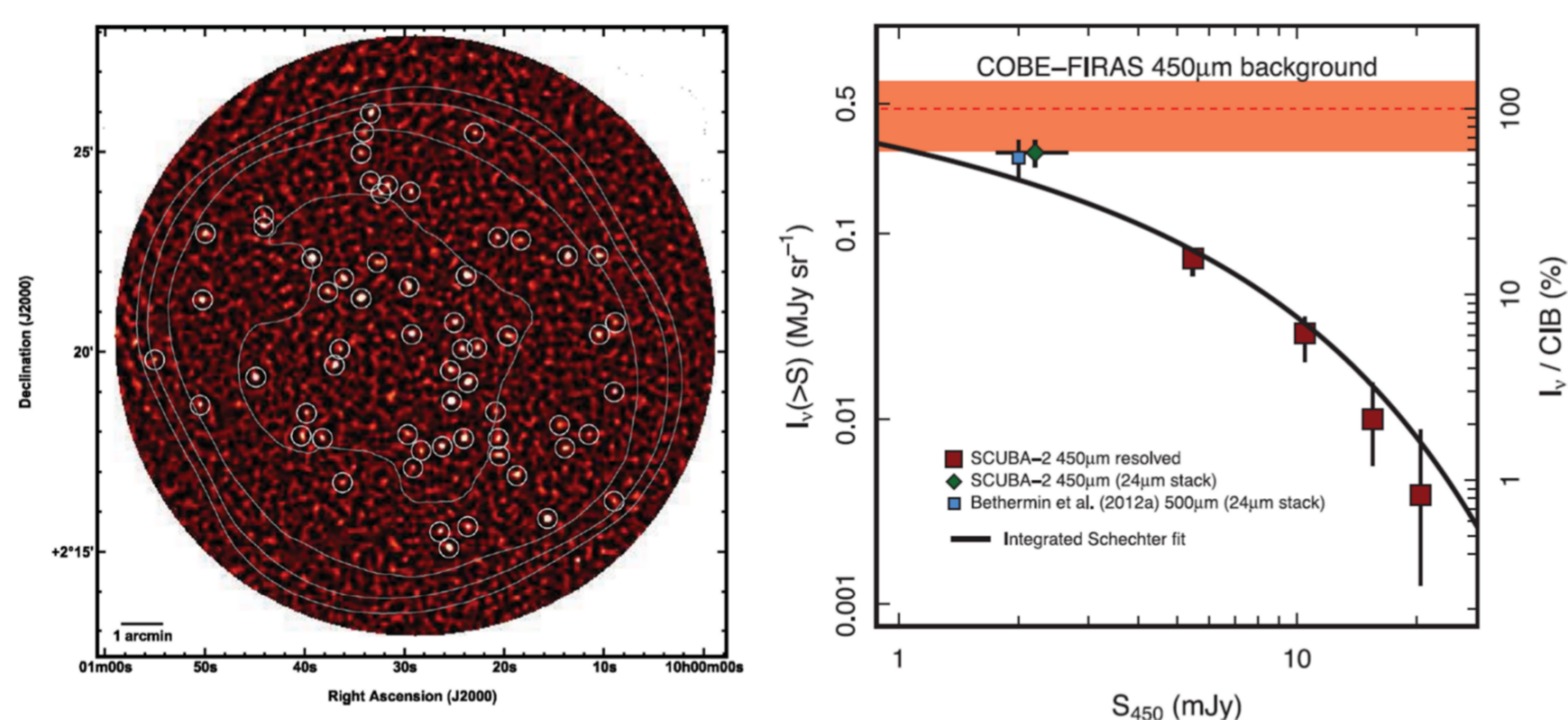} 

\caption{(left) SCUBA-2 450\,$\upmu$m signal-to-noise map of the COSMOS/CANDELS field, showing 60 sources detected (circles) at $>$ 3.75$\sigma$ 
significance. The contours show the variation in noise level, starting with 2\,mJy/beam for the inner contour and increasing to 5\,mJy/beam for the outer; 
(right) The integrated surface brightness of the 450\,$\upmu$m sources relative to the cosmic infrared background measured by \emph{COBE}. Figures from 
Geach et al. \cite{Geach2013}.}

\label{fig:450background_SCUBA2} 

\end{figure}

The results from the S2CLS increased the size sample of 850\,$\upmu$m-selected SMGs by an order of magnitude. Fig.~\ref{fig:S2CLS_SCUBA2} 
shows the first 850\,$\upmu$m maps by Geach and collaborators \cite{Geach2017} from this extensive survey, covering a total area of $\sim$5 degree$^2$ and 
detecting approximately 3000 sources. The average RMS noise in the maps is 1.2\,mJy/beam, close to the expected confusion limit of 0.8\,mJy/beam. Such a 
large survey also allows a comprehensive measurement of the number counts of submillimetre sources and the results show both a distinctive upturn in the 
counts caused by strong gravitational lensing of high redshift galaxies, and a contribution from local sources of submillimetre emission. For the first 
unbiased, blank-field assessment of the number counts of galaxies at 450\,$\upmu$m Geach et al. \cite{Geach2013} showed that 16\% of the cosmic infrared 
background was resolved into individual galaxies (see Fig.~\ref{fig:450background_SCUBA2}), whilst a further $\sim$40\% was recovered in the SCUBA-2 map by 
comparing to \emph{Spitzer}-detected 24\,$\upmu$m emitters. Koprowski et al. \cite{Koprowski2016} utilised multi-frequency data to determine the redshift 
distribution of the 106 galaxies detected in the deepest, central area of COSMOS field and found a median redshift of 2.38 $\pm$ 0.09. Roseboom and 
collaborators \cite{Roseboom2013} explored the physical properties of these galaxies from their spectral energy distributions, revealing correlations, for 
example, between the dust temperature and infrared luminosity. Some 24\% of the 450\,$\upmu$m sources were found to be starbursts, i.e. displaying an 
anomalously high star formation rate.

\vskip 1mm

The nature of SMGs was further explored by Wilkinson and co-workers \cite{Wilkinson2017} who investigated the clustering of galaxies in the S2CLS fields. A 
cross correlation analysis carried out on a sample of $\sim$600 counterparts from the UKIDSS Ultra Deep survey led to an estimation of the halo masses 
of these SMGs and a comparison with passive and star-forming galaxies selected in the same field. It was found that, on average, the SMGs occupy high-mass 
dark matter halos ($M_{halo}$ $>$ 10$^{13}$\,M$_\odot$) at redshifts $z$ $>$ 2.5, consistent with being the progenitors of massive elliptical galaxies 
found in present-day galaxy clusters. It was also found the that SMG clustering strength was consistent with star-forming population and that this appears 
to be the case across all redshifts. Recent work by Bourne and co-workers \cite{Bourne2017} explored the evolution of cosmic star formation in the S2CLS 
data sample. They concluded that the star formation history appears to undergo a transition at $z$ $\sim$ 3 -- 4, as unobscured structure growth in the early 
Universe is surpassed by obscured star formation, driven by the gradual build-up of the most massive galaxies in the Universe during the peak of cosmic 
assembly. The S2CLS catalogue and images have also presented an opportunity for follow-up work, both in terms of the properties of individual sources (e.g. 
with ALMA) and also in the statistical analysis of the entire sample.

\vskip 1mm

\vskip 1mm

\subsubsection{SASSy: The SCUBA-2 all sky survey}

Prior to the start of the legacy survey observing campaign the original ``all-sky'' survey (SASSy) was reborn as the more modest ``SCUBA-2 Ambitious Sky 
Survey'', with the aim of making the largest submillimetre map of the Outer Galaxy to identify the coldest and earliest regions of star formation. As 
described by Thompson et al. \cite{Thompson2013} the survey would cover of order 700 degree$^2$ between longitudes 120$^{\circ}$ and 240$^{\circ}$, 
extending to $\pm$2$^{\circ}$ from the Plane. The primary goal was to detect all the compact sources within the survey bounds above a few times the 
$\sim$40\,mJy noise level at 850\,$\upmu$m. The survey was allocated 480 hours of mainly poor weather (``band 5'') observing time, and was often used a 
fallback project in poor weather conditions. The sources identified are in the process of being compared to the \emph{IRAS} and \emph{Planck} catalogues to 
determine if any new objects have been detected. For example, in the 120 degree$^2$ region of the Galactic Plane covering longitude 120$^{\circ}$ $<$ $l$ 
$<$ 140$^{\circ}$ and latitude $|b|$ $<$ 2.9$^{\circ}$, Nettke et al. \cite{Nettke2017} produced a catalogue of $\sim$300 sources, of which 19 were new 
detections in comparison to \emph{IRAS}, 41 new detections compared to \emph{Planck} and 13 that were not found in either catalogue. Analysis continues of 
this very extensive dataset.

\subsubsection{SCUBA-2 observations of nearby stars survey (SONS)}

Although the main strength of SCUBA-2 is in wide-field mapping, the camera can also image compact sources very quickly and with high image fidelity. The 
SCUBA-2 Observations of Nearby Stars survey (SONS) targeted 100 nearby stars looking for evidence of debris discs -- the extrasolar analogues of the EKB in 
our Solar System. As described by Matthews et al. \cite{Matthews2007} the survey aimed to characterise these discs by: (1) 
providing direct dust masses that could not be obtained from shorter wavelengths; (2) adding to the far-IR/submillimetre spectrum to constrain the dust size 
distribution; (3) using the power of a 15\,m telescope to resolve disc structures around the nearest systems; and (4) looking for evidence of resonant 
clumps and other features in resolved structures that could be indicative of unseen perturbers, such as planets. Of particular importance was to 
investigate the diversity of exo-planetary system architectures, as this represents a key piece of information that will help link the formation and 
evolution of planetary systems with the evolution of planetary building blocks (planetesimals). The survey used 325 hours of observing time and for the 100 
targets reached an average RMS noise of $\sim$1.2\,mJy/beam. A total of 49 discs were detected, many for the first time, and 16 of the nearest discs were 
also spatially resolved by the JCMT.

\vskip 1mm

\begin{figure}[!h] 

\centering 

\includegraphics[width=120mm]{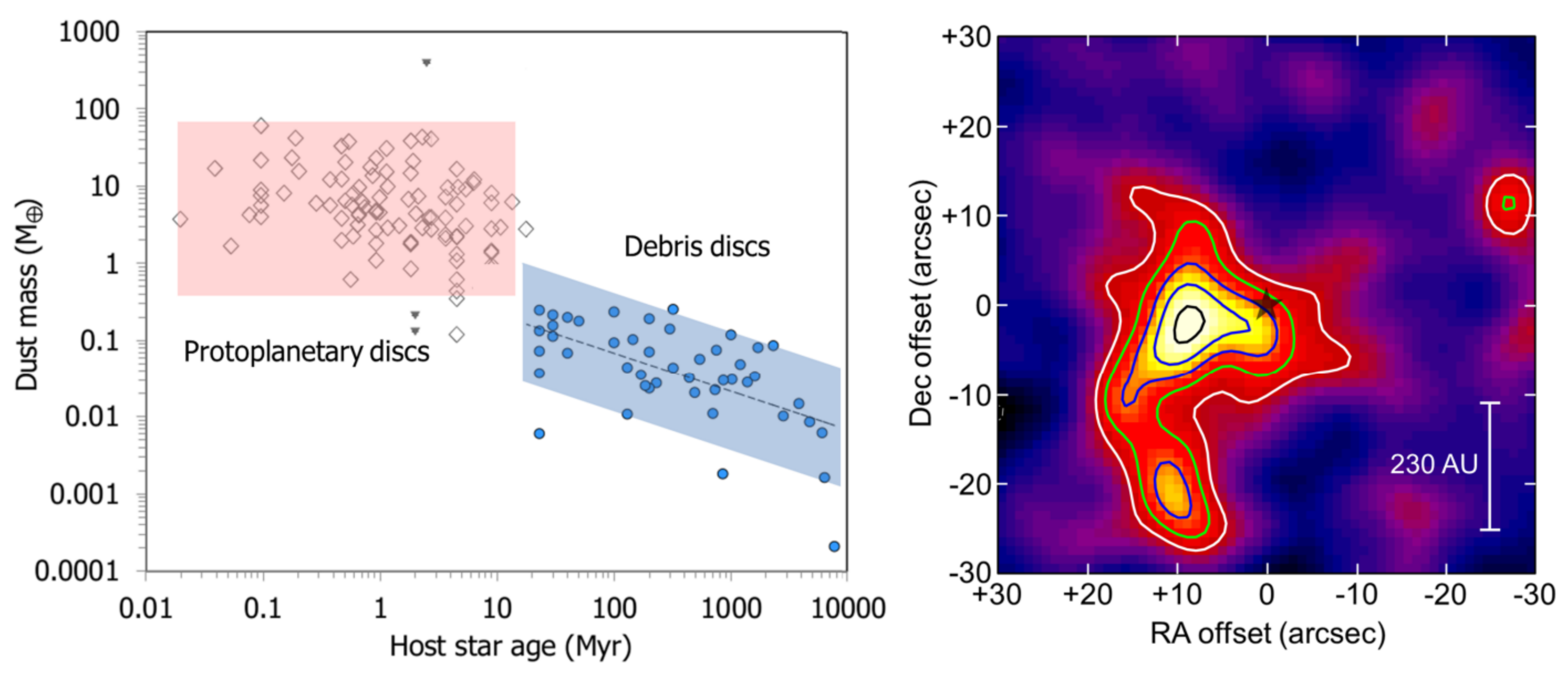} 

\caption{(left) The derived dust mass as a function of host star age, for both protoplanetary and debris discs (the latter exclusively from the SONS 
survey). (right) The SONS 850\,$\upmu$m image of the debris disc surrounding the nearby Sun-like star HD 38858. The dust mass figure was adapted from 
Matthews \& Kavelaars \cite{Matthews&Kavelaars2016} and includes the results from the SONS survey. The HD 38858 figure is from Holland et al. 
\cite{Holland2017}.}

\label{fig:debrisdiscs_SCUBA2} 

\end{figure}

The results from SONS, presented by Holland and collaborators \cite{Holland2017}, more than doubled the number of imaged discs from submillimetre 
observations. The discs are characterised in terms of their flux density, size (radial distribution of the dust) and derived dust properties from their 
spectral energy distributions. The mass of a disc, for particles up to a few millimetres in size, is uniquely obtainable from submillimetre observations, 
and shows a slow decline with age over hundreds of millions of years of stellar evolution (see the left panel of Fig.~\ref{fig:debrisdiscs_SCUBA2}). Many 
individual objects from SONS have also been studied with some surprising results. For example, observations of the nearby Sun-Like star HD 38858 revealed a 
large, extended structure, clearly with a flux peak offset from the star position (right panel of Fig.~\ref{fig:debrisdiscs_SCUBA2}). Kennedy et al. 
\cite{Kennedy2015} used multiple wavelength data, including from \emph{Herschel}, to determine that although the disc is clearly resolved by the SONS 
observations the peak to the south is most likely a background object. The offset nature of the peak emission is still puzzling, but the emission might 
indicate a perturbed disc that could have detectable volatiles. Similarly, observations of the nearby main sequence star 61 Vir, a system which has at 
least two known inner planets, reveal a resolved disc with a diameter of at least 80\,AU, i.e. very similar to the EKB in our Solar System. Marino and 
co-workers \cite{Marino2017} combined the SCUBA-2 data with ALMA observations to conclude that the disc is very likely extended from 60 to over 100\,AU and 
so represents a very broad parent planetesimal belt. The observations of 61 Vir have already shown the legacy of the SONS survey in that it is providing 
comprehensive target list for high angular resolution follow-up observations with submillimetre interferometers, such as SMA an ALMA.

\vskip 1mm

\subsection{The EAO era: A new call for large programs} 

On the 1st March 2015 the East Asian Observatory officially took over responsibility for the operations of the JCMT. Shortly after this (1st July) there 
was also a second call for ``Large Programs'' to be undertaken with the JCMT, covering the period from late 2015 until late 2018. A total of 7 programs 
were approved (6 would use SCUBA-2) and observations began in November 2015. Some 50\% of the total telescope time, amounting to at least 2400 hours, was 
to be dedicated to these programs. A second large program call for either extension to the existing ones or new initiatives was issued in February 2017.

\subsubsection{STUDIES - the SCUBA-2 Ultra Deep Imaging EAO Survey}

The objective of STUDIES is to obtain the first confusion-limited 450\,$\upmu$m map, centred on the COSMOS field at the northern edge of the CANDELS region 
\cite{Wang2016}. The single pointing (``Daisy map'') of around 10 arcminutes in diameter will eventually use 330 hours of the best weather available on 
Mauna Kea. This will take advantage of the high angular resolution offered by the JCMT/SCUBA-2 at 450\,$\upmu$m, compared to \emph{Herschel} at 350 and 
500\,$\upmu$m, and allow detections of faint galaxies with a significant higher surface density. The goal is to reach an RMS noise of 0.6\,mJy/beam at the 
centre of the field and to detect the dominant members in the dusty galaxy population that give rise to the bulk of the far-IR extragalactic background. 
Such a deep map will enable the detection of nearly all L$_{IR}$ > 10$^{12}$ L$_\odot$ galaxies at $z$ $<$ 4, and the majority of L$_{IR}$ $>$ 10$^{11}$ 
L$_\odot$ galaxies at $z$ $<$ 2. The observations will also allow, for the first time, a substantial overlap in the star formation rate range with galaxies 
detected by deep optical surveys. This will provide a more complete census of the cosmic star formation that is both obscured and unobscured by dust. In 
just over a year (Nov 2015 -- Feb 2017) the observations were 40\% complete, and have reached an RMS of 1\,mJy/beam in the centre of the image. A total of 98 
sources have been detected so far at a significance of $>$ 4$\sigma$, with this number expected to dramatically increase as the map goes deeper 
\cite{Wang2016}.

\subsubsection{SC2-COSMOS}

\begin{figure}[!h] 

\centering \includegraphics[width=65mm]{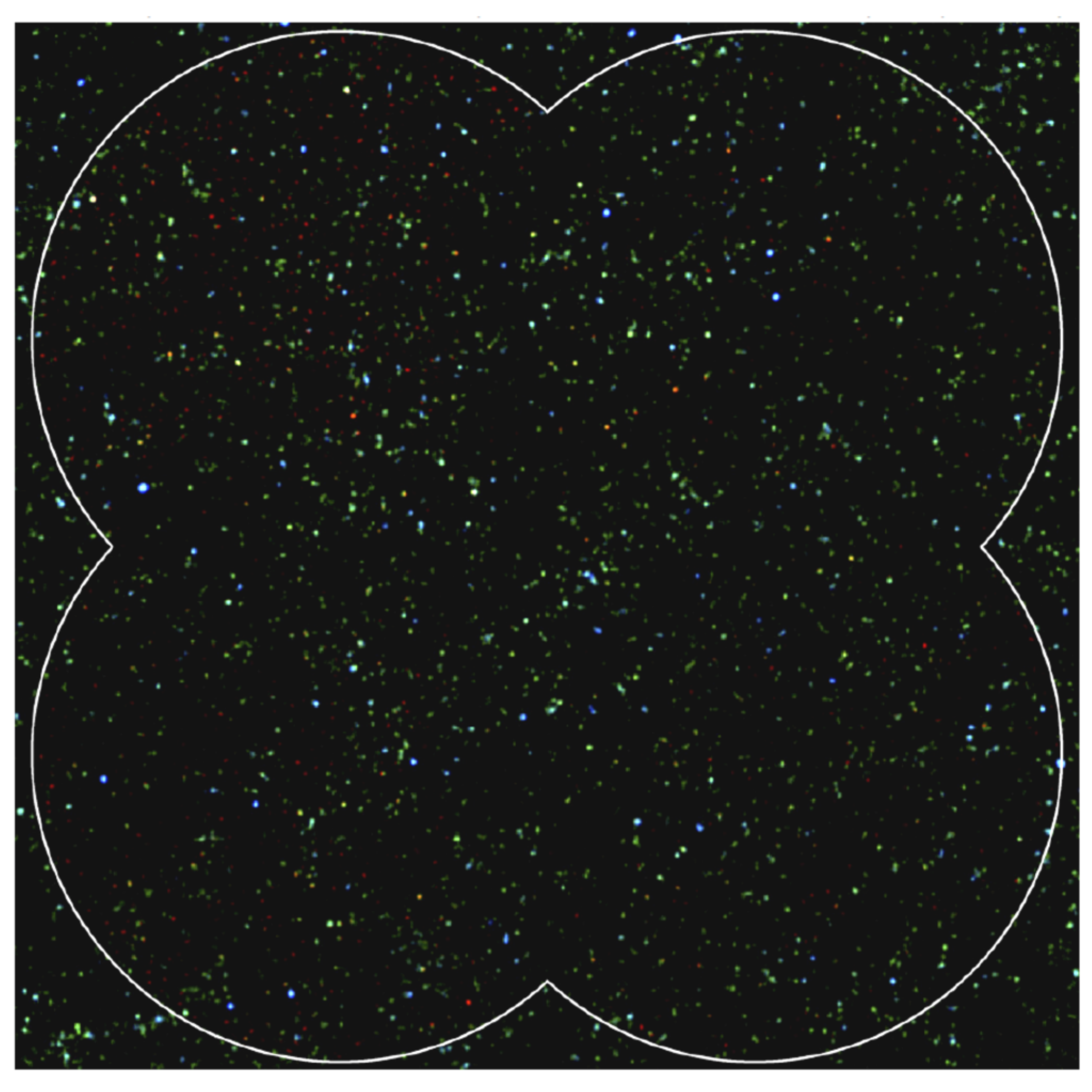} 

\caption{True colour image of the COSMOS field created from SCUBA-2 850\,$\upmu$m (red) and \emph{Herschel}/SPIRE 250\,$\upmu$m (blue) and 350\,$\upmu$m 
(blue). The map highlights the rare dusty sources which are most likely some of the highest redshift galaxies in the sample. The white line represents the 
coverage of the 1.4 $\times$ 1.4 degree SC2-COSMOS map. Figure from Matsuda \cite{Matsuda2016}.}

\label{fig:COSMOS_SCUBA2} 

\end{figure}

COSMOS is a survey of $\sim$1000 submillimetre galaxies in the 2 degree$^2$ COSMOS field \cite{Matsuda2016}. The region is the pre-eminent ALMA-visible, 
degree-scale, extragalactic survey field, and has been studied extensively from the X-ray to the radio. The goal is to first complete the 850\,$\upmu$m map 
of the full COSMOS field (partly covered by SCUBA-2 Cosmology Legacy Survey) to a depth of 1.5\,mJy/beam, and to then increase the depth of this map to 
1.2\,mJy. This map will have twice the area of similar surveys in a single contiguous field, allowing unique tests of the clustering of the submillimetre 
galaxy population on scales up to $\sim$60\,Mpc. By March 2017 the observations were 79\% complete against an allocation of 223 hours, and an RMS noise of 
$<$1.5\,mJy/beam has been achieved across the entire field (as shown in Fig.~\ref{fig:COSMOS_SCUBA2}). Already, some 1400 submillimetre sources have been 
detected in the field and an analysis of the multi-wavelength properties of these galaxies is underway. Follow-up observations with ALMA of the 150 
brightest sources are also planned.

\vskip 1mm

\vskip 1mm

\subsubsection{JINGLE - the JCMT dust and gas in Nearby Galaxies Legacy Exploration}
	
JINGLE is a survey designed to systematically study the cold ISM of galaxies in the local Universe \cite{Xiao2016}. The survey will provide 850\,$\upmu$m 
images with SCUBA-2 for a sample of 192 \emph{Herschel}-selected galaxies, as well as integrated CO(2-1) line fluxes with the heterodyne Receiver A (RxA) 
for a subset of 62 of these galaxies. A total of 780 hours has been allocated to the survey over the 3 year period. The sample builds on multiple surveys 
including \emph{Herschel}/H-ATLAS and the MaNGA optical integral-field spectroscopy surveys. By combining the results from the JCMT observations with all 
these ancillary data, JINGLE will allow for a detailed characterisation of the gas and dust properties of galaxies in the local universe. Scientific 
objectives include studying the dust-to-gas ratio and how it varies across the galaxy population, correlating the molecular gas content with 
spatially-resolved galaxy properties, and investigating the correlation between ISM properties and the dynamics of galaxies. The scaling relations between 
dust, gas, and global physical properties will also provide critical benchmarks for high-redshift studies with JCMT and ALMA. As of early 2017, the survey 
is 36\% complete with 106 galaxies already observed with SCUBA-2.

\subsubsection{BISTRO - B-fields in star-forming region observations}

\begin{figure}[!h] 

\centering \includegraphics[width=65mm]{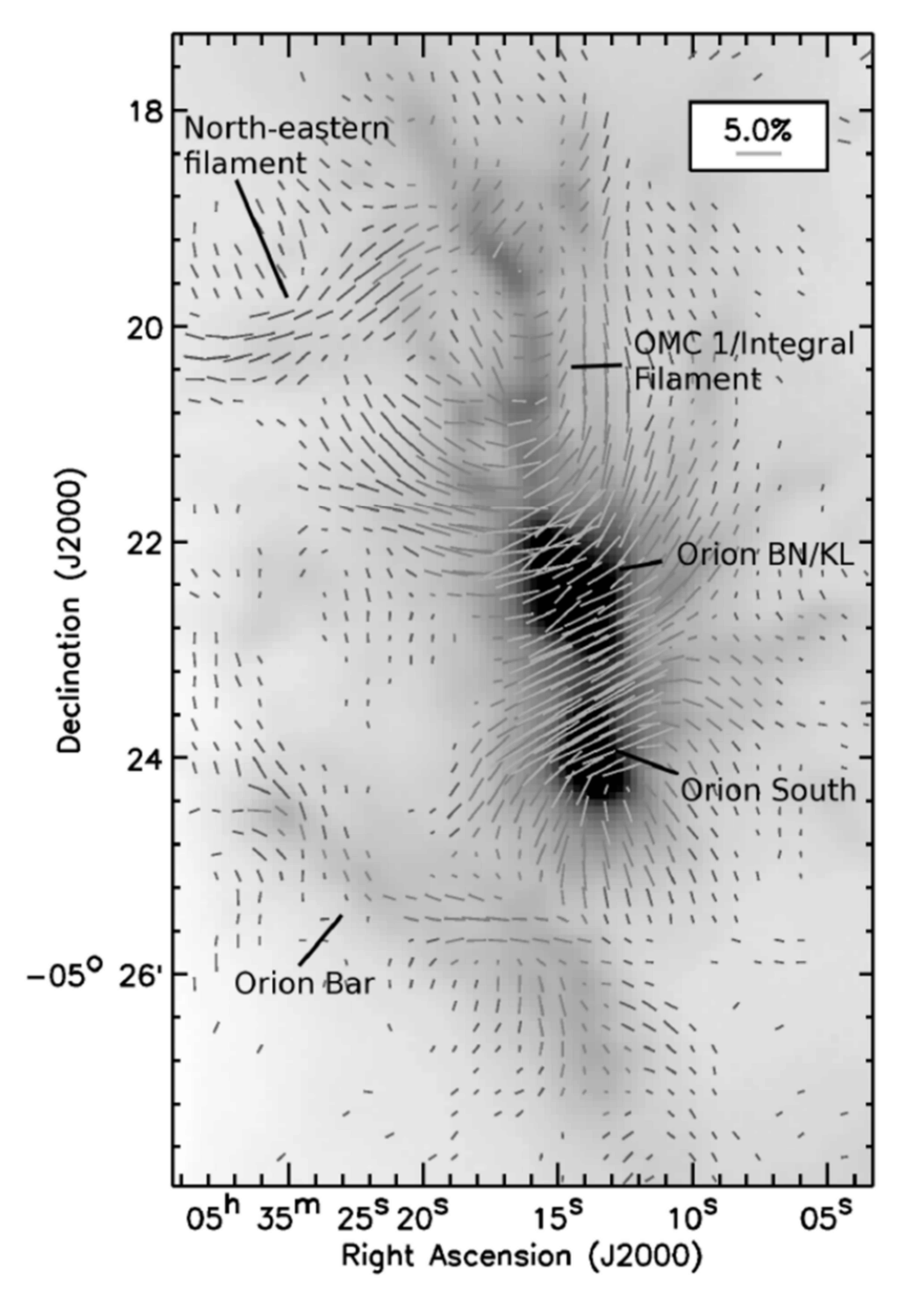} 

\caption{850\,$\upmu$m polarisation map of the central region of the Orion molecular cloud, in which the measured polarisation vectors are rotated by 90 
degrees to show the B-field direction. Figure from Ward-Thompson et al. \cite{Ward-Thompson2017}.}

\label{fig:Bfield_SCUBA2} 

\end{figure}

Without accurate knowledge of the collapse process of molecular clouds, it is not possible to understand how a star forms. The exact role of magnetic 
fields (B-fields) in this process is still open to considerable debate, and so the BISTRO survey will address this by tracing the direction and strength of 
the magnetic field on scales of $\sim$1000 -- 2000\,AU in the central regions of several nearby molecular clouds (e.g. Orion, Ophiuchus, Taurus L1495) 
already observed with the previous Gould Belt legacy survey \cite{Furuya2016}. The scientific objectives are to assess the relative importance of magnetic 
field and turbulence in the star formation process, to test models of magnetic ``funnelling'' of materials onto filaments and to investigate the role of 
B-fields in shaping protostellar evolution (including bipolar outflows from young protostars). The survey uses a rotating half-waveplate polarimeter that 
was developed specifically for use with SCUBA-2 \cite{Friberg2016} and was awarded a total observing time of 224 hours of good weather (``band 2''). As of 
early 2017, some 38\% of the programme has been completed, with regions such as Orion A, Ophiuchus, and Serpens Main already observed. An example of one of 
the early observations is shown in Fig.~\ref{fig:Bfield_SCUBA2} from Ward-Thompson et al. \cite{Ward-Thompson2017} for the central region of the Orion 
molecular cloud. The image shows that magnetic field lies perpendicular to the famous ``integral-shaped filament'' and may be 
responsible for ``funnelling'' matter onto filaments to aid the formation of dense cores that eventually become protostars.

\vskip 1mm

\vskip 1mm

\subsubsection{SCOPE - SCUBA-2 Continuum Observations of Pre-protostellar Evolution}

It is known that stars form in the densest regions within molecular clouds, the earliest phase being linked with the so-called pre-stellar cores. However, 
the formation and early evolution of these cores in different environments is not well known. The SCOPE survey is carrying out an ``all-sky'' survey at 
850\,$\upmu$m of a sample of 2000 cold clumps identified by the \emph{Planck} surveyor \cite{Tatematsu2016}. The JCMT/SCUBA-2 is more sensitive to cold dust 
than \emph{Herschel}, and also has high angular resolution to resolve the substructure of \emph{Planck} cold clumps. The aims of the survey 
include the study of how dense cores form and how star formation varies as a function of environment, the universality of filaments in the cold ISM and 
their roles in generating dense cores, how dust properties change in different environments, and how dust properties affect the chemical evolution of dense 
cores. The survey is being supplemented with observations from other millimetre and radio telescopes (such as Purple Mountain observatory and the Nobeyama 
Radio Observatory 45\,m telescope), and will also form a legacy database for such studies with other instruments, especially ALMA. The survey was awarded 
300 hours of band 3 and 4 time, and as at March 2017 was 69\% complete.

\subsubsection{TRANSIENT - A transient search for variable protostars: How do stars gain their mass?}

The TRANSIENT program is using SCUBA-2 to measure accretion variability in protostars in eight fields within nearby star-forming regions 
\cite{Herczeg2016}. It has been found that an outburst in accretion luminosity heats the dust in the envelope, which is then seen as brighter emission at 
850\,$\upmu$m \cite{Johnstone2013}. Far-IR and submillimetre observations provide a snapshot of accretion rate, averaged over the few-weeks heating 
timescale for the luminosity burst to propagate through the envelope. Given the difficulty of carrying out such long-term observations in the far-IR (the 
need for a space-based observatory) submillimetre monitoring may be the only way to probe the earliest stages of stellar growth, since these stars are so 
heavily embedded they are not visible at optical/near-IR wavelengths. The monitoring programme includes a total of 182 embedded protostars (Class 0/I YSOs 
with envelopes) and 800 disc/flat-spectrum objects. Each region will be observed once a month for a total period of 3 years to search for signs of 
variability across epochs. The program was awarded 150 hours of observing time, split equally in weather bands 1, 2 and 3. As of March 2017 the program is 
27\% complete, and the first variable candidate has already been identified.

\subsection{Other science highlights from SCUBA-2}

\subsubsection{Observations of Comet ISON}

In late 2013 the JCMT launched a campaign to study the chemistry of the sun-grazing comet C/2012 S1 (ISON). Several groups of observers concentrated on 
measuring the production rate of HCN and water as the comet approached perihelion. The comet was also observed multiple times with SCUBA-2, catching the 
final hours before it disintegrated. Keane led the team on these observations \cite{Keane2016} which showed that as ISON approached perihelion the 
continuum emission from the nucleus became an elongated dust column spread out over 60 arcseconds ($\sim$10$^5$\,km) in a direction away from the Sun. One 
of the final images reveals distinct clumps, consistent with the catastrophic disruption of the comet, producing $\sim$5 $\times$ 10$^{10}$\,kg of 
millimetre-sized dust.

\subsubsection{New insights on planet formation}

SCUBA-2 has also been used to explore and place new constraints on the dust and gas mass of protoplanetary discs during the giant planet building phase. 
Williams and collaborators \cite{Williams2013} surveyed a half-degree field towards the $\sigma$ Orionis cluster, which contains almost 300 young stellar 
objects with estimated ages of 3\,Myr. Only nine stars were detected from the observations at 850\,$\upmu$m with these having estimated disc masses of 
between 5 and 17 Jupiter masses. Using a stacking analysis the mean mass for 83 infrared-detected objects that were not detected by SCUBA-2 was determined 
to be 0.5 Jupiter masses, effectively ruling them out of ongoing planet formation. The lack of emission illustrates how little raw material must remain in 
the environs of the vast majority of these young objects. This suggests that planet forming must start very early on, and that the growth of planetary cores 
must be largely complete within a couple of Myr after the host star becomes optically visible.

\subsubsection{Dust surrounding a pulsar}

The nearby Geminga pulsar is believed to have crossed the Galactic Plane in the last 100,000 years. Greaves \& Holland \cite{Greaves&Holland2017} report 
the detection of a shell of material surrounding Geminga that could have formed from compression of the local interstellar medium. A compact source is 
detected from 450\,$\upmu$m observations which may evidence for the existence of a circum-pulsar disc, the first time any such structure has been detected 
in the submillimetre. The inferred mass of dust is expected to exceed 6 Earth masses, and so has the potential to form low-mass planets such as the 
archetypes around PSR B1257+12 \cite{Wolszczan&Frail1992}. Further imaging at high angular resolution is planned for this object.

\subsubsection{Embedded binaries and their dense cores}

The relationship between young, embedded binary stars and their parent cores is not well understood. Sadavoy \& Stahler \cite{Sadavoy&Stahler2017} used VLA 
and SCUBA-2 observations of a number of young stars and cores in the Perseus molecular cloud to explore the origin of binary stars. It was revealed that 
most embedded binaries are found towards the centres of their parent cores. Wide ($>$500\,AU separation) binaries tend to be aligned with the long axes of 
the core, whilst tight systems show no preferred orientation. The authors tested a number of evolutionary models in an attempt to account for the 
populations of both single and binary Class 0 and I sources. The model that best fits the observations suggests that all stars form initially as wide 
binaries, and then either break up into separate stars or shrink into tighter orbits. Future observations will explore whether the high mass fraction of 
dense cores that become stars in Perseus is similar in other star-forming regions.

\subsubsection{SUPER GOODS: ultra-deep imaging of the GOOD-N field}

In addition to the extensive surveys that formed the Cosmology Legacy survey, Cowie and co-workers \cite{Cowie2017} carried out ultra-deep imaging with 
SCUBA-2 of the GOODS-N field. The maps, covering 450 arcmin$^2$, detected 31 and 186 sources at 450\,$\upmu$m and 850\,$\upmu$m, respectively, and reached 
RMS noise levels well below the confusion limit at 850\,$\upmu$m. Using extensive VLA and SMA observations to pinpoint exact galaxy locations, and Keck 
spectra to determine resdhifts, it was shown that the star formation rate of these galaxies reaches a peak at $z$ = 2 -- 3, before dropping at higher 
redshifts. It was also suggested that the shape in the number density of galaxies per unit volume as a function of star formation rate is invariant over 
this particular redshift range.

\subsubsection{The space density of galaxies at $z > 4$}

Until the advent of \emph{Herschel} only a handful of dusty star forming galaxies were known to exist at redshifts greater than 4 and most of these 
were amplified by gravitational lensing. Ivison and co-workers \cite{Ivison2016} selected 109 galaxies for SCUBA-2 imaging based on their extremely red 
far-infrared colours and faint 350\,$\upmu$m and 500\,$\upmu$m fluxes from the \emph{Herschel}-ATLAS imaging survey. The addition of the submillimetre data 
allowed the peak of the spectral energy distribution to be identified and so led to better constraints on the redshifts of these objects. The galaxies were 
determined to be in the redshift range 3.3 to 4.3 (median value of 3.66), with a third lying at $z$ $>$ 4 suggesting a space density of $\sim$6 $\times$ 
10$^{-7}$\,Mpc$^{-3}$. The sample contains some of the most luminous star-forming galaxies and the most overdense cluster of early starburst ellipticals 
known to date.


\section{Science with the JCMT heterodyne instrumentation}

At the time of the dedication of the JCMT in March 1987, two heterodyne receivers were in operation on the telescope; namely a polarisation splitting 
dual-channel 230\,GHz band receiver (RxA) \cite{Padman1989} and a single channel 345 GHz band receiver (RxB) \cite{Parker1991,Avery1992}. Both receivers 
were equipped with Schottky diode mixers. The 230\,GHz receiver, RxA, covered the range 220 -- 280\,GHz using two sets of mixers -- one set for the lower 
part and one for the upper part of the band. RxA was operated for a short-time as a dual-channel receiver but for most of the time operated in a hybrid 
mode with mixers in the opposite polarisation covering different frequency ranges. The 345\,GHz receiver covered the range 320 -- 370\,GHz and used a 
carcinotron as the local oscillator (LO) source.  An early goal was to equip the telescope with state-of-the-art single- or dual-pixel SIS receivers and 
subsequently array receivers with a priority for the 345\,GHz band. This was made more feasible with Canada joining the UK/Netherlands project in the 
spring of 1987, injecting additional resources into the JCMT instrument development fund. A number of single-feed SIS receivers were deployed in the early 
part of the 1990s followed by polarisation-splitting, dual-channel receivers.

\vskip 1mm

\begin{figure}[!h] 

\centering \includegraphics[width=100mm]{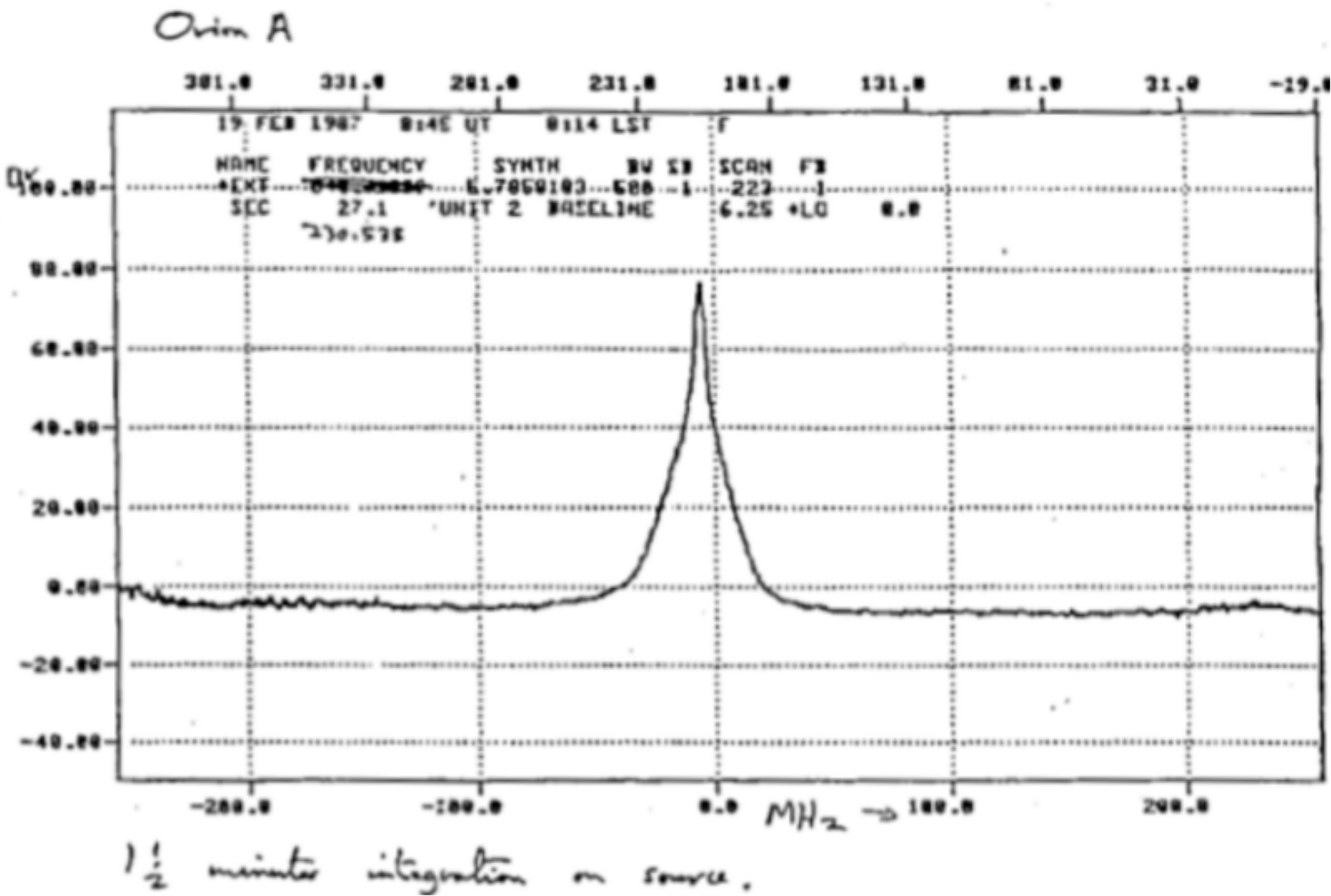} 

\caption[example] {One of the first spectra obtained by the JCMT. A CO J = 2 -- 1 spectrum of OMC1 with RxA from January 1987. Figure from the third issue of 
the JCMT newsletter PROTSTAR (ISSN 0267-1247).}

\label{fig:firstspec}
\end{figure}

In the early years heterodyne science observations were constrained to single-point spectra (as shown in Fig.~\ref{fig:firstspec}) or making small maps up 
to a few arcminutes in size. Observing larger areas was too time consuming for the single pixel instruments with their limited sensitivity and spectrometer 
dump-time. The first change occurred in 1992 with the introduction of the DAS spectrometer \cite{Bos1986} allowing dump-times to be reduced to a second and 
the associated development of ``On-The-Fly'' software for heterodyne mapping. This, combined with more sensitive instruments, caused a noticeable increase 
in the data rate and generated a need for more storage space. In 2007 with the Heterodyne Array Receiver Program (HARP) \cite{Buckle2009} the JCMT became 
the premier observatory for mapping lines in the 350\,GHz band. HARP has a array of 16 mixers, each spaced by 30 arcseconds on the sky, and was supported 
by a 16-channel auto-correlation spectrometer (ACSIS) \cite{Buckle2009} capable of observing a 2\,GHz bandwidth in each pixel. HARP is now used extensively 
for large programmes and is still, after 10 years of operation, a very competitive instrument.

In parallel with the development of receivers by the partners, agreements were also made with external groups to bring receivers to the telescope. A 
345\,GHz SIS receiver from Sutton's group was brought in for two short periods in 1989 and 1990. A 600 and 800\,GHz receiver from Genzel's group was 
used in each year from 1988 to 1996 \cite{Harris1987,Harris1994}, whilst the South Pole Imaging Fabry-P\'{e}rot Instrument (SPIFI) \cite{Bradford2002} visited  
the JCMT for short periods between 1999 -- 2000. Although a few upgrades were carried out, no new heterodyne projects were started after 2000, and 
a high-frequency receiver project (RxE) was cancelled due to budget constraints. The receiver and spectrometer deployments are summarised in 
Table~\ref{tab:tab1} and Table~\ref{tab:tab2}, respectively.

\begin{table}[!h]

\centering

\caption{Spectroscopic receiver instrumentation for the JCMT}

\begin{tabular}{ | l | c | c | l | l |}

\hline

Name   & Frequency range        & Feeds/pixels      & Mixer/detector      & Operation period     \\ 
       & (GHz)                  &                   & technology          &                      \\ \hline \hline
RxA1   & 211 -- 280             & 1$^{(1)}$         & Schottky            & 1987 -- 1992         \\ \hline
RxA2   & 211 -- 280             & 1                 & SIS                 & 1992 -- 1998         \\ \hline
RxA3   & 211 -- 276             & 1                 & SIS                 & 1998 --              \\ \hline
RxB1   & 320 -- 380             & 1                 & Schottky            & 1988 -- 1990         \\ \hline
Sutton & 320 -- 380             & 1                 & SIS                 & 1989 -- 1990         \\ \hline
RxB2   & 320 -- 380             & 2                 & Schottky            & 1990 -- 1991         \\ \hline 
RxB3i  & 320 -- 380             & 1                 & SIS                 & 1991 -- 1997         \\ \hline
RxB3CU & 320 -- 380             & 2                 & SIS                 & 1997 -- 2006         \\ \hline
HARP   & 325 -- 375             & 16                & SIS                 & 2007 --              \\ \hline
RxC1   & 460 -- 495             & 1                 & HEB                 & 1989 -- 1993         \\ \hline
RxC2   & 450 -- 495             & 1                 & SIS                 & 1993 -- 1998         \\ \hline
RxW    & 430 -- 500$^{(2)}$     & 2                 & SIS                 & 1998 -- 2014         \\
       & 630 -- 705             & 2                 & SIS                 & 1998 -- 2014         \\ \hline 
RxG    & 600/800$^{(3)}$        & 1                 & Schottky            & 1988 -- 1994         \\ \hline
RxG2   & 460--495               & 1                 & SIS                 & 1994 -- 1996         \\ \hline
SPIFI  & 630 -- 700             & 25                & Silicon bolometers  & 1999 -- 2000         \\ \hline

\end{tabular}

\begin{flushleft} 

\vskip 0.5mm

$^{(1)}$RxA1 was a dual-channel, polarisation-splitting receiver normally operated with a low frequency mixer (220 -- 240\,GHz) in one channel, and a high 
frequency mixer (240 -- 280\,GHz) in the other. This effectively made the receiver a single feed at any given frequency.

\vskip 0.5mm

$^{(2)}$The 430 -- 500\,GHz mixers were replaced with 325 -- 375\,GHz HARP mixers in 2007 to support operation with the SMA at 345\,GHz.

\vskip 0.5mm

$^{(3)}$RxG operated in the 600\,GHz and 800\,GHz bands with a far-IR laser LO system. Using different gases as the lasing medium allowed tuning to a 
number of scientifically-important frequencies.

\end{flushleft}

\label{tab:tab1}

\end{table}

\begin{table}[!h]

\centering

\caption{Backend spectrometers for the JCMT}

\begin{tabular}{ | l | c | c | l | l |}

\hline

Name      & Type                 & IF inputs       & Operation period         \\ \hline\hline
Kent      & Correlator           & 1               & 1987 -- 1988                   \\ \hline
AOSD      & AOS                  & 1               & 1988 -- 1989                  \\ \hline     
AOSC      & AOS                  & 1               & 1989 -- 1993                   \\ \hline     
DAS       & Correlator           & 8               & 1992 -- 2006                  \\ \hline     
ACSIS     & Correletor           & 16              & 2006 --                         \\ \hline    

\end{tabular}

\label{tab:tab2}

\end{table}

\subsection{Chemistry}

Chemistry is an integral part of heterodyne line observing: without an understanding of the chemistry the data cannot be exploited to the fullest extent. 
An example is the much studied conversion from observed CO intensity to the total molecular mass. Thus, many papers study chemistry in different regions 
such as galaxies, hot cores, protostellar envelopes and discs, shocks and evolved stellar envelopes. This not only gives a better understanding of the 
chemistry but also an improved knowledge of the source morphology and physics. The information from spectral scans aims to catalogue molecules, lines and 
abundances, typically in well-known objects, and often without extensive modelling or analysis of the physical conditions. There have been ten spectral 
line surveys published using JCMT data covering a large fraction of the 230, 345, 460 and 650\,GHz windows. Most of the surveys were conducted in the 1990s 
using single pixel receivers, and groups, for example, led by Greaves \cite{Greaves1991}, Sutton \cite{Sutton1991,Sutton1995} and MacDonald 
\cite{MacDonald1996} studied sources including OMC1, Sgr B2, W3, G34.4, IRAS 16293--2422 and IRC+10216. The citation count is still steadily increasing for 
these papers, demonstrating their great legacy value. The JCMT Spectral Line Legacy Survey (SLS) by Plume and co-workers \cite{Plume2007} was designed to 
study and catalogue the lines in some typical regions -- a low-mass core (NGC 1333 IRAS 4), three high-mass cores spanning a range of star-forming 
environments and evolutionary states (W49, AFGL 2591, and IRAS 20126), and the Orion Bar photo-dissociation region. The SLS used HARP and in contrast to 
most spectral surveys, a region around the central source was also observed, thereby giving additional information about the morphology and chemical 
variations in the local environs. In their SLS paper on the Orion bar, van der Wiel et al. \cite{vanWiel2009} find that the molecular abundances, in 
general, followed the layered structure as predicted by models of photo-dissociation regions, but there were also discrepancies between the models and the 
observations.

\vskip 1mm

Chemistry and physics in particular regions and/or molecular species have been the topic of a large number of papers. Two such papers by van Dishoeck et al. 
\cite{Dishoeck1995} and van der Tak et al. \cite{vanTak2000} have well over 200 citations each. The paper by van Dishoeck is a spectral scan of IRAS 
16293--2422 and uses the observations to isolate the physical regions around the source with different physical and chemical properties. Results for 
sulphur and silicon species in the same source were reported by Blake et al. \cite{Blake1994}, and a similar study of IRAS 16293--2422 using more 
extensive JCMT and IRAM observations was published by Caux and collaborators \cite{Caux2011}. The paper by van der Tak studied the 
structure around high-mass YSOs using a number of spectral lines as well as continuum and mid-IR data. The data were used to delineate the physics and 
chemistry in the different parts of the envelope, where freeze-out of CO was demonstrated as well as grain evaporation in the inner region. The 
chemistry and photo-ionisation of the Orion bar was studied by Hogerheijde et al. \cite{Hogerheijde1995} and van der Werf et al. \cite{vanWerf1996}, and 
these papers resolved not only some of the stratification observed in the earlier mentioned SLS paper, but also concluded that the gas in the bar must be 
clumpy.

\vskip 1mm

Sulphur chemistry in hot cores was studied by Hatchell et el. \cite{Hatchell1998} who found that abundance ratios of the major sulphur species did not 
vary between different hot cores, and, with the exception of carbonyl sulphide (OCS), were in agreement with models. Hence, no evolutionary sequence was 
found for hot cores. Van der Tak and van Dishoeck \cite{vanTak2000b} used the H$^{13}$CO$^+$ abundance to constrain the cosmic ray ionisation rate in the 
envelope of YSOs: cosmic ray ionisation forms H$_3^+$, which is then destroyed by reaction with CO in molecular regions forming HCO$^+$. Models of 
gas-grain chemistry were tested by van der Tak and collaborators \cite{vanTak2000c} by observing H$_2$CO and CH$_3$OH towards massive YSOs; a large 
number of lines allowed excitation temperatures and abundances to be determined. The CH$_3$OH/H$_2$ abundance shows a jump from 10$^{-9}$ to 10$^{-7}$ 
that could be attributed to grain evaporation due to radiation based on a corresponding jump in excitation temperature and correlated (IR measured 
excitation) temperature of C$_2$H$_2$. Sch\"{o}ier et al. \cite{Schoier2002} observed IRAS 16293--2422 and derived a detailed temperature and density 
structure and a detailed comparison of the observations with models strengthened the evidence for infall in the envelope. Furthermore, the molecular 
species are divided into those that have constant abundance in the envelope and those that have increased abundance close to the core. The later 
molecules, like CH$_3$OH, SO and SO$_2$, increase in abundance where the temperature can thermally evaporate molecules from grains. Low mass YSOs, such 
as IRAS 16293--2422, have hot cores but the chemical timescales are much shorter than in the hot cores of massive YSOs.

\vskip 1mm

Organic molecules were observed towards T Tauri and Herbig Ae stars by Thi et al. \cite{Thi2004}. The detections showed that the emission was from dense 
gas at moderate temperature with some species, such as CN, enhanced by photo-dissociation. This is consistent with accretion disc models with a cold 
mid-plane having chemistry affected by freeze-out onto grains, whilst molecules formed by photo-dissociation are enhanced on the disc surface by 
radiation from the central object. The ortho/para ratio in H$_2$ was been studied by Pagani and co-workers \cite{Pagani2009} using N$_2$D$^+$, N$_2$H$^+$ 
and ortho H$_2$D$^+$ lines. The ortho/para ratio in H$_2$ is important for understanding the deuteration amplification in the clouds. Under some 
conditions the abundance of HDCO and CH$_2$DOH has been found to be higher than their un-deuterated analogue species, showing a deuterium enhancement of 
10$^6$ times the D/H ratio. This only occurs if molecules like CO are heavily frozen-out onto grains. Hence, deuterium enrichment is important for 
understanding and studying the freezing-out of molecules onto grains. Frozen-out deuterium-enriched molecules serve as a marker of pristine 
material in comets.

\subsection{Extragalactic sources}

About 150 papers studying galaxies using JCMT heterodyne data have been published, and not unexpectedly, many of the these papers utilise the 346\,GHz CO J 
= 3 -- 2 line, which is easily observable during typical Mauna Kea weather conditions. An early example is the paper by Devereux and co-workers 
\cite{Devereux1994} that reported on observations of CO(3--2) in the centre regions of starburst galaxies. They found that the ratio between the CO(3--2) 
and CO(1--0) lines was higher in the centre of starburst galaxies than in Galactic molecular clouds, whilst the gas mass was typically 10\% of the total 
dynamical mass. To explain the difference to Galactic molecular clouds from earlier studies, the larger ratio required a more complicated model than just 
one in which the gas was hotter. Yao et al. \cite{Yao2003} extended these investigations and observed 60 IR-luminous sources selected from the SCUBA Local 
Universe Galaxy Survey (SLUG). The authors reported an almost identical average CO(3--2) to CO(1--0) line ratio but with a much larger spread in values, 
indicating a large variation in excitation of the gas in IR-luminous galaxies. In parallel, higher level CO lines as well as lines from HCN, HCO$^+$, 
HNC$^+$ and CS were used to study the gas excitation in starburst and IR-bright galaxies. The CO(6--5) transition, observed by Harris et al. in 1991 
\cite{Harris1991}, clearly showed that the gas in nearby starburst galaxies such as M\,82 and NGC\,253 was hotter and denser than in typical Galactic 
clouds. Such studies were extended to high redshift starbursts by Padadopoulous and collaborators \cite{Papadopoulos2000} with the detection of CO(4--3) in 
galaxies at redshifts of 3.79 and 3.53. A number of possible sources for the excitation were discussed in these papers, such as violent turbulence, the 
presence of OB stars, or cosmic rays due to AGNs. Bradford et al. \cite{Bradford2003} concluded that in the case of NGC\,253, the CO(7--6) emission was 
excited by cosmic rays due to the high supernova activity in the region.

\vskip 1mm

Aalto et al. \cite{Aalto2007} discovered an unexpectedly high HNC/HCN line ratio in star-forming galaxies. In Galactic warm dense gas this ratio is lower, 
even in photo-dissociation regions. The observed high line ratio was explained by IR excitation of HNC, which has a much lower energy bending mode than HCN, 
or, alternatively, by X-ray dominated chemistry due to the presence of AGNs. The paper by Greve et al. \cite{Greve2009} studied the starburst galaxies Arp220 
and NGC 6240 in several molecular species and transitions. The authors showed that the emission from these molecules traced different densities and there 
is a size-density relationship for the gas, similar to, but steeper than that observed in Galactic clouds. The bulk of the gas mass \textasciitilde 
(1--2)$\times$10$^{10}$ M$_\odot$ resides in a dense n = 10$^5$ -- 10$^6$ cm$^{-3}$ warm phase. Papadopoulos et al. \cite{Papadopoulos2012} presented 
spectral line energy distributions for 70 LIRGs, with the galaxies covering a range of infrared luminosities and morphologies showing a broad range of ISM 
conditions. On the high excitation side the ISM is dominated by hot ($>100$\,K) and dense (N$>10^4$ cm$^{-3}$) molecular gas with gas mass reservoirs of 
\textasciitilde (few) 10$^9$\,M$_\odot$. The authors conclude that the gas excitation in merger driven ULRIGs is dominated by turbulence and cosmic rays 
rather than UV/optical photons and supernova shocks. This new understanding of the gas phase in massive star-forming galaxies was used to guide later 
observations with the \emph{Herschel} satellite and ALMA. Another noteworthy and highly-cited galaxy paper was presented by Edge \cite{Edge2001} who showed 
that hot gas in galaxy clusters, cooled by X-ray emission, generates a cooling flow of gas onto the galaxies in the clusters. Searches for CO in the 
central galaxies of clusters with cooling flows had only provided one detection before Edge reported 16 more detections of CO in galaxies at the centre of 
clusters with cooling flows.

\vskip 1mm

The NGLS (see Section 4a(iii)) observed an HI-selected sample of 155 galaxies spanning all morphological types with distances less than 25\,Mpc. The 
survey has so far produced 10 papers e.g. Wilson et al. \cite{Wilson2009}. The objective of the heterodyne component of the survey was to study the gas 
properties, gas--to--dust ratio and to compare radial profiles of the dust, HI and CO emission. The authors find a wide range of molecular gas mass 
fractions in the galaxies in the sample. By comparing the NGLS data with merging galaxies at low and high redshift, which have also been observed in the 
CO J = 3--2 line, they show that the correlation of far-IR and CO luminosity shows a significant trend with luminosity. This trend is consistent with a 
molecular gas depletion time that is more than an order of magnitude faster in the merger galaxies than in nearby normal galaxies.  There is also a 
strong correlation of the $L_{\rm{farIR}}$/$L_{\rm{CO(3-2)}}$ ratio with the atomic-to-molecular gas mass ratio. This correlation suggests that some of 
the far-infrared emission originates from dust associated with atomic gas and that its contribution is particularly important in galaxies where most of 
the gas is in the atomic phase.

\subsection{Clouds, Cores and Galactic structure}

Observing and mapping molecular clouds and cores in CO or other lines is a common topic for JCMT heterodyne papers, with around 150 papers having ``cloud'' 
and/or ``core'' in their title. Some well-cited examples include the paper by Davis et al. \cite{Davis1999} who mapped the Serpens molecular cloud in the 
CO 2 -- 1 line and continuum, identifying cores and outflows and estimating their ages. The paper by Kirk and collaborators \cite{Kirk2007} studied the 
kinematics of dense cores in the Perseus molecular cloud with the N$_2$H$^+$ 1 -- 0 and CO 2 -- 1 lines, and found that the internal motion measured by the 
N$_2$H$^+$ line-width in the SCUBA-selected dense cores was more than sufficient to support against gravitational collapse. Whilst many cloud regions were 
mapped early on with the JCMT heterodyne instruments, large-scale mapping did not start until 2007. The change was triggered by the decision to allocate 
large amounts of time to large surveys and the arrival of HARP/ACSIS made this feasible on the heterodyne side. This shift in policy was instigated to keep 
the JCMT competitive in the era of large submillimetre interferometers as the superior angular resolution from the latter would make them much 
better-suited to observe compact objects, such as protostellar accretion discs.

\vskip 1mm

\begin{figure}[!h]

\centering \includegraphics[width=135mm]{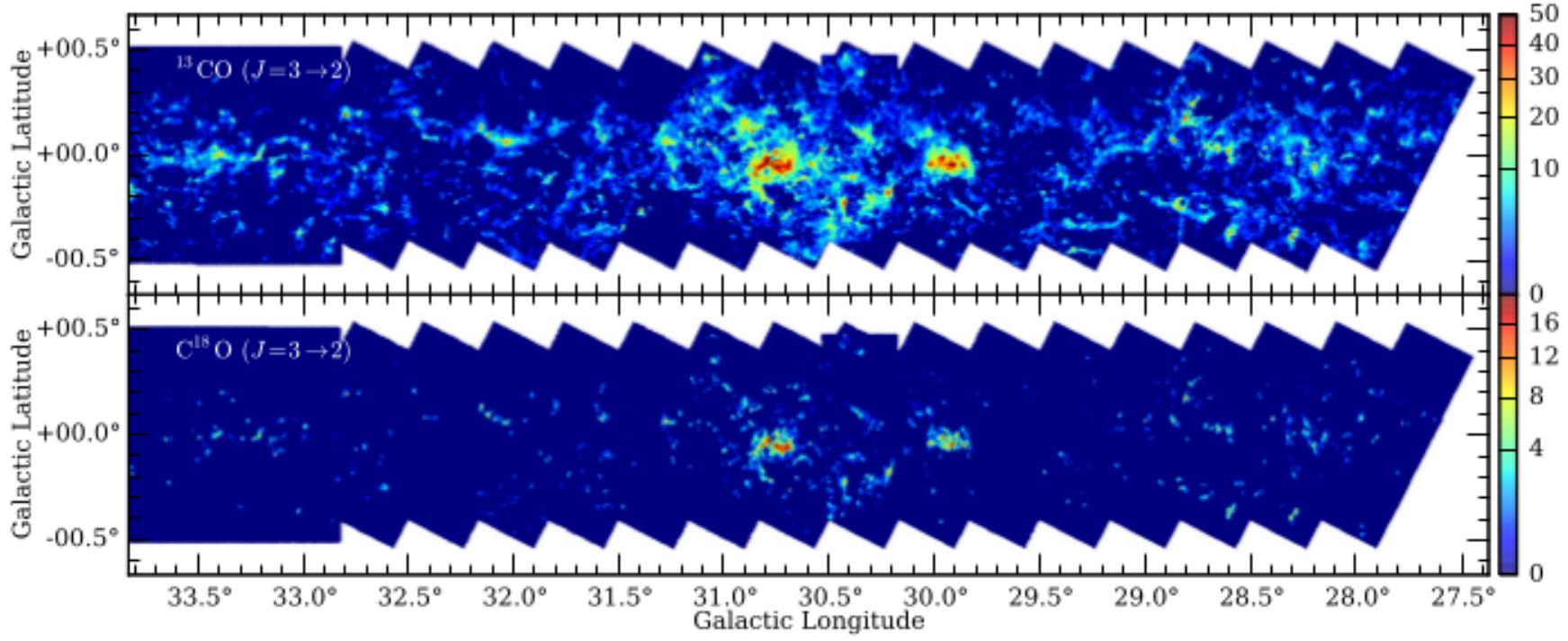}

 \caption{Part of the CHIPMS survey of the Galactic Plane. (top) Velocity integrated emission of $^{13}$CO(3 -- 2), (bottom) The same for C$^{18}$O(3 -- 
2). Figure from Rigby et al. \cite{Rigby2016}.}

\label{fig:Chimps}

\end{figure}

The first generation JCMT Legacy surveys began in 2007, concentrating on heterodyne observations (SCUBA-2 would join the surveys, but not until 2012, as 
discussed in Section 3). Of these surveys, SLS and NGLS have already been described, with a third major survey concentrating on mapping the extent of 
$^{13}$CO and C$^{18}$O(3 -- 2) in a number of molecular clouds in the Gould belt. The results from these surveys have been published in a series of 
papers, including the Orion B region by Buckle et al. \cite{Buckle2010}, the Perseus molecular cloud by Curtis and co-workers \cite{Curtis2010}, Taurus by 
Davis et al. \cite{Davis2010}, Serpens from Graves and co-workers \cite{Graves2010}, Orion A by Buckle et al. \cite{Buckle2012} and the Ophiuchus region by 
White and collaborators \cite{White2015}. As an example of the results, the Buckle et al. paper \cite{Buckle2010} presents temperature, opacity, mass and 
energy content and location of outflow regions in Orion B. Several follow-up papers, including those from Curtis et al. \cite{Curtis2010b} and 
Drabek-Maunder and co-workers \cite{Drabek2016}, went on to analyse the data further in terms of detailing the extent and properties of outflows in the 
regions. In addition, the Perseus region was mapped in the dense gas tracers HCO$^+$ and HCN by Walker-Smith et al. \cite{Walker2014}. Outside of the 
formal legacy surveys, additional surveys included the CO High Resolution Survey (COHRS) by Dempsey et al. \cite{Dempsey2013} that mapped the Galactic 
plane in CO(3 -- 2) within the area 10 < $l$ < 65 and |$b$| < 0.5. The 13CO/C18O(3 -- 2) Heterodyne Inner Milky Way Plane Survey (CHIMPS) published by 
Rigby et al. \cite{Rigby2016} covers the Galactic plane area 28 < $l$ < 46 for |b| < 0.5 (see Fig.~\ref{fig:Chimps}. All of the surveys were obtained with 
a spatial resolution of 14 arcseconds and are publicly available: part of the JCMT's legacy for the future.

\subsection{Star formation: outflows and discs}

Spectroscopy of specific star formation regions (as opposed to entire molecular clouds described above) has also been an extensive area of research for the 
JCMT. Around 200 papers based on JCMT heterodyne data have ``star formation'', ``outflow'' or ``disc'' in the title. The papers cover many aspects from 
compact accretion discs and cores to envelopes and large-scale outflows. Some the most cited papers in the area of star formation with the JCMT are 
discussed below.

\vskip 1mm

Individual outflows have been observed to determine their morphology and physical characteristics. For example, Richer et al. \cite{Richer1992} mapped the 
outflow in Orion B and modelled the outflow as driven by a neutral highly collimated jet, the collimation increasing with velocity in the outflow. Lada and 
Fich 1996 \cite{Lada1996} observed the outflow in NGC\,2264G and again found collimation increasing with outflow velocity. The outflow obeyed a ``Hubble 
law'' with the gas velocity increasing further away from the central source. The NGC 1333/IRAS4 outflow source was studied by Blake et al. (1995) 
\cite{Blake1995} with the authors deducing a depletion of CO and other molecules in the flow, as well as observing the additional presence of SiO in the 
outflow. Indeed, SiO was later identified as an outflow indicator by Nasini et al. \cite{Nisini2007}. Large parts of the NGC1333 cloud complex were 
surveyed by Knee and co-workers \cite{Knee2000} identifying 10 protostellar sources, each of which was found to be driving an outflow. A number of survey 
papers addressed the issue of whether all protostellar and YSOs have outflows. The paper by Parker et al. \cite{Parker1991} surveyed IRAS sources 
representative of low-mass YSOs embedded in dark molecular clouds and found outflows in 70\% of the targets. Other surveys also detected outflows in large 
fractions of the targets, suggesting that many, if not all, such objects have outflows. The survey of the Perseus Cloud complex by Hatchell et al. 
\cite{Hatchell2007} found outflows in 65\% of the 51 SCUBA-identified cores, and all but four of the outflows were also identified by \emph{Spitzer} as 
YSOs. Indeed, only one of the \emph{Spitzer} sources did not have a detected outflow, again showing the almost complete correlation between YSO and 
outflows, and that a large fraction of the cores also have embedded YSOs. Outflows deposit kinetic energy into the circumstellar envelope and cloud, which 
has the potential to stop accretion, disrupt the envelope, and generate turbulence that supports the cloud (or can even disrupt the cloud). Hence, although 
outflows have a clear impact on the star formation process, there is still no clear consensus of just how significant a factor this is.

\vskip 1mm

Protostellar accretion discs were directly studied by using short-baseline interferometry involving the nearby Caltech Submillimeter Observatory (CSO) and 
the Smithsonian Millimeter Array (SMA), as discussed in section 5(f). Without the sub-arcsecond resolution afforded by an interferometer, accretion 
discs are unresolved by the JCMT, and the disc emission needs to be disentangled from that of the ambient cloud. This can be achieved in a number of ways; 
for example, using the chemical or physical characteristics of the region e.g. line line width, or, select sources that have separated from the parent cloud 
either spatially or in velocity space (it is not uncommon that T Tauri stars have left their parent cloud or the cloud has been disrupted). The paper by 
Thi et al. \cite{Thi2001} studied discs around T Tauri and Herbig Ae stars using \emph{ISO} H$_2$ and JCMT CO(3 -- 2) and CO(6 --5) data, selecting sources 
that are spatially separated from their ambient clouds. The H$_2$ emission arises from hot (100 -- 200\,K) gas whilst the CO emission from cooler (20 -- 
80\,K) gas. The lower level CO emission profile was shown to be double peaked, characteristic of a disc in Keplerian rotation. By comparing mass estimates 
from the CO line and continuum, the CO abundance was found to be lower than predicted, which was attributed to freeze-out in the disc centre and 
photo-dissociation on the disc surface. Zadelhoff and co-workers \cite{Zadelhoff2001} observed the sources LkCa 15 and TW Hya in a number of high 
excitation lines, showing that the line emission mainly originated from an intermediate disc layer with high densities of $10^6$ -- $10^8$ cm$^{-3}$ and 
moderately warm temperatures. The authors found evidence for significant freeze-out of CO and HCO$^+$ at low temperatures, but the abundance in the warmer 
upper layer was also low and attributed to photo-dissociation. The first detection of DCO$^+$ in the disc of TW Hya was reported by van Dishoeck et al. 
\cite{Dishoeck2003}. The DCO$^+$/HCO$^+$ ratio was found to be 0.035 $\pm$ 0.015, similar to values in pre-stellar cores. Organic molecules in 
protoplanetary discs surrounding T Tauri and Herbig Ae stars were studied by Thi and collaborators \cite{Thi2004}, using the JCMT for high excitation lines 
and the IRAM 30\,m telescope for low excitation lines. The main conclusions were that abundances were lower compared to the envelopes around protostars. 
The importance of photo-dissociation is shown by the CN/HCN ratio that is found to be higher than in Galactic photo-dominated regions, which has enhanced 
CN/HCN abundance ratio due to photo-dissociation.

\subsection{Solar System: planetary atmospheres and comet chemistry}

The study of Solar System objects using the JCMT heterodyne instruments started with observations of the Sun during the total Solar eclipse over Hawaii on 
11th July 1991. As well as continuum observations of sunspots, spectroscopy of the chromosphere during limb occultation was carried out by Lindsey et al. 
\cite{Lindsey1992} resulting in a measurement of the chromospheric temperature profile in the near-millimetre region. The limited bandwidth of the 
heterodyne instrument reduced the problem with saturation -- sensitivity was not an issue. Other Solar System observations included the first detection of 
CO and HCN in Neptune by Martin et al. \cite{Martin1993}, and of CO in the atmosphere of Pluto at millimetre wavelengths by Bockel\'{e}e-Morvan et al. 
\cite{Bockelee2001} and Greaves et al. \cite{Greaves2011}. The detection of the important catalyst H$_2$O$_2$ in the Martian atmosphere by Clancy and 
co-workers \cite{Clancy2004} was the first such detection in a planetary atmosphere outside that of the Earth. Venus is the planetary atmosphere 
most studied with the JCMT and a dozen papers, including those by Clancy et al. \cite{Clancy2003} and Sandor and collaborators \cite{Sandor2010}, have 
reported observations of temperature structure, wind patterns through the Doppler shifts and atmospheric chemistry and its variability.

\vskip 1mm

Comets are by far the most studied Solar System objects with the JCMT heterodyne instruments, having been observed and monitored from the early 1990s. An 
early influential paper by Senay et al. \cite{Senay1994} highlighted that enough CO gas was observed to explain coma outburst in the most distant comets.  
For comets close to the Sun, the sublimation of water ice is a dominant driver, but the temperature in distant comets is too cold to allow sublimation to 
generate the outbursts. The JCMT has also monitored the gas abundance in a number of comets often in conjunction with other telescopes. The comets 
Hale-Bopp and Hyakutake were studied in a number of highly-cited papers by Biver and collaborators \cite{Biver1997} \cite{Biver2002} \cite{Biver1999}. The 
detection of HNC from Comet Hyakutake by Irvine et al. \cite{Irvine1996} was first seen as evidence for the existence of interstellar ices in comets. The 
HNC/HCN ratio was found to be similar to the interstellar gas phase value and higher than the equilibrium ratio expected in the outermost Solar nebula, 
where comets are thought to be formed. The HNC ratio was later explained by the same authors \cite{Irvine1998} as being due to photo-chemistry in the comet 
coma and not an indication of interstellar origin, whereas isotopic studies supported the view that comets contained pristine unprocessed interstellar 
ices. Meier et al. \cite{Meier1998} reported the third detection of HDO in a long-period, Oort Cloud comet; the data all giving HDO/H$_2$O abundances 
ratios about twice the terrestrial ocean values. Such detections did not support the view that comets supplied the majority of the water for Earth's 
oceans. The detection of DCN in Comet Hale-Bopp by Meier at al \cite{Meier1998b} revealed an even higher deuterium enrichment in HCN of about 6--8 times 
the ratio in H$_2$O. Different deuterium enrichment is a hallmark of interstellar ion-neutral and grain-phase chemistry, whilst it is not expected in 
material processed in denser and warmer part of the Solar nebula. These observations were strong evidence for the presence of interstellar ices in comets.

\subsection{Interferometry: accretion discs and supermassive black holes}

There was early interest in using the JCMT as part of an interferometer, with the first experiment taking place in January 1992. This was a millimetre-wave 
VLBI including the JCMT, Nobeyama, SEST and OVRO. No fringes, however, were found. About the same time the first tests were carried with the Short Baseline 
Interferometer (SBI) between the JCMT and the CSO \cite{Carlstrom1993} \cite{Lay1994}. It operated at the 230, 345 and 460\,GHz bands and was one of the 
first, if not the first interferometer operating at submillimetre wavelengths. The two-element interferometer became a forerunner of the high-frequency 
interferometers operating today, such as the Plateau de Bure, SMA and ALMA. The main scientific contribution of SBI was the detection of accretion discs 
around a number of YSOs. With a resolution of 1 arcsecond or better it could resolve the accretion disc down to a size of $\sim$70\,AU for nearby objects. 
Lay et al. \cite{Lay1994} observed HL Tau and L1551 IRS5, partly resolving the sources with disc major axes of 60\,AU and 80\,AU respectively, whilst the 
minor axis was constrained to be $<$50\,AU. The elongation of the disc was perpendicular to the outflow, as expected. In total, 16 protostellar sources were 
observed including class 0, I and II YSOs, with work by Brown et al. \cite{Brown2000} confirming accretion discs with masses of at least $10^{-2}$\,M$_\odot$. 
Wiedner and collaborators \cite{Weidner2002} also carried out the first submillimetre interferometric observations of Arp 220 in both line and continuum. 
The continuum emission at 342\,GHz clearly was binary in the east--west direction with a separation of 1 arcsecond. The CO(3--2) line showed a binary 
source at some velocities but a more extended structure at other velocities.

\vskip 1mm

The construction of the SMA on Mauna Kea made it possible for both the JCMT and CSO to join the array of eight antennas, with the project becoming known as 
the extended SMA (eSMA). Such an extension would contribute significantly longer baselines whilst affording more sensitivity by doubling the collecting 
area, thus allowing the observation of weaker sources with even higher angular resolution than was possible with the SMA alone. The first science campaign 
occurred in April 2008 with Bottinelli et al. \cite{Bottinelli2009} reporting the first detection of CI absorption towards the lensed system PKS1830--211 
at $z$ = 0.866. The results also showed that it was possible to resolve regions with different CI/CO ratios in the image. Other projects included 
observations of the inner envelope of IRC+10216 by Shinnaga and collaborators \cite{Shinnaga2009} where the HCN maser emission was resolved and 
vibrationally excited KCl masers were detected for the first time. Two regions were observed in the inner envelope; the acceleration zone R < 5 R$_*$ and a 
shell zone with a velocity close to the terminal expansion velocity. The shell zone extends to 30 R$_*$ and has a clumpy structure in HCN(3 -- 2) emission 
in the $v$ = (0, 1$^{1e}$, 0) state. Other papers reported on the formation of circumstellar discs around protostellar objects. One example was the 
dynamical velocity field of IRAS 16293-2422, investigated by Favre et al. \cite{Favre2014} to within 50\,AU from the central object, with the rotation 
deviating from Keplerian due to the disc mass being dynamically significant. Technical development at the SMA together with the pending arrival of ALMA led 
to a gradual decline in the use of eSMA.

\vskip 1mm

\begin{figure}[!h]

\centering \includegraphics[width=70mm]{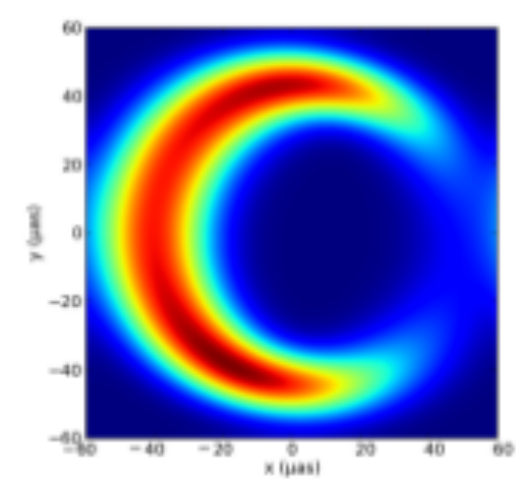}

  \caption{The results from a simple crescent model for the shadow of a black hole, one of a number of models being used to provide a statistical 
description of the existing EHT observations of SgrA*. Figure from Kamruddin \& Dexter \cite{Kamruddin2013}.}

  \label{fig:EHTmodel}

\end{figure}

In April 2007 three telescopes, namely the JCMT, one antenna of the CARMA array in California and the SMT telescope in Arizona, observed Sgr A* at 230\,GHz 
in the Galactic Centre. Doeleman and collaborators \cite{Doeleman2008} reported, for the first time, on resolved structures of the size of the event 
horizon around the super-massive black hole (SMBH) at the centre of the Milky Way. The detected source size of $\sim$40 micro-arcseconds is slightly 
smaller than the expected size of the event horizon of the (presumed) black hole, suggesting that the bulk of the SgrA$^*$ emission may not be centred on 
the black hole, but instead arises in the surrounding accretion flow. The project to image the shadow of the black hole at the Galactic Centre was later 
named the Event Horizon Telescope (EHT). Up until 2013 only the three original telescopes participated in the EHT, but subsequently a phased SMA joined in, 
and currently eight telescope are involved. Observations are now dual-polarisation, whilst other SMHB candidate sources, such as the centre of the Virgo 
cluster M87, have also been observed \cite{Doeleman2012}. Current observations are being interpreted by a range of different models including the geometric 
crescent model proposed by Kamruddin \& Dexter \cite{Kamruddin2013} (see, for example, Fig.~\ref{fig:EHTmodel}), which qualitatively provides an excellent 
statistical description of the existing data. The first real resolved image of the shadow of the black hole at the centre of our Galaxy is expected in the 
next few years.


\vskip 1mm

\begin{figure}[!h] 
\centering \includegraphics[width=100mm]{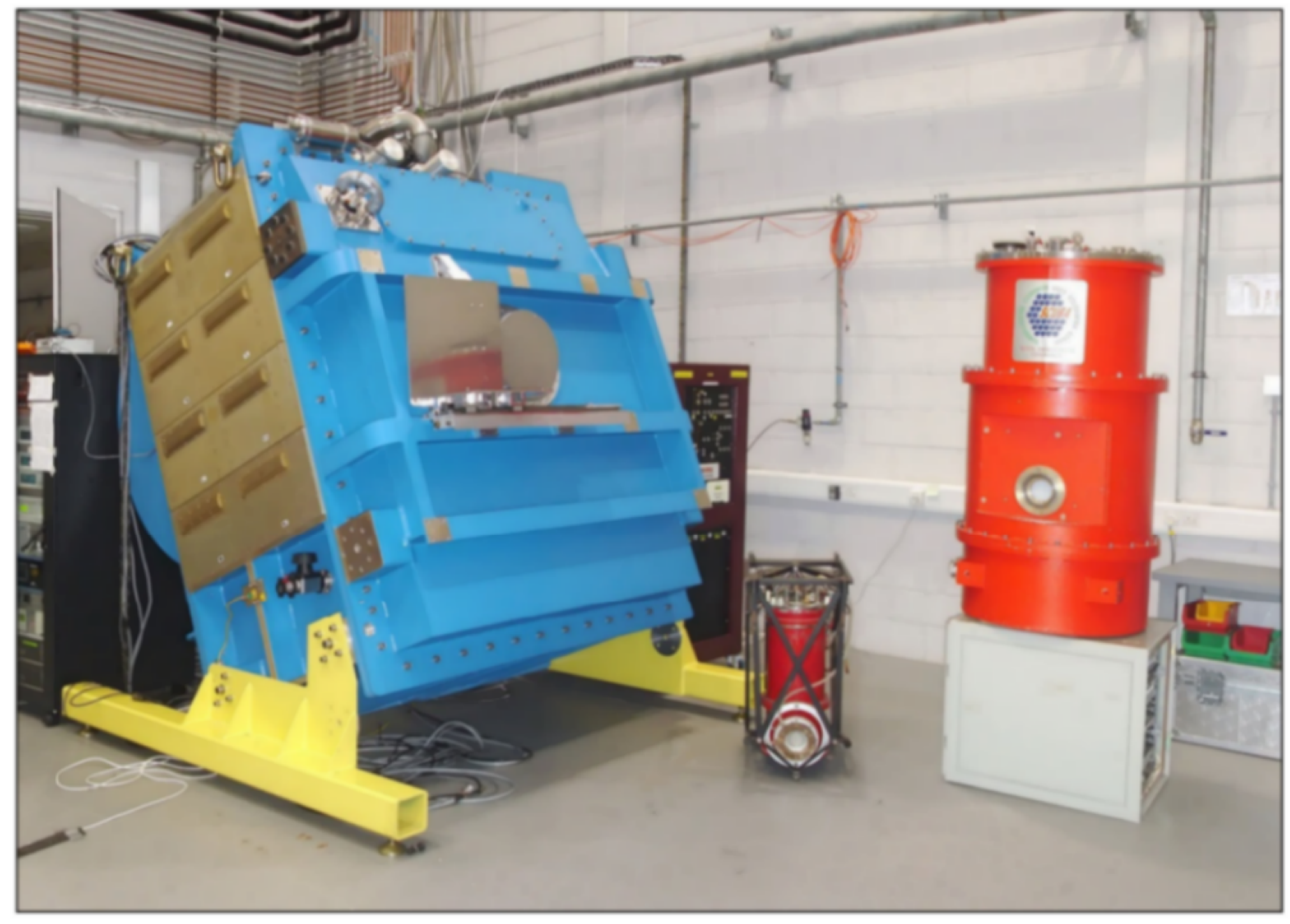} 

\caption{The JCMT continuum instruments together in the ROE Crawford laboratory, prior to the delivery of SCUBA-2 to JCMT (left) SCUBA-2 (10,000 pixels) 
(centre) UKT14 (1 pixel) and (right) SCUBA (128 pixels). SCUBA now resides in the National Museum of Scotland in Edinburgh. Photograph courtesy of the Royal 
Observatory Edinburgh.}

\label{fig:Insts_in_the_lab} 

\end{figure}

\section{Concluding remarks}

\begin{figure}[!h] 
\centering \includegraphics[width=100mm]{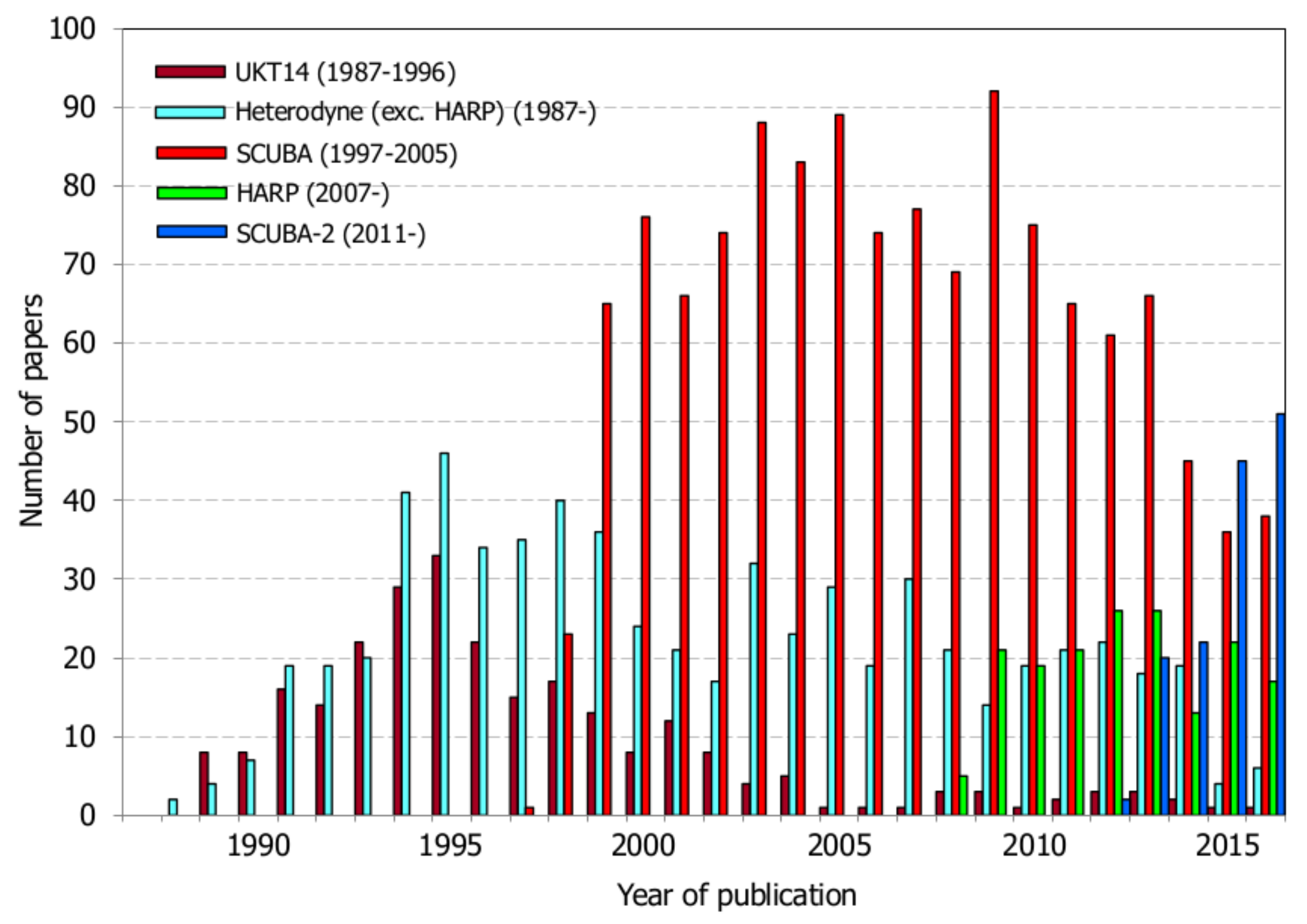} 

\caption{The number of peer-reviewed papers per year until the end of 2016 (information has been gathered from the SAO/NASA Astrophysics Data System). 
The dates in brackets refer to the operational period for the instrument (or range of instruments).}

\label{fig:papers} 

\end{figure}

Looking back on the past history of the telescope it is clear that the scientific impact relied on a number of timely technological innovations in new 
instrumentation. Fig.~\ref{fig:papers} shows the number of (peer-reviewed) papers over the past 30 years split between the three continuum instruments, the 
single/dual pixel heterodyne receivers and the HARP array. The plot is dominated by SCUBA with just over 50\% of the total number of papers. In terms of an 
overall legacy, SCUBA had, without a doubt, one of the biggest impacts of any instrument built for an astronomical telescope. In the period 1997 -- 2005 it 
revolutionised our knowledge in a number of areas of astronomy. In particular, it led to a major advance in the understanding of the astronomical origin 
questions: how planets, stars, and galaxies form. SCUBA revealed discs of cold dust around nearby stars that are evidence that planet formation is ongoing 
or has already occurred. It has detected large numbers of young protostars and ``pre-stellar cores'' -- objects on the brink of becoming stars, making 
possible the first statistical studies of the earliest stage of star formation. Finally, and perhaps most significantly, very shortly after the detection 
of a strong submillimetre background, SCUBA showed that this background is composed of high-redshift, ultraluminous, dusty galaxies. These galaxies have 
all the properties expected for elliptical galaxies in their formation stage, objects which have been looked for in vain for over a decade with optical 
telescopes. Indeed, the paper describing this seminal discovery now has over 1000 citations, making it by far the most cited scientific paper in the 
history of the JCMT. Further evidence of the scientific impact of the JCMT came from an analysis of the productivity/impact of 36 
radio/millimetre/submillimetre telescopes carried out by Trimble \& Zaich \cite{Trimble2006} for the year 2001, which showed that the JCMT was the most 
successful facility with 21.1 citations per paper.

\vskip 1mm

\begin{figure}[!h] 
\centering \includegraphics[width=100mm]{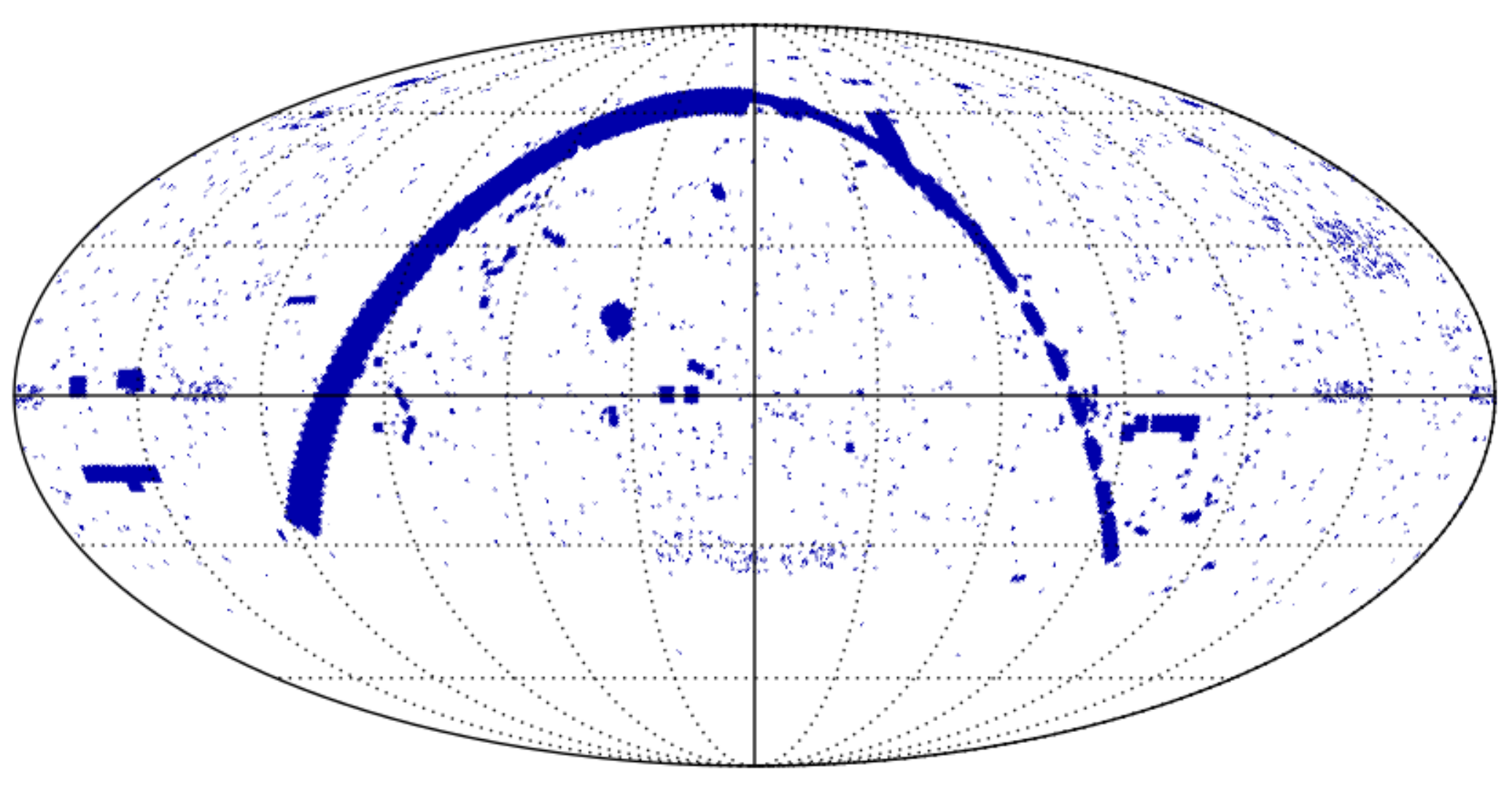} 

\caption{The areas of sky observed by SCUBA-2 up until Dec 2016, dominated by areas in or near the Galactic Plane. Figure courtesy of Graham Bell (EAO).}

\label{fig:sky_coverage} 

\end{figure}

Now in its 30th year of operation, the JCMT continues to produce world-leading science. As of mid-2017 a number of new large scientific programmes have 
recently been awarded time on the telescope, including an extension to the BISTRO magnetic field survey of Gould belt clouds, a new survey to resolve star 
formation in the Galactic plane with HARP, a dust and gas survey of nearby evolved stars, and an extensive study of the Andromeda galaxy. The East Asian 
Observatory is now looking for opportunities to expand the capabilities of the telescope with a series of instrument upgrades over the next five years 
\cite{Dempsey2016}. Of particular importance in this regard is to capitalise on the key strengths of single-aperture telescopes in an era that is becoming 
increasingly dominated by multi-element interferometers (such as the SMA and ALMA). There are initial design plans for a much larger (of order 100 pixel) 
850\,$\upmu$m heterodyne array to replace the current HARP system. Despite the relatively large field-of-view, particularly compared to the predecessor 
instruments on the JCMT, SCUBA-2 has still only covered some 5.3\% of the total sky visible from Mauna Kea (as shown in Fig.~\ref{fig:sky_coverage}). New 
technologies are also emerging that could see SCUBA-2 upgraded with new detectors or indeed replaced by an even larger format (100,000+ pixel) imaging 
camera \cite{Bintley2016}. Finally, a full replacement is planned for the Receiver A (operating at 1.3\,mm) to allow science to continue even when the 
weather is not suitable for submillimetre observations. It is clear from these ambitious plans that the JCMT will have a bright and relevant future.

\section*{Acknowledgment}

For the period 1987 until February 2015 the JCMT was operated by the Joint Astronomy Centre on behalf of the UK Science and Technologies Facilities 
Council (STFC), the Netherlands Organisation for Pure Research, and the National Research Council of Canada. From March 2015 the telescope has been 
operated by the East Asian Observatory on behalf of the National Astronomical Observatory of Japan, the Academia Sinica Institute of Astronomy and 
Astrophysics of Taiwan, the Korea Astronomy and Space Science Institute, the National Astronomica Observatories of China and the Chinese Academy of 
Sciences. Additional operational support funding is also provided by STFC and participating universities in the UK and Canada. This paper has made use of 
NASA's Astrophysics Data System.


\end{document}